\documentclass[preprints,review,accept,universe,moreauthors,10pt,a4paper]{Definitions/mdpi}
\usepackage{amsfonts}
\usepackage{amsmath}
\usepackage{amssymb}
\usepackage{mathrsfs}
\usepackage{mathtools}

\firstpage{1} 
\pubvolume{8}
\issuenum{6}
\articlenumber{318}
\pubyear{2022}
\copyrightyear{2022}
\datereceived{25 April 2022} 
\dateaccepted{4 June 2022} 
\datepublished{7 June 2022} 
\hreflink{https://doi.org/10.3390/universe8060318} 

\usepackage{xspace}

\def\aj{Astron.~J.}

\def\apj{Astrophys.~J.}

\def\phrv{Phys.~Rev.}

\def\prd{Phys.~Rev.~D}

\def\mpla{Mod.~Phys.~Lett.~A}

\def\ijmpd{Int.~J.~Mod.~Phys.~D}

\def\cqgra{Class.~Quant.~Grav. }
\def\gregr{Gen.~Rel.~Grav.}
\def\epjc{Eur.~Phys.~J.~C}
\def\ejph{Eur.~J.~Phys.}

\def\phla{Phys.~Lett.~A}

\def\anphy{Annals~Phys.}
\def\anp{Annalen~Phys.}
\def\cmaph{Commun.~Math.~Phys.}

\def\jmp{J.~Math.~Phys.}

\def\prl{Phys.~Rev.~Lett.}
\def\rvmp{Rev.~Mod.~Phys.}

\def\zap{Zeits.~f.~Astrophys.}

\def\physrep{Phys.~Rep.}

\Title{Covariant Evolution of Gravitoelectromagnetism}

\TitleCitation{Gravitoelectromagnetism}

\Author{A. Danehkar$^{1,2}$\orcidA{}}

\AuthorNames{A. Danehkar}

\AuthorCitation{Danehkar, A.}

\address{%
$^{1}$\quad Department of Astronomy, University of Michigan, Ann Arbor, MI 48109, USA\\
$^{2}$\quad Eureka Scientific, 2452 Delmer St, Oakland, CA 94602, USA; danehkar@eurekasci.com}

\abstract{The long-range gravitational terms associated with tidal forces, frame-dragging effects, and  gravitational waves are described by the Weyl conformal tensor, the traceless part of the Riemann curvature that is not locally affected by the matter field. The Ricci and Bianchi identities provide a set of dynamical and kinematic equations governing the matter coupling and evolution of the electric and magnetic parts of the Weyl tensor, so-called gravitoelectric and gravitomagnetic fields. A detailed analysis of the Weyl gravitoelectromagnetic fields can be conducted using a number of algebraic and differential identities prescribed by the $1+3$ covariant formalism. In this review, we consider the dynamical constraints and propagation equations of the gravitoelectric/-magnetic fields and covariantly debate their analytic properties. We discuss the special conditions under which gravitational waves can propagate, the inconsistency of a Newtonian-like model without gravitomagnetism, the nonlinear generalization to multi-fluid models with different matter species, as well as observational effects caused by the Weyl fields via the kinematic quantities. The $1+3$ tetrad and $1+1+2$ semi-covariant methods, which can equally be used for gravitoelectromagnetism, are briefly explained, along with their correspondence with the covariant formulations.
}


\begin{document}


\section{Introduction}
\label{intro}

In general relativity, the properties of curvatures are described by the Riemann curvature, which can be split into terms including the Ricci tensor defined by the Einstein equations \cite{Einstein1918}, and the traceless part, called the Weyl tensor \cite{Weyl1918,Jordan1960,Jordan2009} constrained by the Ricci and Bianchi identities \cite{Kundt1962,Kundt2016,Truemper1964,Truemper2021}.
The Weyl tensor can also be split into an electric part, called the gravitoelectric field, and a magnetic part, called the gravitomagnetic field, owing to some similarities to their electromagnetic counterparts \cite{Pirani1962a}. 
The analogies between electromagnetism and gravitational fields had been demonstrated in other works prior to the 1960s \cite{Matte1953,Pirani1957,Bel1958a,Bel1958b,Penrose1960,Bel1962,Bel2000}, which enabled the spacetime perturbations to be expressed in terms of the electric and magnetic parts of the traceless Riemann curvature with respect to timelike vectors. 

The Bianchi identities, which are held by the Riemann curvature, can be decomposed into a set of constraint and propagation equations governing the dynamics of the gravitoelectric and gravitomagnetic fields in a form that rather reminds us of the Maxwell equations \cite{Truemper1964,Truemper2021,Hawking1966,Ellis1971,Ellis1971,Ellis1973}. 
However, the Weyl gravitoelectromagnetic fields constrained by the Bianchi identities cannot completely describe a solution of the Einstein equations, so we also need the kinematic quantities that are subject to the Ricci identities \cite{Ehlers1961,Ellis1971,Ehlers1993}. Accordingly, the kinematic constraints of the gravitoelectric/-magnetic fields are provided by the Ricci identities.
In this way, the Bianchi and Ricci identities serve as the main equations for analyzing the dynamics and kinematic evolution of the tidal and frame-dragging fields encoded in the Weyl tensor.

The $1 + 3$ covariant formalism contains algebraic and differential identities that allow us to express exact (non-linear) solutions to the dynamical and kinematic evolution of the Weyl gravitoelectromagnetic fields.
The development of the $1 + 3$ covariant approach began with the work of Heckmann and Sch\"{u}cking \cite{Heckmann1955,Heckmann1956}, Ehlers \cite{Ehlers1961,Ehlers1993}, Kundt and Tr\"{u}mper \cite{Kundt1962,Kundt2016}, and other early works \cite{Truemper1964,Truemper2021,Truemper1965,Ehlers1968}.   
The first covariant analysis of perturbations in this context was carried out by Hawking \cite{Hawking1966}. An elegant covariant form of the Bianchi identities governing the gravitoelectric-/magnetic fields and matter evolution was considered by Ellis \cite{Ellis1971}, which was earlier proposed by Tr\"{u}mper \cite{Truemper1967}. The covariant formulation was then used to study density perturbations \cite{Ellis1989}, the nonlinear dynamics of comic microwave background (CMB) anisotropies \cite{Maartens1999}, and other cosmological models (see the review by \cite{Ellis1999a}). 

The covariant approach has extensively been employed in the studies of the gravitoelectromagnetic fields, whose dynamics and kinematics are constricted by the Bianchi and Ricci identities. The covariant formulations allow the classification of cosmological models, a fluid description of the matter field, and a kinematic description of perturbations in almost-FLRW universes  \cite{Ellis1973,Ellis1999a}. In this literature, we have a covariant study of the evolution of
the gravitoelectric/-magnetic tensors in comparison to Newtonian theory \cite{Ellis1997}, as well as
a proof of the inconsistency of a universe with a purely gravitomagnetic field \cite{Maartens1998} and a covariant analogy between the Bianchi equations and the Maxwell equations \cite{Maartens1998b}.
An improved covariant method was used to demonstrate that the silent universe, where the gravitomagnetic field vanishes, is inconsistent with the exact nonlinear theory \cite{Maartens1997b}. 

It is the aim of this paper to review the applications of the $1 + 3$ covariant formalism for gravitoelectromagnetism. The paper is organized as follows. In Section~\ref{sec2}, we describe the covariant forms of the derivatives of projected vectors and rank-2 tracefree tensors, the kinematic quantities of the fluid, and the dynamic quantities of the matter.
In \S ~\ref{sec3}, we covariantly express the gravitoelectromagnetic fields and other algebraic terms of the curved spacetime. In \S ~\ref{sec4}, we see that the gravitoelectric/-magnetic fields, which are not locally affected by the matter field, are indeed coupled to the dynamic quantities of the matter field via the Bianchi
identities. In \S ~\ref{sec5}, we show how the curls and distortions of the electric and magnetic parts of the Weyl tensor characterize gravitational waves. Section~\ref{sec6} discusses an irrotational purely gravitoelectric dust model, where the gravitomagnetic field vanishes, as well as a purely gravitomagnetic dust model in the absence of the gravitoelectric field.
Section~\ref{sec7} summarizes the covariant formulations for multi-fluid models, followed by a discussion of the tetrad formulation in \S ~\ref{sec8} and the $1 +1 + 2$ semi-covariant formalism in \S ~\ref{sec9}.
In \S ~\ref{sec10}, we discuss observational effects of the Weyl fields induced by the kinematic quantities on electromagnetic waves.

\section{1 + 3 Covariant Formalism}
\label{sec2}

In the $1+3$ covariant approach to general relativity, we replace the spacetime metric with the projected vectors and projected symmetric tracefree (PSTF) tensors, along with the kinematic quantities of the fluid, and the dynamic quantities of the energy-momentum tensor. This formalism started with the
works of Heckmann, Sch\"{u}cking \cite{Heckmann1955}, Raychaudhuri \cite{Raychaudhuri1957}, and Ehlers \cite{Ehlers1961,Ehlers1993} and has been employed for numerous applications in cosmology (see e.g. \cite{Hawking1966,Ellis1989,Ellis1997,Maartens1997b,Maartens1998,Maartens1998b,Maartens1999,Ellis1999a}, also the recent book by \cite{Ellis2012}). 
In this section, we introduce the $1+3$ covariant  mathematics that is necessary for discussing the dynamical and kinematic equations of the gravitoelectric/-magnetic fields. We follow the notations and conventions adopted in the literature \cite{Ellis1971,Lesame1996,Bruni1992,Maartens1997b,Maartens1997c}, in particular, $8\pi G = 1 = c$, round brackets enclosing indices associated with
symmetrization, and square brackets around indices for antisymmetrization.

Various 4-velocity vector fields are typically present in a given region of relativistic $1+3$ spacetime and cosmological models. We choose a timelike 4-velocity field $u^{a}$ to be a unit vector field, i.e. $u^{a}u_{a}=-1$, and use it to perform a $1 + 3$ decomposition of spacetime $g_{ab}$ into the time direction and the 3-dimensional space without questioning the appropriate choice of such a 4-velocity. Accordingly, the metric $g_{ab}$ is projected parallel and orthogonal to $u^{a}$ as follows (see e.g. \cite%
{Ehlers1961,Ehlers1993,Ellis1973}):
\begin{equation}
g_{ab} = h_{ab} - u_{a}u_{b},  \label{eq:e_2_31}
\end{equation}%
where $h_{ab}$ is the \textit{projector tensor} yielding the spatial metric containing 3 space quantities, and $g_{ab}$ and $h_{ab}$ have the following properties:
\begin{align}
& \begin{array}{ccc}
{g_{ab} u^b = u_a,} & {g_{ac} g^{cb}= \delta_{a}{}^{b} ,} & {\delta_a {}^a =
4,}
\end{array} \label{eq:e_2_35} \\
& \begin{array}{ccc}
{h_{ab} u^b = 0,} & {h_a {}^c h_{cb} = h_{ab} ,} & {h_a {}^a = 3.}%
\end{array} \label{eq:e_2_27}
\end{align}%
The spatially \textit{projected alternating tensor} is then defined as follows (also compare with the notations used by \cite%
{Ellis1967,Maartens1995}):
\begin{equation}
\begin{array}{ccc}
{\varepsilon _{abc}=\eta _{abcd}u^{d},} & {\varepsilon_{abc}=\varepsilon_{\lbrack
abc]},}  & {\varepsilon_{abc}u^{c}=0,} 
\end{array} \label{eq:e_2_1}
\end{equation}%
where $\eta _{abcd}$ is the spacetime alternating tensor defined by 
\begin{equation}
\begin{array}{cc}
{\eta _{abcd}=-4!\sqrt{|g|}\delta ^{0}{}_{[a}\delta ^{1}{}_{b}\delta
^{2}{}_{c}\delta ^{3}{}_{d]},} & {g = \mathrm{det}g_{ab},}
\end{array} \label{eq:e_2_2}
\end{equation}%
The above expressions are the basis mathematics for covariant irreducible decomposition of tensors and derivatives, while we have the following practical identities and contractions: 
\begin{align}
& \begin{array}{cc}
{\eta _{abcd}=2u_{[a}\varepsilon _{b]cd}-2\varepsilon_{ab[c}u_{d]},} & {
\varepsilon_{abc}\varepsilon ^{def}=3!h_{[a}{}^{d}h_{b}{}^{e}h_{c]}{}^{f},}
\end{array} \label{eq:e_2_3} \\
& \begin{array}{ccc}
{ \varepsilon _{abf}\varepsilon ^{def}=2!h_{[a}{}^{d}h_{b]}{}^{e}, } & 
{ \varepsilon _{aef}\varepsilon ^{def}=2h_{a}{}^{d}, } & 
{ \varepsilon _{aef}\varepsilon ^{aef}=6.}
\end{array} \label{eq:e_2_28} 
\end{align}%

The projection of any tensor is denoted by a $\bot $ symbol, i.e. ${}_{\bot }T_{ab\cdots d}\equiv h_{a}{}^{c}h_{b}{}^{e}\cdots
h_{d}{}^{f}T_{ce\cdots f}$, where $u^{a}{}_{\bot }T_{ab\cdots d}=0$. The spatially \textit{projected vectors} and  \textit{projected symmetric tracefree} (PSTF) rank-2 tensors are defined as (see Appendix~\ref{app_1} for higher-rank PSTF tensors):
\begin{equation}
\begin{array}{cc}
{V_{\left\langle a\right\rangle }\equiv h_{a}{}^{b}V_{b},} & {%
S_{\left\langle {ab}\right\rangle }\equiv \left\{ {h_{(a}{}^{c}h_{b)}{}^{d}-{%
{\frac{1}{3}}}h^{cd}h_{ab}}\right\} S_{cd},}%
\end{array}
\label{eq:e_2_4}
\end{equation}%
A spatially projected rank-2 tensors $S_{ab}$ can be split into a scalar trace, a projected vector being spatially
dual to the skew part, and a PSTF part: 
\begin{equation}
S_{ab}={{\frac{1}{3}}}Sh_{ab}+\varepsilon _{abc}S^{c}+S_{\left\langle {ab}%
\right\rangle }, \label{eq:e_2_5}
\end{equation}%
where $S\equiv S_{cd}h^{cd}$ is the spatial trace,  and $S_{a} \equiv h_{ab}S^{b} ={{\frac{1}{2}}}\varepsilon _{abc}S^{[bc]}$ is the projected vector dual to the skew part. 

We may also define a vector product and its generalization to rank-2 tensors \cite{Danehkar2009}: 
\begin{equation}
\begin{array}{ccc}
{[V,W]_{a}\equiv \varepsilon _{abc}V^{b}W^{c},} & {[S,Q]_{a}\equiv
\varepsilon _{abc}S^{b}{}_{d}Q^{cd},} & {[V,S]_{ab}\equiv \varepsilon _{cd(a}S_{b)}{}^{c}V^{d}.}
\end{array}
\label{eq:e_2_7}
\end{equation}%

The covariant (spacetime) derivative $\nabla _{a}$ of any tensor can be split into the following \textit{time derivative} and \textit{spatial derivative}, respectively,
\begin{equation}
\begin{array}{cc}
{\dot{T}_{a\cdots }=u^{b}\nabla _{b}T_{a\cdots }, } & {\mathrm{D}_{b}T_{a\cdots }=h_{b}{}^{d}h_{a}{}^{c}\cdots \nabla
_{d}T_{c\cdots }.} 
\end{array}
\label{eq:e_2_8}
\end{equation}%
Following \cite{Maartens1997b}, $\mathrm{D}_{a}$ symbolizes the spatially projected part of the covariant derivative.\footnote{In \cite{Ellis1989,Bruni1992,Challinor1998}, it is shown by ${}^{(3)}\nabla
_{a}$, whereas $\hat{\nabla}_{a}$ in \cite{Maartens1995,Dunsby1997}.} The \textit{Fermi derivatives},
orthogonal projections of time derivatives along $u^{a}$, are then denoted by 
\begin{equation}
\begin{array}{cc}
{\dot{V}_{\left\langle a\right\rangle }\equiv h_{a}{}^{b}\dot{V}_{b},} & {%
\dot{S}_{\left\langle {ab}\right\rangle }\equiv \left\{ {%
h_{(a}{}^{c}h_{b)}{}^{d}-{{\frac{1}{3}}}h^{cd}h_{ab}}\right\} \dot{S}_{cd}.}%
\end{array}
\label{eq:e_2_33}
\end{equation}

The \textit{spatial divergences and curls} are defined as, respectively, \cite{Maartens1997b}
\begin{align}
& \begin{array}{cc}
{\mathrm{div}V\equiv \mathrm{D}^{a}V_{a},} & {\mathrm{(div}S)_{a}\equiv 
\mathrm{D}^{b}S_{ab},}
\end{array} \label{eq:e_2_10} \\
& \begin{array}{cc}
{ \mathrm{curl}V_{a}\equiv \varepsilon _{abc}\mathrm{D}^{b}V^{c},} & 
{ \mathrm{(curl}S)_{ab}\equiv \varepsilon _{cd(a}\mathrm{D}^{c}S_{b)}{}^{d}.} 
\end{array} \label{eq:e_2_11}
\end{align}%
If $S_{ab}=S_{(ab)}$, then $\mathrm{(curl}S)_{ab}=\mathrm{(curl}%
S)_{\left\langle {ab}\right\rangle }$. If $S_{ab}=S_{[ab]}$, we get ($S_{a} \equiv {{\frac{1}{2}}}\varepsilon _{abc}S^{[bc]}$): 
\begin{equation}
\begin{array}{cc}
{\mathrm{curl}S_{ab}=\mathrm{D}_{\left\langle a\right. }S_{\left.
b\right\rangle }-{{\frac{2}{3}}}\mathrm{D}^{c}S_{c}h_{ab},} & {\mathrm{D}%
^{b}S_{ab}=\mathrm{curl}S_{a}}.%
\end{array}
\label{eq:e_2_20}
\end{equation}%
The $\mathrm{div}~\mathrm{curl}$ does not typically vanish for vectors or
rank-2 tensors (see \cite%
{Maartens1997b,Maartens1997a,vanElst1998,Maartens1998b}). We should remind that $\mathrm{D}_{c}h_{ab}=0=\mathrm{D}_{d}\varepsilon
_{abc} $, $\dot{h}_{ab}=2u_{(a}\dot{u}_{b)}$, $\dot{\varepsilon}%
_{abc}=3u_{[a}\varepsilon _{bc]d}\dot{u}^{d}$, and $u^{a}\dot{\varepsilon}%
_{abc}=-\dot{u}^{a}\varepsilon _{abc}$. 
We may also employ the produce and its generalization to define the \textit{temporal rotation} relative to $u^{a}$ as follows 
\begin{equation}
\begin{array}{ccc}
{[\dot{u},V]_{a} \equiv  -u^{c}\dot{\varepsilon}_{abc}V^{b},} & {[\dot{u}%
,S]_{ab} \equiv  -u^{c}\dot{\varepsilon}_{cd(a}S_{b)}{}^{d}.}
\end{array}
\label{eq:e_2_32}
\end{equation}

The \textit{spatial distortions} of vectors and rank-2 tensors are also expressed by \cite{Maartens1997c}
\begin{align}
& \mathrm{D}_{\left\langle a\right. }V_{\left. b\right\rangle }=\mathrm{D}%
_{(a}V_{b)}-{{\frac{1}{3}}}(\mathrm{div}V)h_{ab},  \label{eq:e_2_12} \\
& \mathrm{D}_{\left\langle a\right. }S_{\left. {bc}\right\rangle }=\mathrm{D}%
_{(a}S_{bc)}-{{\frac{2}{5}}}h_{(ab}(\mathrm{divS})_{c)}.  \label{eq:e_2_13}
\end{align}%
As demonstrated by \cite{Maartens1997c}, the covariant derivatives of scalars, vectors, and rank-2 tensors can irreducibly be decomposed into projected algebraic terms: 
\begin{align}
\nabla _{a}f = & -\dot{f}u_{a}+\mathrm{D}_{a}f,  \label{eq:e_2_14} \\
\nabla _{b}V_{a} = &-u_{b}\left\{ {\dot{V}_{\left\langle a\right\rangle }+%
\dot{u}_{c}V^{c}u_{a}}\right\}  \notag \\
&+u_{a}\left\{ {{{\frac{1}{3}}}\Theta
V_{b}+\sigma _{bc}V^{c}+[\omega ,V]_{b}}\right\}  \notag \\
&{+{{\frac{1}{3}}}(\mathrm{div}V)h_{ab}-{{\frac{1}{2}}}\varepsilon _{abc}%
\mathrm{curl}V^{c}+\mathrm{D}_{\left\langle a\right. }V_{\left.
b\right\rangle },}  \label{eq:e_2_15} \\
\nabla _{c}S_{ab} = &-u_{c}\left\{ {\dot{S}_{\left\langle {ab}\right\rangle
}+2u_{(a}S_{b)d}\dot{u}^{d}}\right\} \notag \\
& +2u_{(a}\left\{ {{{\frac{1}{3}}}\Theta
S_{b)c}+S_{b)}{}^{d}(\sigma _{cd}+\varepsilon _{cde}\omega ^{e})}\right\} 
\notag \\
&{+{{\frac{3}{5}}}(\mathrm{div}S)_{\left\langle a\right. }h_{\left.
b\right\rangle c}-{{\frac{2}{3}}}\varepsilon _{dc(a}\mathrm{curl}%
S_{b)}{}^{d}+\mathrm{D}_{\left\langle a\right. }S_{\left. {bc}\right\rangle
},}  \label{eq:e_2_16}
\end{align}%
where the various algebraic terms prescribed by the kinematic quantities, defined below, emerge from the relative motion of comoving observers. 

The \textit{kinematic quantities} of the fluid are defined by \cite{Ehlers1961,Ehlers1993,Ellis1973} 
\begin{align}
\nabla _{b}u_{a} &\equiv \mathrm{D}_{b}u_{a}-\dot{u}_{a}u_{b},  \label{eq:e_2_18} \\
\mathrm{D}_{b}u_{a} &\equiv {{\frac{1}{3}}}\Theta h_{ab}+\sigma _{ab}+\omega _{ab}, 
\label{eq:e_2_19}
\end{align}%
where $\dot{u}_{a} \equiv u^{b}\nabla _{b}u_{a}$ is the acceleration
vector of the fluid motion under forces, $\Theta \equiv \mathrm{D}^{a}u_{a}$ is the expansion scalar, $\sigma _{ab} \equiv \mathrm{D}_{\left\langle a\right. }u_{\left.
b\right\rangle }$ is the traceless symmetric ($\sigma _{ab}=\sigma _{(ab)}$, 
$\sigma _{a}{}^{a}=0$) shear tensor describing the distortion of the matter flow, and $\omega _{ab} \equiv \mathrm{D}_{[a}u_{b]}$ is
the skew-symmetric ($\omega _{ab}=\omega _{\lbrack ab]}$, $\omega
_{a}{}^{a}=0$) vorticity tensor describing the rotation of the matter
relative to a non-rotating frame. The vorticity vector $\omega _{a}$ is also
defined as $\omega _{a}=-{{\frac{1}{2}}}\varepsilon _{abc}\omega ^{bc}$, 
where $\omega _{a}u^{a}=0$, $\omega _{ab}\omega ^{b}=0$ and the
magnitude $\omega ^{2}={{\frac{1}{2}}}\omega _{ab}\omega ^{ab}\geq 0$.
Accordingly, we may also express the vorticity vector as $\omega _{a}=-{{\frac{1}{2}}}\varepsilon _{abc}\mathrm{D}^{b}u^{c}.$\footnote{%
Note that the vorticity vector and tensor here have an
opposite sign compared to definitions often found in some papers (e.g. \cite{vanElst1996a}). Following \cite{Ehlers1961,Ehlers1993,Ellis1971}, the sign convention is similar to the Newtonian definition $%
\vec{\omega}=-{{\frac{1}{2}}}\vec{\nabla}\times \vec{v}$.} 
In the frame of instantaneously comoving observers, we have $\dot{u}_{a}=\dot{u}%
_{\left\langle {a}\right\rangle }$ and $\dot{u}_{a}=-\dot{h}_{ab}u^{b}$. The shear and vorticity products of a rank-2 PSTF tensor can also be written using the produce and its generalization, ${[\sigma ,S]_{a}=\varepsilon _{abc}\sigma ^{b}{}_{d}S^{cd},}$ and $[\omega,S]_{\left\langle {ab}\right\rangle }=\varepsilon _{cd\left\langle a\right.
}S_{\left. b\right\rangle }{}^{c}\omega ^{d}$. 

The vorticity vector can be split into some algebraic terms encoded by the kinematic quantities:
\begin{align}
\nabla _{b}\omega _{a} & =-\omega _{c}\dot{u}^{c}u_{a}u_{b}-\dot{\omega}%
_{\left\langle a\right\rangle }u_{b}-u_{a}\left\{ {\sigma _{bc}\omega ^{c}+{{%
\frac{1}{3}}}\Theta \omega _{b}}\right\} +\mathrm{D}_{b}\omega _{a},
\label{eq:e_2_21} \\
\mathrm{D}_{b}\omega _{a} & =\mathrm{D}_{\left\langle a\right. }\omega _{\left.
b\right\rangle }+{{\frac{1}{3}}}\mathrm{D}^{c}\omega _{c}h_{ab}-{{\frac{1}{2}%
}}\mathrm{curl}\omega _{c}\varepsilon _{ab}{}^{c},  \label{eq:e_2_22}
\end{align}%
where $\dot{\omega}_{\left\langle a\right\rangle }\equiv h_{a}{}^{b}\dot{%
\omega}_{b}$. The relative motion of the fluid, expansion, rotation, and local distortion are described by the kinematic quantities, while the dynamical effects of the energy and momentum are expressed by the dynamic quantities.

The total energy-momentum tensor $T_{ab}$ is decomposed in terms of the \textit{dynamic quantities} as follows \cite{Ehlers1957,Ellis1964}
\begin{equation}
T_{ab}=\rho u_{a}u_{b}+ph_{ab}+2q_{(a}u_{b)}+\pi _{ab},  \label{eq:e_2_17}
\end{equation}%
where $\rho \equiv T_{ab}u^{a}u^{b}$ is the energy density relative
to $u^{a}$, $p \equiv {{\frac{1}{3}}}T_{ab}h^{ab}$ is the pressure, $%
q_{a} \equiv -T_{\left\langle {a}\right\rangle b}u^{b}=-h_{a}{}^{c}T_{cb}u^{b}$ is
the energy flux ($q_{a}u^{a}=0$), and $\pi _{ab} \equiv T_{\left\langle {ab}\right\rangle
}=T_{cd}h^{c}{}_{\left\langle a\right. }h^{d}{}_{\left. b\right\rangle
}=\left( {h^{c}{}_{(a}u^{d}{}_{b)}-{{\frac{1}{3}}}h_{ab}h^{cd}}\right)
T_{cd} $ is the anisotropic stress ($\pi ^{a}{}_{a}=0$, $\pi _{ab}=\pi _{(ab)}$, and $\pi
_{ab}u^{b}=0$). Taking $q^{a}=\pi _{ab}=0$, we have a \textit{perfect fluid%
}, while additionally $p=0$ leads to a pressure-free matter (\textit{dust} or cold
dark matter). A detailed study of analytic equations governing the gravitoelectric/-magnetic fields requires
the use of a number of algebraic and differential identities written in terms of the kinematic and dynamic quantities.

\section{Algebraic Identities of the Curvature}
\label{sec3}

In this section, we discuss the algebraic properties of the Riemann curvature. The gravitational fields are locally described by a set of algebraic relations between the Ricci tensor and the energy-momentum tensor, the so-called Einstein equations. However, the tidal forces and frame-dragging effects, which are not directly influenced by the matter field, are described by the traceless part of the Riemann curvature, called the Weyl tensor. 

The Einstein field equations \cite{Einstein1918} describe how the Ricci curvature is affected by the nearby matter field, which, in the absence of a cosmological constant, take the following form 
\begin{equation}
R_{ab}=T_{ab}-{{\frac{1}{2}}}Tg_{ab},  \label{eq:e_3_38}
\end{equation}%
where $R_{ab} \equiv R^{c}{}_{acb}$ is the Ricci tensor, $R_{abcd}$
is the Riemann curvature, and $T_{ab}$ is the energy-momentum tensor of the matter field ($T \equiv T_{c}{}^{c}$). 
Taking the definition (\ref{eq:e_2_17}) of $T_{ab}$, a successive contraction of the Einstein field equations leads to the following relations:
\begin{align}
R_{ab}u^{a}u^{b} & ={{\frac{1}{2}}}(\rho +3p),  \label{eq:e_3_53} \\
h_{a}{}^{b}R_{bc}u^{c} & =-q_{a},  \label{eq:e_3_54} \\
h_{a}{}^{c}h_{b}{}^{d}R_{cd} & ={{\frac{1}{2}}}(\rho -p)h_{ab}+\pi _{ab}.
\label{eq:e_3_55}
\end{align}%
Additionally, the trace of (\ref{eq:e_3_38}) implies $R=-T$, where $%
R=R_{a}{}^{a} $ and $T=T_{a}{}^{a}=-\rho +3p$.

The Riemann curvature tensor $R_{abcd}$ is decomposed into the Ricci tensor terms and the \textit{Weyl
conformal tensor} $C_{abcd}$ as follows \cite{Weyl1918,Jordan1960,Jordan2009}: 
\begin{equation}
R_{abcd}=C_{abcd}-g_{a[d}R_{c]b}-g_{b[d}R_{c]a}-{{\frac{1}{3}}}%
Rg_{a[c}g_{d]b},  \label{eq:e_3_1}
\end{equation}%
where the Weyl tensor has the following properties
\begin{equation}
\begin{array}{cc}
{C_{abcd}=C_{[ab][cd]},} & {C^{a}{}_{bca}=0=C_{a[bcd]}.}
\end{array}
\label{eq:e_3_3}
\end{equation}
The part of the Riemann curvature that is not directly affected by gravitational sources is represented by the Weyl curvature tensor, which describes tidal forces, frame-dragging effects, and gravitational waves. The Weyl tensor is irreducibly split into the gravitoelectric and gravitomagnetic
fields \cite{Pirani1962a,Pirani1962b} (see also \cite%
{Hawking1966,Ellis1971,Hawking1973}): 
\begin{equation}
\begin{array}{cc}
{E_{ab}=C_{acbd}u^{c}u^{d},} & {H_{ab}={{%
\frac{1}{2}}}\varepsilon _{acd}C^{cd}{}_{be}u^{e}.}%
\end{array}
\label{eq:e_3_4}
\end{equation}%
The gravitoelectric tensor represents a covariant Lagrangian description of tidal forces, while frame-dragging effects are provided by the gravitomagnetic tensor. Both of them support the propagation of gravitational waves. These fields are spacelike and traceless symmetric: 
\begin{equation}
\begin{array}{cccc}
{E_{ab}=E_{(ab)},} & {H_{ab}=H_{(ab)},} & {E_{a}{}^{a}=H_{a}{}^{a}=0,} & {%
E_{ab}u^{b}=H_{ab}u^{b}=0.}%
\end{array}%
\end{equation}%
The gravitoelectric/-magnetic tensors are irreducible, and each has five independent
components. They are, in principle, physically measurable in the frame of a comoving observer in such a way that they are encoded in the Weyl tensor: 
\begin{equation}
C_{ab}{}^{cd}=4\{u_{[a}u^{[c}+h_{[a}{}^{[c}\}E_{b]}{}^{d]}+2\varepsilon
_{abe}u^{[c}H^{d]e}+2\varepsilon ^{cde}u_{[a}H_{b]e}.  \label{eq:e_3_5}
\end{equation}%
In this way, they enable gravitational action at a distance (tidal forces, frame-dragging, and waves) and affect the motion of matter via the geodesic deviation equation \cite%
{Levi-Civita1927,Pirani1956,Pirani1957,Szekeres1965}. 
Along with the Ricci tensor $R_{ab}$ constrained locally by the matter
via the Einstein field equations, the gravitoelectric and gravitomagnetic fields completely describe
the algebraic properties of the Riemann curvature. Accordingly, we may decompose 
the Riemann curvature into its perfect/imperfect matter terms, and gravitoelectric/-magnetic parts, i.e. $R^{ab}{}_{cd}=R_{\mathrm{P}}^{ab}{}_{cd}+R_{\mathrm{I}}^{ab}{}_{cd}+R_{%
\mathrm{E}}^{ab}{}_{cd}+R_{\mathrm{H}}^{ab}{}_{cd}$, as follows \cite{Ellis1999a}
\begin{align}
R_{\mathrm{P}}^{ab}{}_{cd} & ={{\frac{2}{3}}}(\rho
+3p)u^{[a}u_{[c}h^{b]}{}_{d]}+{{\frac{2}{3}}}\rho h^{a}{}_{[c}h^{b}{}_{d]},
\label{eq:e_3_7} \\
R_{\mathrm{I}%
}^{ab}{}_{cd} & =-2u^{[a}h^{b]}{}_{[c}q_{d]}-2u_{[c}h^{[a}{}_{d]}q^{b]}-2u^{[a}u_{[c}\pi {}^{b]}{}_{d]}+2h^{[a}{}_{[c}\pi {}^{b]}{}_{d]},
\label{eq:e_3_8} \\
R_{\mathrm{E}}^{ab}{}_{cd} & =4u^{[a}u_{[c}E^{b]}{}_{d]}+4h^{[a}{}_{[c}E^{b]}{}_{d]},
\label{eq:e_3_9} \\
R_{\mathrm{H}}^{ab}{}_{cd} & =2\varepsilon ^{abe}u_{[c}H_{d]e}+2\varepsilon
_{cde}u^{[a}H^{b]e},  \label{eq:e_3_10}
\end{align}%
where $R_{\mathrm{P}}^{ab}{}_{cd} $ (and $R_{\mathrm{I}}^{ab}{}_{cd}$) is the perfect (and imperfect) matter term, and $R_{\mathrm{E}}^{ab}{}_{cd}$ (and $R_{\mathrm{H}}^{ab}{}_{cd}$) is the gravitoelectric (and gravitomagnetic) curvature part.

\subsection{Spatial Curvature}

We define the \textit{spatial Riemann curvature} as follows \cite{Ellis1990} (see also \cite%
{ZelManov1956a,ZelManov1956b,Cattaneo-Gasperini1961,Ferrarese1963,Ferrarese1965,Lottermoser1988}
for alternative definitions):
\begin{align}
\mathcal{R}^{ab}{}_{cd} &=\bot \left( {R^{ab}{}_{cd}}\right)
-k^{[a}{}_{[c}k^{b]}{}_{d]}  \notag \\
&=h_{q}{}^{a}h_{s}{}^{b}h_{c}{}^{f}h_{d}{}^{p}R^{qs}{}_{fp}-k^{[a}{}_{[c}k^{b]}{}_{d]},
\label{eq:e_3_30}
\end{align}%
where $k_{ab}=\mathrm{D}_{b}u_{a}$ is the relative flow tensor between two
neighboring observers as defined by (\ref{eq:e_2_19}).
 
In irrotational spacetime ($\omega _{a}=0$), $\mathcal{R}^{ab}{}_{cd}$ represents the 3-curvature of the space orthogonal to 
$u_{a}$ with the typical Riemann curvature symmetries. 
In the presence of an irrotational matter fluid, the tangent planes of fundamental observers closely fit together to create spacelike hypersurfaces orthogonal to their
worldlines \cite{Barrow2007}. They are typical of the $u_{a}$-congruence and simultaneously represent the hypersurfaces for all comoving observers.
However, according to Frobenius' theorem, a rotational spacetime does not have such an integrable hypersurface \cite{Wald1984,Poisson2004}, so the observers' planes cannot smoothly go along with each other.

Assuming non-vanishing vorticity, the spatial curvature has the following algebraic properties: 
\begin{equation}
\begin{array}{cc}
{\mathcal{R}_{abcd}=\mathcal{R}_{[ab][cd]},} & {\mathcal{R}%
^{a}{}_{[bcd]}=2k^{a}{}_{[b}\omega _{cd]},}%
\end{array}
\label{eq:e_3_31}
\end{equation}%
\begin{align}
\mathcal{R}_{abcd}-\mathcal{R}_{cdab} =&-{{\frac{2}{3}}}\Theta \left( {%
h_{ac}\omega _{bd}+\omega _{ac}h_{bd}-h_{ad}\omega _{bc}-\omega _{ad}h_{bc}}%
\right)  \notag \\
&-2\left( {\sigma _{ac}\omega _{bd}+\omega _{ac}\sigma _{bd}-\sigma
_{ad}\omega _{bc}-\omega _{ad}\sigma _{bc}}\right) .  \label{eq:e_3_32}
\end{align}%
In the absence of vorticity, we see that $\mathcal{R}_{abcd}=\mathcal{R}_{cdab}$, where the spatial Riemann curvature have the same symmetries of its 4-dimensional counterpart. 

From the spatial Riemann curvature, one may also define the corresponding \textit{spatial Ricci tensor} \cite{Ellis1990}: 
\begin{align}
\mathcal{R}_{ac} &=\mathcal{R}_{a}{}^{b}{}_{cb}=\mathcal{R}%
^{b}{}_{abc}=h^{bd}\mathcal{R}_{abcd}  \notag \\
&=h^{bd}h_{a}{}^{q}h_{b}{}^{s}h_{c}{}^{f}h_{d}{}^{p}R_{qsfp}-\Theta
k_{ac}+k_{ab}k^{b}{}_{c},  \label{eq:e_3_34}
\end{align}%
with the following property \cite{Ellis1990}:
\begin{equation}
\mathcal{R}_{[cb]}={{\frac{1}{3}}}\omega _{bc}\Theta +\left( {\omega
_{db}\sigma _{c}{}^{d}-\omega _{dc}\sigma _{b}{}^{d}}\right) ,
\label{eq:e_3_35}
\end{equation}%
as well as the \textit{spatial Ricci scalar} describing the local scalar of the
space orthogonal to $u_{a}$ \cite{Ellis1990}: 
\begin{equation}
\mathcal{R}=\mathcal{R}^{a}{}_{a}=h^{ab}\mathcal{R}_{ab}=R+2R_{bd}u^{b}u^{d}-%
{{\frac{2}{3}}}\Theta ^{2}+2\sigma ^{2}-2\omega ^{2}.  \label{eq:e_3_36}
\end{equation}%
Substituting them into the Einstein field equations yields the following generalized Friedmann \cite%
{Friedmann1922,Friedmann1999} equation: 
\begin{equation}
\mathcal{R}=2\left( {\rho -{{\frac{1}{3}}}\Theta ^{2}+\sigma ^{2}-\omega ^{2}%
}\right) ,  \label{eq:e_3_37}
\end{equation}%
which depicts how the matter field is associated with the spatial curvature.
In the absence of vorticity ($\omega _{ab}=0$), $\mathcal{R}$ is the Ricci
scalar of the hypersurfaces orthogonal to the fluid lines.

Considering the decomposition of the Riemann curvature, Eqs. (\ref{eq:e_3_7})--(\ref{eq:e_3_10}), 
one could also rewrite the spatial Riemann curvature in terms of the gravitoelectric tensor, along with kinematic and dynamic quantities \cite{Barrow2007},
\begin{align}
\mathcal{R}_{abcd} = &-\varepsilon _{abq}\varepsilon _{cds}E^{qs}+{{\frac{1}{%
3}}}\left( {\rho -{{\frac{1}{3}}}\Theta ^{2}}\right) \left( {%
h_{ac}h_{bd}-h_{ad}h_{bc}}\right)  \notag \\
&+{{\frac{1}{2}}}\left( {h_{ac}\pi _{bd}+\pi _{ac}h_{bd}-h_{ad}\pi
_{bc}-\pi _{ad}h_{bc}}\right)  \notag \\
&-{{\frac{1}{3}}}\Theta \left( {h_{ac}(\sigma _{bd}+\omega _{bd})+(\sigma
_{ac}+\omega _{ac})h_{bd}-h_{ad}(\sigma _{bc}+\omega _{bc})-(\sigma
_{ad}+\omega _{ad})h_{bc}}\right)  \notag \\
&-(\sigma _{ac}+\omega _{ac})(\sigma _{bd}+\omega _{bd})+(\sigma
_{ad}+\omega _{ad})(\sigma _{bc}+\omega _{bc}).  \label{eq:e_3_51}
\end{align}%
Contracting (\ref{eq:e_3_51}) on the first and third indices, we derive the spatial Ricci tensor expressed by 
the so-called \textit{Gauss-Codacci equation} \cite{Barrow2007}: 
\begin{align}
\mathcal{R}_{ab} =&E_{ab}+{{\frac{2}{3}}}\left( {\rho -{{\frac{1}{3}}}%
\Theta ^{2}+\sigma ^{2}-\omega ^{2}}\right) h_{ab}+{{\frac{1}{2}}}\pi _{ab} 
\notag \\
&-{{\frac{1}{3}}}\Theta (\sigma _{ab}+\omega _{ab})+\sigma _{c\left\langle
a\right. }\sigma _{\left. b\right\rangle }{}^{c}-\omega _{c\left\langle
a\right. }\omega _{\left. b\right\rangle }{}^{c}+2\sigma _{c[a}\omega
_{b]}{}^{c}.  \label{eq:e_3_52}
\end{align}%
An additional contraction of the above equation results in the generalized Friedmann equation (\ref{eq:e_3_37}).

\subsection{Commutation Laws}

Using the spatial derivative definition (\ref{eq:e_2_8}), one can arrive at the commutators of the
spatial derivatives for scalars, vectors, and rank-2 tensors. The key point to be considered is that for any scalar function $f$ we have, 
\begin{equation}
\mathrm{D}_{a}\mathrm{D}_{b}f=h_{a}{}^{c}h_{b}{}^{d}\nabla _{c}\mathrm{D}%
_{d}f=h_{a}{}^{c}h_{b}{}^{d}\nabla _{c}\left( {h_{d}{}^{e}\nabla _{e}f}%
\right) .  \label{eq:e_3_39}
\end{equation}%
Applying the Leibniz rule of $\nabla _{a}$ to the last bracket
and using Eq. (\ref{eq:e_2_19}), one obtains the following purely relativistic expression \cite%
{Ellis1990} 
\begin{equation}
\mathrm{D}_{[a}\mathrm{D}_{b]}f=-\omega _{ab}\dot{f}.  \label{eq:e_3_40}
\end{equation}%
The above scalar commutator corresponds to the relativistic behavior of rotating spacetimes in general relativity. 

Similarly, for any vector field $V_{a}$ orthogonal to $u^{a}$ ($V_{a}u^{a}=0$), we derive the following vector commutator,
\begin{equation}
\mathrm{D}_{[a}\mathrm{D}_{b]}V_{c}=-\omega _{ab}\dot{V}_{\left\langle
c\right\rangle }+{{\frac{1}{2}}}\mathcal{R}_{dcba}V^{d}.  \label{eq:e_3_41}
\end{equation}%
Furthermore, for any rank-2 tensor field $S_{ab}$ orthogonal to $u_{a}$ ($%
S_{ab}u^{a}=S_{ab}u^{b}=0$), we obtain 
\begin{equation}
\mathrm{D}_{[a}\mathrm{D}_{b]}S_{cd}= -\omega _{ab}\dot{S}_{\left\langle {cd}\right\rangle }+{{\frac{1}{2}}}\left( {\mathcal{R}_{ecba}S^{e}{}_{d}+%
\mathcal{R}_{edba}S_{c}{}^{e}}\right) .  \label{eq:e_3_42}
\end{equation}%
The above fully nonlinear commutators hold at all perturbative levels.
In the absence of rotation ($\omega _{ab}=0$), $\mathcal{R}_{abcd}$ corresponds to the Riemann curvature of
3-D hypersurfaces orthogonal to the $u_{a}$-congruence.

However, time derivatives do not typically commute with their
spacelike counterparts. In particular, for any scalars, we have the following expression at all perturbative levels \cite{Maartens1999} 
\begin{equation}
\mathrm{D}_{a}\dot{f}-h_{a}{}^{b}\left( {\mathrm{D}_{b}f}\right) ^{\cdot }=%
\dot{f}\dot{u}_{a}+{{\frac{1}{3}}}\Theta \mathrm{D}_{a}f+\mathrm{D}%
_{b}f\left( {\sigma ^{b}{}_{a}+\omega ^{b}{}_{a}}\right) ,  \label{eq:e_3_43}
\end{equation}%
which is the key equation for solving the dynamical evolution of spatial gradients that are covariantly associated with inhomogeneities \cite{Ellis1989}. Similarly, for any vector field $V_{a}$ and tensor field $S_{ab}$  orthogonal to $u^{a}$ to first order, we get:
\begin{align}
\mathrm{D}_{a}\dot{V}_{b}-\left( {\mathrm{D}_{a}V_{b}}\right) _{\bot
}^{\cdot } & ={{\frac{1}{3}}}\Theta \mathrm{D}_{a}V_{b},  \label{eq:e_3_44} \\
\mathrm{D}_{a}\dot{S}_{bc}-\left( {\mathrm{D}_{a}S_{bc}}\right) _{\bot
}^{\cdot } & ={{\frac{1}{3}}}\Theta \mathrm{D}_{a}S_{bc}.  \label{eq:e_3_45}
\end{align}%
Contracting the above equations leads to their divergences to first order: 
\begin{align}
\mathrm{D}^{b}\dot{V}_{b}-\left( {\mathrm{D}^{b}V_{b}}\right) _{\bot
}^{\cdot } & ={{\frac{1}{3}}}\Theta \mathrm{D}^{b}V_{b},  \label{eq:e_3_46} \\
\mathrm{D}^{c}\dot{S}_{bc}-\left( {\mathrm{D}^{c}S_{bc}}\right) _{\bot
}^{\cdot } & ={{\frac{1}{3}}}\Theta \mathrm{D}^{c}S_{bc}.  \label{eq:e_3_47}
\end{align}%
In an almost-FLRW background, the orthogonally projected gradient and time derivative of the
first-order vector $V_{a}$ and spacelike tensor $\dot{S}_{ab}$ possess the linear commutation laws $a\mathrm{D}_{a}\dot{V}_{b}=\left( {a\mathrm{D}_{a}V_{b}}\right) ^{\cdot }$ and $a\mathrm{D}_{a}\dot{S}_{bc}=\left( {a\mathrm{D}_{a}S_{bc}}\right) ^{\cdot }$.

From Eq. (\ref{eq:e_3_40}), for any scalar, it also follows
\begin{equation}
\mathrm{curl}\left( {\mathrm{D}_{a}f}\right) \equiv \varepsilon _{abc}%
\mathrm{D}^{[b}\mathrm{D}^{c]}f=-2\omega _{a}\dot{f},  \label{eq:e_3_48}
\end{equation}%
which depicts the relation between vorticity and non-integrability, implying that in the case of non-zero vorticity there is no constant-time 3-surfaces everywhere orthogonal to $u^{a}$, since the instantaneous rest spaces cannot smoothly mesh each other.

\subsection{Identities in Irrotational Dust Spacetimes}

One of the most commonly used fluids in cosmology is the dust model ($p = q_a
=\pi _{ab}=0$), while the irrotational dust model ($\omega _{ab}=0$) also appears to be useful for the late universe (see \cite{Maartens1997b} for relevant calculations). Here we summarize some properties of irrotational dust spacetimes.

In the case of dust spacetimes, the Ricci tensor is $R_{ab}={{\frac{1}{2}}}\rho (u_{a}u_{b}+h_{ab})$, so the Riemann curvature is written as
\begin{equation}
R^{ab}{}_{cd}=R_{\mathrm{E}}^{ab}{}_{cd}+R_{\mathrm{H}}^{ab}{}_{cd}+{{\frac{2%
}{3}}}\rho u^{[a}u_{[c}h^{b]}{}_{d]}+{{\frac{2}{3}}}\rho
h^{a}{}_{[c}h^{b}{}_{d]}.  \label{eq:e_3_12}
\end{equation}%
The Ricci identities $\nabla _{\lbrack a}\nabla _{b]}f=0$, $2\nabla
_{\lbrack a}\nabla _{b]}V_{c}=R_{abcd}V^{d}$, and $2\nabla _{\lbrack
a}\nabla _{b]}S_{cd}=R^{e}{}_{cab}S_{ed}+R^{e}{}_{dab}S_{ce}$, together with (\ref{eq:e_3_12}) in irrotational spacetime ($\nabla _{b}u_{a}={{\frac{1}{3}}}\Theta h_{ab}+\sigma _{ab}$ and ${\dot{u}_{a}=0}=\omega _{ab}$) lead to the following key identities \cite{Maartens1997b}:
\begin{align}
& (\mathrm{D}_{a}f)^{\cdot } =\mathrm{D}_{a}\dot{f}-{{\frac{1}{3}}}\Theta 
\mathrm{D}_{a}f-\sigma _{a}{}^{b}\mathrm{D}_{b}f,  \label{eq:e_3_14} \\
& \begin{array}{cc}
{\mathrm{D}_{[a}\mathrm{D}_{b]}f=0,} & {\mathrm{curl(D}_{a}f)=0,}%
\end{array}
\label{eq:e_3_28}
\end{align}%
\begin{align}
(\mathrm{D}_{a}V_{b})^{\cdot }= & \mathrm{D}_{a}\dot{V}_{b}-{{\frac{1}{3}}}%
\Theta \mathrm{D}_{a}V_{b}-\sigma _{a}{}^{c}\mathrm{D}_{c}V_{b}+H_{a}{}^{d}%
\varepsilon _{dbc}V^{c},  \label{eq:e_3_16} \\
\mathrm{D}_{[a}\mathrm{D}_{b]}V_{c} =&({{\frac{1}{9}}}\Theta ^{2}-{{\frac{1%
}{3}}}\rho )V_{[a}h_{b]c}-\sigma _{c[a}\sigma _{b]d}V^{d}  \notag \\
&+V_{[b}\left\{ {E_{c]a}-{{\frac{1}{3}}}\Theta \sigma _{c]a}}\right\}
+h_{c[a}\left\{ {E_{b]d}-{{\frac{1}{3}}}\Theta \sigma _{b]d}}\right\} V^{d},
\label{eq:e_3_17} 
\end{align}%
\begin{align}
(\mathrm{D}_{a}S_{bc})^{\cdot }= &\mathrm{D}_{a}\dot{S}_{bc}-{{\frac{1}{3}}}%
\Theta \mathrm{D}_{a}S_{bc}-\sigma _{a}{}^{d}\mathrm{D}%
_{d}S_{bc}+H_{a}{}^{d}\varepsilon _{be(b}S_{c)}{}^{e},  \label{eq:e_3_18} \\
\mathrm{D}_{[a}\mathrm{D}_{b]}S^{cd} =&2({{\frac{1}{9}}}\Theta ^{2}-{{\frac{%
1}{3}}}\rho )S_{[a}{}^{(c}h_{b]}{}^{d)}-2\sigma {}_{[a}{}^{(c}\sigma
_{b]e}S^{d)e}  \notag \\
&-2S_{[b}{}^{(c}\left\{ {E_{c]}{}^{d)}-{{\frac{1}{3}}}\Theta \sigma
_{c]}{}^{d)}}\right\} +h_{[a}{}^{(c}\left\{ {E_{b]e}-{{\frac{1}{3}}}\Theta
\sigma _{b]e}}\right\} S^{d)e}.  \label{eq:e_3_20}
\end{align}
Eq. (\ref{eq:e_3_20}) generalizes the Ricci identities for the commutation of spatial derivatives of rank-2 tensors. There are the further important identities \cite{Maartens1997b}: 
\begin{align}
\mathrm{D}^{2}(\mathrm{D}_{a}f)= & \mathrm{D}_{a}(\mathrm{D}^{2}f)+\mathcal{R}%
_{a}{}^{b}\mathrm{D}_{b}f,  \label{eq:e_3_25} \\
\mathrm{curl}(fS_{ab})= & f\mathrm{curl}(S_{ab})+\varepsilon _{cd(a}S_{b)}{}^{d}%
\mathrm{D}^{c}f,  \label{eq:e_3_22} \\
\mathrm{D}^{b}\mathrm{curl}S_{ab} =&{{\frac{1}{2}}}\varepsilon _{abc}%
\mathrm{D}^{b}\mathrm{(D}_{d}S^{cd})+\varepsilon _{abc}S^{b}{}_{d}\left( {{{%
\frac{1}{3}}}\Theta \sigma ^{cd}-E^{cd}}\right)  \notag \\
&-\sigma _{ab}\varepsilon ^{bcd}\sigma _{ce}S^{e}{}_{d},  \label{eq:e_3_29} \\
\mathrm{curl}(S_{ab})^{\cdot }= & \mathrm{curl}\dot{S}_{ab}-{{\frac{1}{3}}}%
\Theta \mathrm{curl}S_{ab}-\sigma _{e}{}^{c}\varepsilon _{cd(a}\mathrm{D}%
^{e}S_{b)}{}^{d}+3H_{c\left\langle a\right. }S_{\left. b\right\rangle
}{}^{c},  \label{eq:e_3_23} \\
\mathrm{curlcurl}S_{ab} =&-\mathrm{D}^{2}S_{ab}+{{\frac{3}{2}}}\mathrm{D}%
_{\left\langle a\right. }\mathrm{D}^{c}S_{\left. b\right\rangle c}+(\rho -{{%
\frac{1}{3}}}\Theta ^{2})S_{ab}  \notag \\
&+3S_{c\left\langle a\right. }\left\{ {E_{\left. b\right\rangle }{}^{c}-{{%
\frac{1}{3}}}\Theta \sigma _{\left. b\right\rangle }{}^{c}}\right\} +\sigma
_{cd}S^{cd}\sigma _{ab}
\notag \\
&-S^{cd}\sigma _{ca}\sigma _{bd}+\sigma ^{cd}\sigma
_{c(a}S_{b)d},  \label{eq:e_3_24} \\
\varepsilon _{abc}S^{b}{}_{p}\mathrm{curl}Q^{cp}= & 2S^{bc}\mathrm{D}%
_{[a}Q_{b]c}-{{\frac{1}{2}}}S_{ab}\mathrm{D}_{c}Q^{bc},  \label{eq:e_3_21}
\end{align}%
where $\mathcal{R}_{ab}$ is the spatial Ricci tensor defined by (\ref{eq:e_3_52}), and $\mathrm{D}^{2}=\mathrm{D}^{a}\mathrm{D}_{a}$ is the spatial Laplacian operator. Linearized forms ($\sigma_{ab}=E_{ab} = H_{ab} = 0$) of (\ref{eq:e_3_29})--(\ref{eq:e_3_24}) for
the spatial and time derivative of the curl, as well as the curl of the curl can be written, respectively,
\begin{align}
\mathrm{D}^{b}\mathrm{curl}S_{ab}&={{\frac{1}{2}}}\varepsilon _{abc}\mathrm{D}%
^{b}\mathrm{(D}_{d}S^{cd}),  \label{eq:e_3_56} \\ 
\mathrm{curl}(S_{ab})^{\cdot } &=\mathrm{curl}\dot{S}_{ab}-{{\frac{1}{3}}}%
\Theta \mathrm{curl}S_{ab},  \label{eq:e_3_26} \\
\mathrm{curlcurl}S_{ab} & =-\mathrm{D}^{2}S_{ab}+{{\frac{3}{2}}}\mathrm{D}%
_{\left\langle a\right. }\mathrm{D}^{c}S_{\left. b\right\rangle c}+(\rho -{{%
\frac{1}{3}}}\Theta ^{2})S_{ab}.  \label{eq:e_3_27}
\end{align}%
For a dust congruence, we impose $\dot{u}_{a}\equiv \mathrm{D}_{a}\Phi =0$, so
Eq. (\ref{eq:e_3_40}) leads to $\mathrm{D}_{[a}\mathrm{D}_{b]}\Phi =-\omega
_{ab}\dot{\Phi}=0$. Therefore, for the fundamental observers seeing an isotropic radiation field in a dust spacetime, the spacetime is locally FLRW (EGS theorem \cite{Ehlers1968}, see also \cite{Clarkson2000,Stephani1967a,Stephani1967b,Krasinski1997}).\footnote{Clarkson \cite{Clarkson2000} generalized the EGS theorem to an irrotational perfect-fluid model permitting the acceleration, which is not the FLRW spacetime. Stephani \cite{Stephani1967a,Stephani1967b} also considered an irrotational perfect-fluid model that admits an isotropic radiation field for the fundamental observers. These models are FLRW if the acceleration of the
fundamental observers vanishes.}

\section{Analytic Identities for the Weyl Tensor}
\label{sec4}

In this section, we describe the analytic identities that provide dynamical and kinematic constraints for the gravitoelectric/-magnetic parts of the Weyl tensor. The Ricci and Bianchi
identities are the main equations that govern the gravitoelectric and gravitomagnetic fields, while Einstein's equations correspond to the algebraic relation between the Ricci curvature and the matter field. We see that the decomposition of Bianchi identities facilitated by the covariant formalism leads to the dynamical and evolutionary equations for the gravitoelectric/-magneticc tensors.

We consider the \textit{Bianchi identities} as field equations for the free gravitational field,
\begin{equation}
\nabla _{\lbrack e}R_{ab]cd}=0.  \label{eq:e_4_1}
\end{equation}%
As shown by Kundt
and Tr\"{u}mper \cite{Kundt1962,Kundt2016}, the Bianchi identities are equivalent to the following form
(see also \cite{Szekeres1965,Ellis1971,Ellis1973,Hawking1973}) 
\begin{equation}
\nabla ^{d}C_{abcd}=-\nabla _{\lbrack a}(R_{b]c}-{{\frac{1}{6}}}%
g_{b]c}R)=-\nabla _{\lbrack a}(T_{b]c}-{{\frac{1}{3}}}g_{b]c}T_{d}{}^{d}).
\label{eq:e_4_2}
\end{equation}%
Substituting the dynamic quantities (\ref{eq:e_2_17}) into the Bianchi identities and decomposing them
according to the formalism introduced in the previous sections, the following equations are obtained \cite{Truemper1964,Truemper2021,Ellis1971,Ellis1973,vanElst1996a,Ellis1999a}:
\begin{align}
h_{a}{}^{c}h_{b}{}^{d}\nabla ^{b}E_{cd}-3\omega ^{b}H_{ab} &\notag \\
-\eta_{abcd}\sigma ^{b}{}_{e}H^{ec}u^{d} & = {{\frac{1}{3}}}h_{a}{}^{b}\nabla
_{b}\rho -{{\frac{1}{3}}}\Theta q_{a}+{{\frac{1}{2}}}\sigma _{ab}q^{b}  \notag \\
&-{{\frac{3}{2}}}\eta _{abcd}\omega ^{b}q^{c}u^{d}-{{\frac{1}{2}}}%
h_{a}{}^{c}h_{b}{}^{d}\nabla ^{b}\pi _{cd},  \label{eq:e_4_1_1}\\
h_{a}{}^{c}h_{b}{}^{d}\nabla ^{b}H_{cd}+3\omega ^{b}E_{ab} &\notag \\
+\eta_{abcd}\sigma ^{b}{}_{e}E^{ec}u^{d} & = -(\rho +p)\omega _{a}  
-{{\frac{1}{2}}}h_{a}{}^{c}\eta _{bcde}\nabla ^{b}q^{d}u^{e}  \notag \\
&-{{\frac{1}{2}}}\eta _{abcd}\sigma ^{c}{}_{e}\pi ^{ce}u^{d}+{{\frac{1}{2}}}%
\omega ^{b}\pi _{ab},  \label{eq:e_4_1_2}
\end{align}%
\begin{align}
h_{(a}{}^{c}h_{b)}{}^{d}\eta _{cefg}\nabla^{e}H_{d}{}^{f}u^{g}+2h_{(a}{}^{c}\eta _{b)def}\dot{u}%
^{d}H_{c}{}^{e}u^{f} &\notag \\
-h_{a}{}^{c}h_{b}{}^{d}\dot{E}_{cd}-\Theta E_{ab} +h_{(a}{}^{c}\eta _{b)def}\omega ^{d}E_{c}{}^{e}u^{f} & \notag
\\
+3\left( {\sigma
_{c(a}E_{b)}{}^{c}-{{\frac{1}{3}}}h_{ab}\sigma _{cd}E^{cd}}\right) & ={{\frac{1%
}{2}}}\sigma _{ab}(\rho +p)  \notag \\
&+{{\frac{1}{2}}}\left( {h_{(a}{}^{c}h_{b)}{}^{d}\nabla _{c}q_{d}-{{\frac{1%
}{3}}}h_{ab}h_{c}{}^{d}\nabla _{c}q^{d}}\right) \notag \\
& +\left( {\dot{u}_{(a}q_{b)}-{%
{\frac{1}{3}}}h_{ab}\dot{u}_{c}q^{c}}\right) +{{\frac{1}{2}}}%
h_{a}{}^{c}h_{b}{}^{d}\dot{\pi}_{cd}  \notag \\
&+{{\frac{1}{6}}}\Theta \pi _{ab}+{{\frac{1}{2}}}\left( {\sigma _{c(a}\pi
_{b)}{}^{c}-{{\frac{1}{3}}}h_{ab}\sigma _{cd}\pi ^{cd}}\right) \notag \\
& -{{\frac{1}{2}%
}}h_{(a}{}^{c}\eta _{b)def}\omega ^{d}\pi _{c}{}^{e}u^{f},
\label{eq:e_4_1_3} \\
h_{(a}{}^{c}h_{b)}{}^{d}\eta _{cefg}\nabla
^{e}E_{d}{}^{f}u^{g}+2h_{(a}{}^{c}\eta _{b)def}\dot{u}%
^{d}E_{c}{}^{e}u^{f}&\notag \\
+h_{a}{}^{c}h_{b}{}^{d}\dot{H}_{cd}+\Theta H_{ab} -h_{(a}{}^{c}\eta _{b)def}\omega ^{d}H_{c}{}^{e}u^{f} &\notag \\
-3\left( {\sigma_{c(a}H_{b)}{}^{c}-{{\frac{1}{3}}}h_{ab}\sigma _{cd}H^{cd}}\right) &={{\frac{3%
}{2}}}\left( {\omega _{(a}q_{b)}-{{\frac{1}{3}}}h_{ab}\omega _{c}q^{c}}%
\right)  \notag \\
&+{{\frac{1}{2}}}h_{(a}{}^{f}\eta _{b)cdg}u^{c}\sigma _{f}{}^{d}q^{g}+{{%
\frac{1}{2}}}h_{(a}{}^{f}\eta _{b)cde}u^{c}\nabla ^{e}\pi _{f}{}^{d}
\label{eq:e_4_1_4}
\end{align}%
The decomposition of $R_{abcd}$ into $R_{ab}$ and $C_{abcd}$ combined with the once-contracted Bianchi identities yield two constraints and two propagation equations for the gravitoelectric and gravitomagneticc tensors, which are analogous to the Maxwell equations in an expanding spacetime (see \cite{Hawking1966,Ellis1973}).

The \textit{twice contracted Bianchi identities} describe the conservation
of the total energy momentum tensor, 
\begin{equation}
\nabla ^{b}T_{ab}=\nabla ^{b}(R_{ab}-{{\frac{1}{2}}}g_{ab}R)=0, 
\label{eq:e_4_25}
\end{equation}%
which are decomposed according to the covariant formalism as follows \cite{Ellis1973,Ellis1999a}: 
\begin{align}
\dot{\rho}+(\rho +p)\Theta +h_{a}^{b}\nabla _{b}q^{a}+2\dot{u}%
_{a}q^{a}+\sigma _{ab}\pi ^{ab} &=0,  \label{eq:e_4_1_5} \\
(\rho +p)\dot{u}_{a}+h_{a}{}^{b}\nabla _{b}p+h_{a}{}^{b}\dot{q}_{b}+{{\frac{4%
}{3}}}\Theta q_{a}+\sigma _{ab}q^{b} & \notag \\
-\eta _{abcd}\omega ^{b}q^{c}u^{d}+h^{c}{}_{a}h^{d}{}_{b}\nabla ^{b}\pi
_{cd}+\dot{u}^{b}\pi _{ab}& =0.  \label{eq:e_4_1_6}
\end{align}%
In a perfect-fluid model, they are reduced to the typical energy-density and momentum-density conservation laws: 
\begin{align}
\dot{\rho}+(\rho +p)\Theta &=0,  \label{eq:e_4_26} \\
(\rho +p)\dot{u}_{a}+h^{b}{}_{a}\nabla _{b}p &=0.  \label{eq:e_4_27}
\end{align}%
Eq. (\ref{eq:e_4_25}), which is the conservation of total energy-momentum law,
describes how the matter field determines the geometry, i.e., the motion of the matter. Considering the Einstein field equations defined by $G_{ab}\equiv R_{ab}-{{\frac{1}{2}}}g_{ab}R=T_{ab}-%
\Lambda g_{ab}$, the twice-contracted Bianchi identities (\ref{eq:e_4_25}) leads to $\nabla ^{b}G_{ab}=0$ ($\nabla ^{b}\Lambda =0$ if $\Lambda$ is constant in space and time) that describes the dynamical evolution of the spacetime and the matter in it.

The kinematic equations are provided by the \textit{Ricci identities} of the
vector field $u_{a}$, i.e. 
\begin{equation}
2\nabla _{\lbrack a}\nabla _{b]}u_{c}=R_{abcd}u^{d}.  \label{eq:e_4_7}
\end{equation}%
Substituting Eqs. (\ref{eq:e_2_18}) and (\ref{eq:e_2_19}), using
Einstein field equations, and decomposing the orthogonally projected part
into the trace, symmetric trace-free, and skew symmetric terms, as well as the parallel
part, the following constraints are obtained ($\omega ^{2}\equiv {{\frac{1}{2}}}\omega _{ab}\omega ^{ab}$ and $\sigma ^{2}\equiv {{\frac{1}{2}}}\sigma _{ab}\sigma ^{ab}$) \cite{Ellis1999a}: 
\begin{align}
\dot{\Theta}+{{\frac{1}{3}}}\Theta ^{2}= &2(\omega ^{2}-\sigma
^{2})+h_{a}{}^{b}\nabla _{b}\dot{u}^{a}+\dot{u}^{a}\dot{u}_{a}-{{\frac{1}{2}}%
}(\rho +3p),  \label{eq:e_4_8} \\
h_{a}{}^{b}\dot{\omega}_{b}+{{\frac{2}{3}}}\Theta \omega _{a}= & \sigma
_{a}{}^{b}\omega _{b}-{{\frac{1}{2}}}\varepsilon _{abcd}h^{c}{}_{e}\nabla
^{b}\dot{u}^{e}u^{d},  \label{eq:e_4_9} \\
E_{ab} =&\left( {\dot{u}_{a}\dot{u}_{b}-{{\frac{1}{3}}}h_{ab}\dot{u}_{c}%
\dot{u}^{c}}\right) +\left( {h^{c}{}_{(a}h^{d}{}_{b)}\nabla _{c}\dot{u}_{d}-{%
{\frac{1}{3}}}h_{ab}\nabla _{c}\dot{u}^{c}}\right) \notag \\
&-h^{c}{}_{a}h^{d}{}_{b}\dot{\sigma}_{cd}  -{{\frac{2}{3}}}\Theta \sigma _{ab}-\left( {\sigma _{ae}\sigma ^{e}{}_{b}-%
{{\frac{2}{3}}}h_{ab}\sigma ^{2}}\right) \notag \\
& -\left( {\omega _{a}\omega _{b}-{{%
\frac{1}{3}}}h_{ab}\omega _{c}\omega ^{c}}\right) +{{\frac{1}{2}}}\pi _{ab}.
\label{eq:e_4_10}
\end{align}%
Eq. (\ref{eq:e_4_8}) is called the \textit{Raychaudhuri propagation equation} \cite{Raychaudhuri1955} that describes gravitational attraction 
\cite{Ehlers1961,Ehlers1993,Ellis1971,Ellis1973}.
Eq. (\ref{eq:e_4_9}) is called the \textit{vorticity propagation
equation} \cite{Ellis1999a} represents the
vorticity conservation in a perfect-fluid model admiting the acceleration \cite{Ehlers1961,Ehlers1993,Ellis1971,Ellis1973,Ellis1999a}.
Eq.(\ref{eq:e_4_10}) demonstrates how the gravitoelectric tensorial field (i.e., tidal forces \cite{Ellis1971})
directly induces the shear that is then passed into Eqs. (\ref{eq:e_4_8}) and (\ref{eq:e_4_9}) and modifies the fluid flow.

Further kinematic constraints are derived from the Ricci identities \cite{Ellis1999a}: 
\begin{align}
{{\frac{2}{3}}}h^{b}{}_{a}\nabla _{b}\Theta = & h^{c}{}_{a}h^{d}{}_{b}\nabla
^{b}\sigma _{cd}-\eta _{abcd}\nabla ^{b}\omega ^{c}u^{d} \notag \\ 
&-2\eta _{abcd}\dot{u}%
^{b}\omega ^{c}u^{d}+q_{a},  \label{eq:e_4_16} \\
h^{b}{}_{a}\nabla _{b}\omega ^{a}= & \omega _{a}\dot{u}^{a},  \label{eq:e_4_17} \\
H_{ab} =&h^{c}{}_{(a}h^{d}{}_{b)}\eta _{cefg}\nabla ^{e}\sigma
^{f}{}_{d}u^{g}\notag \\
& -\left( {h^{c}{}_{(a}h^{d}{}_{b)}\nabla _{c}\omega _{d}-{{%
\frac{1}{3}}}h_{ab}\nabla _{c}\omega ^{c}}\right)  \notag \\
&{-2\left( {h^{c}{}_{(a}h^{d}{}_{b)}\dot{u}_{c}\omega _{d}-{{\frac{1}{3}}}%
h_{ab}\dot{u}_{c}\omega ^{c}}\right) .}  \label{eq:e_4_18}
\end{align}%
In Eq. (\ref{eq:e_4_16}), we see how the spatial gradient of the expansion
is linked to the spatial curl of the vorticity  and the spatial divergence of the shear. Eq. (\ref{eq:e_4_17}) is called the \textit{vorticity divergence identity} \cite{Ellis1971}. Moreover, the gravitomagnetic tensorial field, describing frame-dragging effects, in Eq. (\ref{eq:e_4_18}) induces the spatial distortion of the vorticity and the spatial curl of the shear.

\subsection{Perfect-fluid Models}
\label{sec4_1}

Using the $1+3$ covariant notations introduced in Sec.\,\ref{sec2}, Eqs. (\ref{eq:e_4_1_1})--(\ref{eq:e_4_1_4}) for a perfect-fluid matter can simply be rewritten as follows: 
\begin{align}
\mathrm{D}^{b}E_{ab}= & 3\omega ^{b}H_{ab}+[\sigma ,H]_{a}+{{\frac{1}{3}}}%
\mathrm{D}_{a}\rho ,  \label{eq:e_4_12} \\
\mathrm{D}^{b}H_{ab}= &-3\omega ^{b}E_{ab}-[\sigma ,E]_{a}-\omega _{a}(\rho
+p),  \label{eq:e_4_13} \\
\mathrm{curl}H_{ab}+2[\dot{u},H]_{\left\langle {ab}\right\rangle }= & \dot{E}%
_{\left\langle {ab}\right\rangle }+\Theta E_{ab}-3\sigma _{c\left\langle
a\right. }E_{\left. b\right\rangle }{}^{c}-[\omega ,E]_{\left\langle {ab}%
\right\rangle }\notag \\
& +{{\frac{1}{2}}}\sigma _{ab}(\rho +p),  \label{eq:e_4_14} \\
\mathrm{curl}E_{ab}+2[\dot{u},E]_{\left\langle {ab}\right\rangle }= & -\dot{H}%
_{\left\langle {ab}\right\rangle }-\Theta H_{ab}+3\sigma _{c\left\langle
a\right. }H_{\left. b\right\rangle }{}^{c}+[\omega ,H]_{\left\langle {ab}%
\right\rangle }.  \label{eq:e_4_15}
\end{align}%
Based on Eq. (\ref{eq:e_4_12}), the spatial gradient of the energy density covariantly emerges as a source for the gravitoelectric field, while the shear and vorticity products of the gravitomagnetic field are associated with the inhomogeneity in the general-relativistic fluid model.
This implies that the gravitoelectric field represents a generalization of the Newtonian tidal force 
\cite{Ellis1971}. In Eq. (\ref{eq:e_4_13}), the gravitomagnetic field originates from the angular momentum density, i.e., $(\rho+p)\omega _{a}$, as well as the shear and vorticity products of the gravitoelectric field.

In a perfect-fluid model, the Ricci equations (\ref{eq:e_4_8})--(\ref{eq:e_4_18}) can also be simplified using the covariant conventions: 
\begin{align}
\dot{\Theta}+{{\frac{1}{3}}}\Theta ^{2}= & \mathrm{D}_{a}\dot{u}^{a}+\dot{u}^{a}%
\dot{u}_{a}+2(\omega ^{2}-\sigma ^{2})-{{\frac{1}{2}}}(\rho +3p),
\label{eq:e_4_19} \\
\dot{\omega}_{\left\langle a\right\rangle }+{{\frac{2}{3}}}\Theta \omega
_{a}= & \sigma _{a}{}^{b}\omega _{b}-{{\frac{1}{2}}}\mathrm{curl}\dot{u}_{a},
\label{eq:e_4_20} \\
E_{ab}= & \mathrm{D}_{\left\langle a\right. }\dot{u}_{\left. b\right\rangle }+%
\dot{u}_{\left\langle a\right. }\dot{u}_{\left. b\right\rangle }-\dot{\sigma}%
_{\left\langle {ab}\right\rangle }-\sigma _{c\left\langle a\right. }\sigma
_{\left. b\right\rangle }{}^{c} \notag \\ 
& -{{\frac{2}{3}}}\sigma _{ab}\Theta -\omega
_{\left\langle a\right. }\omega _{\left. b\right\rangle },  \label{eq:e_4_21} \\
{{\frac{2}{3}}}\mathrm{D}_{a}\Theta = & \mathrm{D}^{b}\sigma _{ab}-\mathrm{curl}%
\omega _{a}-2[\dot{u},\omega ]_{a},  \label{eq:e_4_22} \\
\mathrm{D}^{a}\omega _{a}= & \omega _{a}\dot{u}^{a},  \label{eq:e_4_23} \\
H_{ab}= & \mathrm{curl}\sigma _{ab}-\mathrm{D}_{\left\langle a\right. }\omega
_{\left. b\right\rangle }-2\dot{u}_{\left\langle a\right. }\omega _{\left.
b\right\rangle }.  \label{eq:e_4_24}
\end{align}%
According to Eq. (\ref{eq:e_4_21}), the gravitoelectric tensorial field can directly induce the acceleration distortion and the time derivative of the shear.
It can be seen in Eq. (\ref{eq:e_4_24}) that the vorticity distortion and the shear curl can directly originate from the gravitomagnetic tensorial field (see \cite{Maartens1998b} for more discussions). Although other equations do not involve the gravitoelectric and gravitomagnetic fields, they are important constraints for the kinematic quantities. The Raychaudhuri equation (\ref{eq:e_4_19}) is the only equation that includes the dynamic quantities, which can be employed to describe the gravitational attractive force of the nearby matter (see e.g. \cite{Ehlers1961,Ehlers1993,Ellis1971,Ellis1973}).

\subsubsection{Nonperturbative Shearless Spacetimes}
\label{sec4_1_1}

We now consider small perturbations of the fluid motion and the Weyl gravitoelectric/-magnetic fields in a way that neglects products of small quantities and performs derivatives relative to the undisturbed metric.
In a nonperturbative shearless ($\sigma_{ab}=0$) perfect-fluid model, we avoid perturbations that are merely associated with coordinate transformation and have no physical significance, so Eq. (\ref{eq:e_4_12})--(\ref{eq:e_4_21}) can be rewritten in first order as follows (also compare with \cite{Danehkar2020}):
\begin{align}
& \begin{array}{ccc}
{\mathrm{D}^{b}E_{ab}={{\frac{1}{3}}}\mathrm{D}_{a}\rho,} & {\mathrm{D}^{b}H_{ab}=-\omega _{a}(\rho+p),}
\end{array} \label{eq:e_4_30} \\
&\mathrm{curl}H_{ab}=\dot{E}_{\left\langle {ab}\right\rangle }+\Theta E_{ab},  \label{eq:e_4_32} \\
&\mathrm{curl}E_{ab}=-\dot{H}_{\left\langle {ab}\right\rangle }-\Theta H_{ab},
\label{eq:e_4_33}\\
&\dot{\Theta}+{{\frac{1}{3}}}\Theta ^{2}=\mathrm{D}_{a}\dot{u}^{a}+\dot{u}^{a}%
\dot{u}_{a}-{{\frac{1}{2}}}(\rho +3p),  \label{eq:e_4_34} \\
&\dot{\omega}_{\left\langle a\right\rangle }+{{\frac{2}{3}}}\omega _{a}\Theta
=-{{\frac{1}{2}}}\mathrm{curl}\dot{u}_{a},  \label{eq:e_4_35} \\
&E_{ab}=\mathrm{D}_{\left\langle a\right. }\dot{u}_{\left. b\right\rangle }+%
\dot{u}_{\left\langle a\right. }\dot{u}_{\left. b\right\rangle } -\omega_{\left\langle a\right. }\omega _{\left. b\right\rangle }.
\label{eq:e_4_36}
\end{align}%
It can be seen that the evolution of the gravitoelectric/-magnetic fields or their curls do not
lead to perturbations of the expansion, vortecity, or density according to the first-order dynamical equations in nonperturbative shearless spacetimes (see also \cite{Hawking1966}).

\subsubsection{Irrotational Dust Spacetimes}
\label{sec4_1_2}

Irrotational dust spacetimes have widely been used to study the late universe, as well as for the evolution of density perturbations \cite{Lesame1996} and gravitational waves \cite{Maartens1997b}. Here we rewrite the Bianchi and Ricci equations for an irrotational ($\omega_{ab}=0=\dot{u}_{a}$) dust ($p=0$) model using the covariant formulations, and similar to the nonperturbative shearless model we also perform derivatives with respect to the undisturbed metric, so we have
\begin{align}
&\begin{array}{ccc}
{\mathrm{D}^{b}E_{ab}= [\sigma ,H]_{a}+{{\frac{1}{3}}}\mathrm{D}_{a}\rho,} & {\mathrm{D}^{b}H_{ab}=-[\sigma ,E]_{a},}
\end{array}\label{eq:e_5_1} \\
&\mathrm{curl}H_{ab}=\dot{E}_{\left\langle {ab}\right\rangle }+\Theta
E_{ab}-3\sigma _{c\left\langle a\right. }E_{\left. b\right\rangle }{}^{c}+{{%
\frac{1}{2}}}\rho \sigma _{ab},  \label{eq:e_5_3} \\
&\mathrm{curl}E_{ab}=-\dot{H}_{\left\langle {ab}\right\rangle }-\Theta
H_{ab}+3\sigma _{c\left\langle a\right. }H_{\left. b\right\rangle }{}^{c},
\label{eq:e_5_4} \\
&E_{ab}=-\dot{\sigma}_{\left\langle {ab}\right\rangle }-\sigma
_{c\left\langle a\right. }\sigma _{\left. b\right\rangle }{}^{c}-{{\frac{2}{3%
}}}\sigma _{ab}\Theta ,  \label{eq:e_5_5}\\
& \begin{array}{ccc}
{H_{ab}=\mathrm{curl}\sigma _{ab},} & {\mathrm{D}^{b}\sigma _{ab}={{\frac{2}{3}}}\mathrm{D}_{a}\Theta,}
\end{array} \label{eq:e_5_6}\\
& \begin{array}{ccc}
{\dot{\rho}+\rho \Theta =0,} & {\dot{\Theta}+{{\frac{1}{3}}}\Theta ^{2}=-\sigma ^{ab}\sigma _{ab}-{{\frac{1}{2}}}\rho.}
\end{array}  \label{eq:e_5_8}
\end{align}%
Irrotational dust models with realistic inhomogeneities accommodate both the gravitoelectric and gravitomagnetic fields. In the linearized theory, where the model is almost close to an FLRW spacetime, gravitational waves
in irrotational dust spacetime are typically characterized by transverse traceless tensor modes. We should note that an FLRW spacetime is covariantly described by $\mathrm{D}_{a}\rho =0=\mathrm{D}_{a}\Theta$ and $\sigma
_{ab}=E_{ab}=H_{ab}=0$, whereas the density and expansion gradients and the gravitoelectric/-magnetic tensorial fields are first order of smallness in almost-FLRW spacetimes such as a linearized irrotational model. In the linearized theory, imposing $E_{ab}=0 $ leads to the vanishing of anisotropy and
inhomogeneity, resulting in an FLRW spacetime, where $H_{ab}=0$, so a linearized purely
gravitomagnetic irrotational dust spacetime deos not exist \cite{Maartens1998}.

Covariant calculations can be conducted to show that the constraint equations (\ref{eq:e_5_1}) and (\ref{eq:e_5_6}) are preserved under the evolution in irrotational dust spacetimes.
Let us assign $\mathcal{C}^{A}=0$ (where $A=1,\ldots 
,4 $) to the following constraints:
\begin{align}
\mathcal{C}^{1}{}_{a}\equiv& \mathrm{D}^{b}\sigma _{ab}-{{\frac{2}{3}}}%
\mathrm{D}_{a}\Theta =0,  \label{eq:e_5_11} \\
\mathcal{C}^{2}{}_{ab}\equiv& \mathrm{curl}\sigma _{ab}-H_{ab}=0,
\label{eq:e_5_12} \\
\mathcal{C}^{3}{}_{a}\equiv& \mathrm{D}^{b}E_{ab}-[\sigma ,H]_{a}-{{\frac{1}{3%
}}}\mathrm{D}_{a}\rho =0,  \label{eq:e_5_13} \\
\mathcal{C}^{4}{}_{a}\equiv& \mathrm{D}^{b}H_{ab}+[\sigma ,E]_{a}=0.
\label{eq:e_5_14}
\end{align}%
The time derivatives of the above constraints (${\mathcal{C}}^{A}$) generate a set of
equations that can be denoted by $\dot{\mathcal{C}}^{A}=\mathcal{F}^{A}(\mathcal{C}^{A})$, where the terms $%
\mathcal{F}^{A}$ do not contain any time derivatives, since they are removed
by using the propagation equations and relevant identities \cite%
{Maartens1997b,Maartens1998}: 
\begin{align}
\dot{\mathcal{C}}^{1}{}_{a}=&-\Theta \mathcal{C}^{1}{}_{a}+2[\sigma ,\mathcal{%
C}^{2}]_{a}-\mathcal{C}^{3}{}_{a},  \label{eq:e_5_15} \\
\dot{\mathcal{C}}^{2}{}_{ab}=&-\Theta \mathcal{C}^{2}{}_{ab}-2[\sigma ,%
\mathcal{C}^{1}]_{\left\langle {ab}\right\rangle },  \label{eq:e_5_16} \\
\dot{\mathcal{C}}^{3}{}_{a} =&-{{\frac{4}{3}}}\Theta \mathcal{C}^{3}{}_{a}+{%
{\frac{1}{2}}}\sigma {}_{a}{}^{b}\mathcal{C}^{3}{}_{b}-{{\frac{1}{2}}}\rho 
\mathcal{C}^{1}{}_{b}  \notag \\
&+{{\frac{3}{2}}}E_{a}{}^{b}\mathcal{C}^{1}{}_{b}-[E,\mathcal{C}^{2}]_{a}+{{%
\frac{1}{2}}}\mathrm{curl}\mathcal{C}^{4}{}_{a},  \label{eq:e_5_17} \\
\dot{\mathcal{C}}^{4}{}_{a} =&-{{\frac{4}{3}}}\Theta \mathcal{C}^{4}{}_{a}+{%
{\frac{1}{2}}}\sigma {}_{a}{}^{b}\mathcal{C}^{4}{}_{b}  \notag \\
&+{{\frac{3}{2}}}H_{a}{}^{b}\mathcal{C}^{1}{}_{b}-[H,\mathcal{C}^{2}]_{a}+{{%
\frac{1}{2}}}\mathrm{curl}\mathcal{C}^{3}{}_{a}.  \label{eq:e_5_18}
\end{align}%
Assuming an initial spatial surface $\{t=t_{0}\}$, where $t$ is a proper time along the worldlines, is satisfied by the constraints ${\mathcal{C}}^{A}$, i.e. $\mathcal{C}^{A}|_{t_{0}}=0$, the evolution of the constraints ($\dot{\mathcal{C}}^{A}$) implies that the constraints should be satisfied for all time, as $\mathcal{C}^{A}=0$ is a solution for the initial condition. The constraint set is linear, so the solution should be unique. Thus, the constraints ${\mathcal{C}}^{A}$ are preserved under evolution. For example, if we provide $\sigma _{ab}(t_{0})$ and $\mathrm{D}_{a}\rho (t_{0})$, $\mathcal{C}%
^{1}{}_{a}$ produces $\mathrm{D}_{a}\Theta (t_{0})$, $\mathcal{C}%
^{2}{}_{ab}$ generates $H_{ab}(t_{0})$, and $\mathcal{C}^{3}{}_{a}$
presents $\mathrm{D}^{b}E_{ab}(t{}_{0})$. A consistency condition is expected to be imposed by $\mathcal{C}^{4}{}_{a}$, though this is not the case, since we have 
\begin{equation}
\mathcal{C}^{4}{}_{a}={{\frac{1}{2}}}\mathrm{curl}\mathcal{C}^{1}{}_{a}-%
\mathrm{D}^{b}\mathcal{C}^{2}{}_{ab}.  \label{eq:e_5_19}
\end{equation}%
Let us consider that $\mathrm{D}_{a}\Theta $ is determined by $\mathcal{C}^{1}{}_{a}$, 
$H_{ab}$ by $\mathcal{C}^{2}{}_{ab}$, and $\mathrm{D}_{a}\rho $ by $\mathcal{C}^{3}{}_{a}$, so the constraint equations are consistent with each other owing to the presence of $\mathcal{C}^{4}{}_{a}$. Therefore, if we have a solution to the constraints on $\{t=t_{0}\}$, it is consistent and evolves consistently. Eq. (\ref{eq:e_5_19}) implies that no new
vectorial constraint emerges from the divergence of the tensorial constraint $\mathcal{C}^{2}{}_{ab}$, so the tensorial constraint is characteristically \textit{transverse traceless}.

Supposing the gravitomagnetic tensorial field that is divergence-free (a condition for gravitational waves), the constraint (\ref{eq:e_5_14}) leads to
\begin{equation}
\begin{array}{ccc}
{\mathrm{D}^{b}H_{ab}=0} & \Longleftrightarrow & {[\sigma ,E]_{a}=0.}%
\end{array}
\label{eq:e_5_20}
\end{equation}%
The covariant formulations can be employed to show that the condition (\ref{eq:e_5_20}) is
satisfied under the evolution without considering further conditions. In
particular, the condition (\ref{eq:e_5_20}) does not impose $H_{ab}=0$, so it establishes
spacetimes with $\mathrm{D}^{b}H_{ab}=0\neq H_{ab}$.

\subsection{Imperfect-fluid Models}

We now covariantly write the exact (nonlinear) Bianchi equations in imperfect-fluid models. Substituting the total energy-momentum tensor (\ref{eq:e_2_17}) into the Bianchi equations (\ref{eq:e_4_1_1})--(\ref{eq:e_4_1_4}) and applying the $1+3$ covariant notations yield: 
\begin{align}
\mathrm{D}^{b}E_{ab}=&3\omega ^{b}H_{ab}+[\sigma ,H]_{a}+{{\frac{1}{3}}}%
\mathrm{D}_{a}\rho -{{\frac{1}{3}}}\Theta q_{a}\notag \\
&+{{\frac{1}{2}}}\sigma
_{ab}q^{b}-{{\frac{3}{2}}}[\omega ,q]_{a}-{{\frac{1}{2}}}\mathrm{D}^{b}\pi
_{ab},  \label{eq:e_4_2_1} \\
\mathrm{D}^{b}H_{ab}=&-3\omega ^{b}E_{ab}-[\sigma ,E]_{a}-\omega _{a}(\rho
+p) \notag \\
& -{{\frac{1}{2}}}\mathrm{curl}(q_{a})-{{\frac{1}{2}}}[\sigma ,\pi ]_{a}+{{%
\frac{1}{2}}}\omega ^{b}\pi _{ab},  \label{eq:e_4_2_2} \\
\mathrm{curl}(H_{ab})+2[\dot{u},H]_{\left\langle {ab}\right\rangle }= &\dot{E}%
_{\left\langle {ab}\right\rangle }+\Theta E_{ab}-[\omega ,E]_{\left\langle {%
ab}\right\rangle }-3\sigma _{c\left\langle a\right. }E_{\left.
b\right\rangle }{}^{c} \notag \\
&+{{\frac{1}{2}}}\mathrm{D}_{\left\langle a\right. }q_{\left. b\right\rangle
}+\dot{u}_{\left\langle a\right. }q_{\left. b\right\rangle }+{{\frac{1}{2}}}%
\dot{\pi}_{\left\langle {ab}\right\rangle }+{{\frac{1}{6}}}\Theta \pi _{ab} \notag \\
& -{{\frac{1}{2}}}[\omega ,\pi ]_{\left\langle {ab}\right\rangle }+{{\frac{1}{2}}%
}\sigma ^{e}{}_{\left\langle a\right. }\pi _{\left. b\right\rangle e} +{{\frac{1}{2}}}\sigma _{ab}(\rho +p),
\label{eq:e_4_2_3} \\
\mathrm{curl}(E_{ab})+2[\dot{u},E]_{\left\langle {ab}\right\rangle }=&-\dot{H}%
_{\left\langle {ab}\right\rangle }-\Theta H_{ab}+[\omega ,E]_{\left\langle {%
ab}\right\rangle }+3\sigma _{c\left\langle a\right. }H_{\left.
b\right\rangle }{}^{c} \notag \\ 
&+{{\frac{3}{2}}}\omega _{\left\langle a\right. }q_{\left. b\right\rangle }+{{%
\frac{1}{2}}}[\sigma ,q]_{\left\langle {ab}\right\rangle }+{{\frac{1}{2}}}%
\mathrm{curl}(\pi _{ab}).%
\label{eq:e_4_2_4}
\end{align}%
It can be seen in Eqs. (\ref{eq:e_4_2_1}) and (\ref{eq:e_4_2_2}) that the shear and vorticity coupled with the energy flux and gravitomagnetic field act like sources for the gravitoelectric field, whereas the shear and vorticity coupled with the anisotropic stress and gravitoelectric field appear as sources for the gravitomagnetic field. The long-range gravitational fields, associated with tidal forces, frame-dragging effects, and gravitational waves, especially making tensorial contributions to CMB anisotropies \cite{Maartens1999}, are ruled by Eqs. (\ref{eq:e_4_2_1})--(\ref{eq:e_4_2_4}). The implications of inhomogeneous coupling terms in the nonlinear Bianchi equations are beyond the scope of this review and deserve further discussion (see e.g. \cite{Maartens1998b}).

In irrotational spacetimes ($\omega _{a}=0$) under homogeneous and isotropic conditions ($\sigma
_{ab}=\pi _{ab}=0$), Eqs. (\ref{eq:e_4_2_1})--(\ref{eq:e_4_2_4}) simply reduce to:
\begin{align}
\mathrm{D}^{b}E_{ab}=&{{\frac{1}{3}}}\mathrm{D}_{a}\rho -Hq_{a},
\label{eq:e_4_2_5} \\
\mathrm{D}^{b}H_{ab}=& -{{\frac{1}{2}}}\mathrm{curl}(q_{a}),
\label{eq:e_4_2_6} \\
\mathrm{curl}(H_{ab})+2[\dot{u},H]_{\left\langle {ab}\right\rangle }=&\dot{E}%
_{\left\langle {ab}\right\rangle }+(3H)E_{ab}+{{\frac{1}{2}}}\mathrm{D}%
_{\left\langle a\right. }q_{\left. b\right\rangle }+\dot{u}_{\left\langle
a\right. }q_{\left. b\right\rangle },  \label{eq:e_4_2_7} \\
\mathrm{curl}(E_{ab})+2[\dot{u},E]_{\left\langle {ab}\right\rangle }=&-\dot{H}%
_{\left\langle {ab}\right\rangle }-(3H)H_{ab}.  \label{eq:e_4_2_8}
\end{align}%
where $H=\textstyle{\frac{1}{3}} \Theta$ is the Hubble parameter. 
We see that the gradient of the energy density and the energy flux constrained by the Hubble expansion rate act as sources for the gravitoelectric divergence, while the curl of the energy flux is a source for the gravitomagnetic divergence. Additionally, Eqs (\ref{eq:e_4_2_7}) and (\ref{eq:e_4_2_8}) present wave solutions
if we take the curl of $\mathrm{curl}(H_{ab})$ and the time derivative of $\mathrm{curl}(E_{ab})$, and applying the linearized identities (\ref{eq:e_3_26}) and (\ref{eq:e_3_27}) to them, resulting in $\ddot{H}_{ab}-\mathrm{D}^{2}H_{ab}$. Similarly, the covariant calculations of $\mathrm{curlcurl}(E_{ab})$ and $\mathrm{curl}(\dot{H}_{ab})$ in the linearized forms give a wave solution for the gravitoelectric tensor, i.e., $\ddot{E}_{ab}-\mathrm{D}^{2}E_{ab}$.

\subsubsection{Linearization Scheme}

The exact dynamics of the gravitoelectric/-magnetic tensorial fields with the full matter including flux and anisotropic terms are covariantly governed by the nonlinear Bianchi equations (\ref{eq:e_4_2_1})--(\ref{eq:e_4_2_4}). We may also linearize these equations around any chosen background such as an FLRW
metric. In particular, in the FLRW background, all the inhomogeneous and anisotropic terms ($q_{a}$ and $\pi_{ab}$) vanish, and only the first order of the quantities appear in the linearized Bianchi equations.

Linearization of Eqs. (\ref{eq:e_4_2_1})--(\ref%
{eq:e_4_2_4}) leads to the following system of constraint and propagation equations: 
\begin{align}
\mathrm{D}^{b}E_{ab}= &{{\frac{1}{3}}}\mathrm{D}_{a}\rho -{{\frac{1}{3}}}%
\Theta q_{a}-{{\frac{1}{2}}}\mathrm{D}^{b}\pi _{ab},  \label{eq:e_4_2_11} \\
\mathrm{D}^{b}H_{ab}=&-\omega _{a}(\rho +p)-{{\frac{1}{2}}}\mathrm{curl}%
(q_{a}),  \label{eq:e_4_2_12} \\
\mathrm{curl}(H_{ab})=&\dot{E}_{\left\langle {ab}\right\rangle }+\Theta
E_{ab}+{{\frac{1}{2}}}\sigma _{ab}(\rho +p)\notag \\
&+{{\frac{1}{2}}}\mathrm{D}%
_{\left\langle a\right. }q_{\left. b\right\rangle }+{{\frac{1}{2}}}\dot{\pi}%
_{ab}+{{\frac{1}{6}}}\Theta \pi _{ab},  \label{eq:e_4_2_13} \\
\mathrm{curl}(E_{ab})=&-\dot{H}_{\left\langle {ab}\right\rangle }-\Theta
H_{ab}+{{\frac{1}{2}}}\mathrm{curl}(\pi _{ab}).  \label{eq:e_4_2_14}
\end{align}%
Furthermore, the Ricci identities (\ref{eq:e_4_10}) and (\ref{eq:e_4_18}) according to the linearization scheme are written as follows: 
\begin{align}
E_{ab}=&\mathrm{D}_{\left\langle a\right. }\dot{u}_{\left. b\right\rangle }-%
\dot{\sigma}_{\left\langle {ab}\right\rangle }-{{\frac{2}{3}}}\sigma
_{ab}\Theta +{{\frac{1}{2}}}\pi _{ab},  \label{eq:e_4_2_15} \\
H_{ab}=&\mathrm{curl}\sigma _{ab}-\mathrm{D}_{\left\langle a\right. }\omega
_{\left. b\right\rangle } . \label{eq:e_4_2_16}
\end{align}%
Once again, the gravitoelectric and gravitomagnetic fields originate from the energy density gradient and the angular momentum density in Eqs. (\ref{eq:e_4_2_11}) and (\ref{eq:e_4_2_12}), respectively.
Moreover, we see that  the anisotropic stress gradient and the energy flux constrained by the expansion scalar are additional sources for $(\mathrm{div}E)_{a}$, while the energy ﬂux curl acts as a source for $(\mathrm{div}H)_{a}$. The anisotropic stress $\pi _{ab}$ still remains in the propagation equations (\ref{eq:e_4_2_13}) and (\ref{eq:e_4_2_14}) after linearization. We also have the energy flux distortion in the $\mathrm{curl}(H_{ab})$-equation. The full energy-momentum tensor that includes the flux and stress allows us to investigate inhomogeneities and anisotropies in cosmological models. In particular, the anisotropic stress can be used to analyze the CMB (see e.g. \cite{Maartens1999,Challinor1999,Tsagas2008}).

\subsubsection{Purely Tensorial Perturbations under Linearization}

We now consider the so-called \textit{purely tensorial perturbations%
} \cite{Carloni2007} under a linearization regime, which are characterized by vanishing vector and
scalar variables, i.e. $f=\mathrm{D}_{a}f=\mathrm{D}_{a}\mathrm{D}_{b}f=0$ and $V_{a}=\mathrm{D}_{b}V_{a}=0$ for any scalars $f$ and vectors $V_{a}$. Imposing only tensor perturbations, the linearized equations (\ref{eq:e_4_2_11})--(\ref{eq:e_4_2_16}) shall be: 
\begin{align}
&\begin{array}{cc}
{\mathrm{D}^{b}E_{ab}=-{{\frac{1}{2}}}\mathrm{D}^{b}\pi _{ab},} & {\mathrm{D}^{b}H_{ab}=0,}%
\end{array}
\label{eq:e_4_3_3} \\
& \mathrm{curl}(H_{ab})=\dot{E}_{\left\langle {ab}\right\rangle }+\Theta
E_{ab}+{{\frac{1}{2}}}\sigma _{ab}(\rho +p)+{{\frac{1}{2}}}\dot{\pi}_{ab}+{{%
\frac{1}{6}}}\Theta \pi _{ab},  \label{eq:e_4_3_4} \\
&\mathrm{curl}(E_{ab})=-\dot{H}_{\left\langle {ab}\right\rangle }-\Theta
H_{ab}+{{\frac{1}{2}}}\mathrm{curl}(\pi _{ab}),  \label{eq:e_4_3_5} \\
&\begin{array}{cc}
{E_{ab}=-\dot{\sigma}_{\left\langle {ab}\right\rangle }-{{\frac{2}{3}}}%
\sigma _{ab}\Theta +{{\frac{1}{2}}}\pi _{ab},} & {H_{ab}=\mathrm{curl}\sigma
_{ab}.}%
\end{array}
\label{eq:e_4_3_6}
\end{align}%
The time derivative and curl of equations (\ref{eq:e_4_3_4}) and (\ref{eq:e_4_3_5}%
), as well as the time derivative of (\ref{eq:e_4_3_6}a), together with the conditions $\mathrm{D}^{b}E_{ab}=0$ and $\mathrm{D}^{b}H_{ab}=0$, lead to \cite{Carloni2007} (see \cite{Challinor2000} for detailed calculation):
\begin{align}
\ddot{H}_{ab}-\mathrm{D}^{2}H_{ab}+&{{\frac{7}{3}}}\Theta \dot{H}_{ab}+{{%
\frac{2}{3}}}\left( {\Theta ^{2}-3w\rho }\right) H_{ab}=\left( {\mathrm{curl}%
\pi }\right) _{ab}^{\cdot }+{{\frac{2}{3}}}\Theta \left( {\mathrm{curl}\pi }%
\right) _{ab},  \label{eq:e_4_3_7} \\
\ddot{E}_{ab}-\mathrm{D}^{2}E_{ab}+&{{\frac{7}{3}}}\Theta \dot{E}_{ab}+{{%
\frac{2}{3}}}\left( {\Theta ^{2}-3w\rho }\right) E_{ab}-{{\frac{1}{6}}}%
\Theta \rho \left( {1+w}\right) \left( {1+2c_{s}^{2}}\right) \sigma _{ab} \notag \\ 
&=-{{\frac{1}{2}}}\ddot{\pi}_{ab}+{{\frac{1}{2}}}\mathrm{D}^{2}\pi _{ab}-{{%
\frac{5}{6}}}\Theta \dot{\pi}_{ab}-{{\frac{1}{3}}}\left( {\Theta ^{2}-\rho }%
\right) \pi _{ab},%
\label{eq:e_4_3_8} \\
\ddot{\sigma}_{ab}-\mathrm{D}^{2}\sigma _{ab}+&{{\frac{5}{3}}}\Theta \dot{%
\sigma}_{ab}+\left( {{{\frac{1}{9}}}\Theta ^{2}+{{\frac{1}{6}}}\left( {1-9w}%
\right) \rho }\right) \sigma _{ab}=-\dot{\pi}_{ab}-{{\frac{2}{3}}}\Theta \pi
_{ab},  \label{eq:e_4_3_9}
\end{align}%
where $c_{s}$ is the adiabatic sound speed of the fluid flow, and $w$ is the effective barotropic index \cite{Carloni2007} ($c_{s}^{2}=\dot{p}/\dot{\rho}$ and $w=p/\rho$ in FLRW spacetimes \cite{Bruni1992}).

Solving Eqs. (\ref{eq:e_4_3_7})--(\ref{eq:e_4_3_9}) yields the
evolution of tensor perturbations. Note that Eq. (\ref{eq:e_4_3_8}) is effectively third order due to the presence of a term with $\sigma _{ab} $ coupled with the energy density, which makes it impossible to write a closed wave equation for $E_{ab}$. However, taking $\pi _{ab} = 0$ for consistency, Eqs. (\ref{eq:e_4_3_7}) and (\ref{eq:e_4_3_8} ) should result in wave solutions since the shear also maintains a wave solution provided by Eq. (\ref{eq:e_4_3_9}).

\section{Gravitational Waves}

\label{sec5}

Gauge-invariant transverse tensor perturbations in homogeneous and isotropic FLRW models are known as gravitational waves. The wave solutions of the gravitoelectric/-magnetic tensorial fields were first introduced by Hawking \cite{Hawking1966} in a model, where perturbations of the Weyl tensor do not originate from rotational or density perturbations, and were later covariantly extended by Ellis and Bruni \cite{Ellis1989} in the gauge-invariant perturbation approach. 

Considering the divergence equations (\ref{eq:e_4_30}), to have a wave solution, we require the condition: 
\begin{equation}
\begin{array}{cc}
{\mathrm{D}^{b}E_{ab}=0,} & {\mathrm{D}^{b}H_{ab}=0.}%
\end{array}
\label{eq:e_4_37}
\end{equation}%
Taking a time derivative from the $\mathrm{curl}H_{ab}$ equation (\ref{eq:e_4_32}) and a spatial curl from the $\mathrm{curl}E_{ab}$ equation (\ref{eq:e_4_33}), after simplification, we have \cite{Hogan1997,Dunsby1997a}
\begin{align}
\ddot{E}_{ab}-\mathrm{D}^{2}E_{ab}+{{\frac{7}{3}}}\Theta \dot{E}%
_{ab}+E_{ab}\left( {\dot{\Theta}+{{\frac{4}{3}}}\Theta ^{2}+{{\frac{1}{3}}}%
(\rho -3p)}\right) & \notag \\
+\sigma _{ab}\left( {{{\frac{1}{3}}}\Theta (\rho +p)+{{\frac{1}{2}}}(\dot{%
\rho}+\dot{p})}\right) & =0.  \label{eq:e_4_38}
\end{align}%
In empty, non-expanding ($\Theta=0$) space, it reduces to the typical form seen in the
linearized theory, i.e. $\nabla^{2}E_{ab}\equiv \nabla_{a}\nabla^{a}E_{ab}=0$.
From Eq. (\ref{eq:e_4_38}), it follows that that a necessary covariant condition for the propagation of gravitational waves should be: 
\begin{equation}
\mathrm{curl}E_{ab} \neq 0 \neq \mathrm{curl}H_{ab}.
\label{eq:e_4_40}
\end{equation}%
Therefore, the vanishing divergences and non-vanishing curls of the gravitoelectric and gravitomagnetic tensors are the necessary covariant conditions for the existence of gravitational
waves. The gravitomagnetic field obviously has an essential role in supporting gravitational waves.  
Based on the similarity with electromagnetism, where neither the electric nor magnetic fields independently provide a full description of electromagnetic waves, it was argued that both the gravitoelectric and gravitomagnetic fields are simultaneously necessary for supporting tensor perturbations \cite{Ellis1997a}.
Indeed, as seen in Eq. (\ref{eq:e_4_38}), gravitational waves are exactly characterized by the spatial curls of the electric and magnetic parts of the Weyl tensor.

The nonlinear evolution of $E_{ab}$ and $H_{ab}$, with a perfect-fluid source are determined from Eqs. (\ref{eq:e_4_14}) and (\ref{eq:e_4_15}). The only differences between these two equations 
are the signs of the right sides, besides the energy density and pressure coupled with the shear in the $\mathrm{curl}H_{ab}$ equation. Once they are linearized about an FLRW background, they reduce to Eqs. (\ref{eq:e_4_32}) and (\ref{eq:e_4_33}). On taking the time-derivative of the former and substituting
the curl of the latter, the $\mathrm{curl}E_{ab}$ and $\mathrm{curl}H_{ab}$ terms support the propagation of gravitational waves. However, Eqs. (\ref{eq:e_4_14}) and (\ref{eq:e_4_15}) do not close up in general, owing to the presence of the ${{\frac{1}{2}}}(\rho +p)\sigma _{ab}$ term in the $\mathrm{curl}H_{ab}$ equation.
It is necessary to add the shear evolution equation to the $\mathrm{curl}H_{ab}$ equation to obtain a third-order equation for $E_{ab}$, once its time derivative is calculated (see e.g. \cite{Dunsby1997a}).

In the linearization scheme, purely tensorial perturbations are obtained by requiring the density perturbations and rotational perturbations vanish to first order. The condition that the terms on the right hand side vanish from the constraint (\ref{eq:e_4_30}) is similar to the transverse condition of tensorial perturbations in the metric approach. We remind that the Weyl tensor is the traceless part of the Riemann tensor, so both $E_{ab}$ and $H_{ab}$ are traceless, again similar to the tensorial perturbations of the Bardeen formalism \cite{Bardeen1980}.\footnote{%
In the Bardeen formalism \cite{Bardeen1980}, the equation of motion is $\ddot{H}%
_{T}^{(2)}+{\textstyle{\frac{{2\dot{\ell}}}{\ell }}}\dot{H}%
_{T}^{(2)}+(k^{2}+2K)H_{T}^{(2)}=0$ that is derived for the first-order gauge-invariant amplitude of the tensor perturbation $H_{T}^{(2)}$ in the absence of anisotropic stress perturbations, where $\ell $ is the
cosmological scale factor, $K$ is the spatial curvature of the FLRW background, and $k$ is the physical wavenumber with $K=0$ \cite{Harrison1967}. The metric tensor perturbation is $H_{T}^{(2)}Q_{\mu \nu }^{(2)}$,
where $Q_{\mu \nu }^{(2)}$ are polarization tensors (eigenfunctions) of the Helmholtz equation. Requiring the
transverse ($\nabla ^{\nu }Q_{\mu \nu }^{(2)}=0$) and traceless ($Q^{(2)\mu}{}_{\mu }=0$) polarization conditions permit two degrees of freedom for $Q_{\mu \nu }^{(2)}$.} 

As the condition (\ref{eq:e_4_37}) is required for gravitational waves, one might expect that both $E_{ab}$ and $H_{ab}$ possess the SO(2) electric-magnetic symmetry (see the review by \cite{Danehkar2019}), so their propagation equations should be invariant under the transformation $E_{ab}\Leftrightarrow H_{ab}$. However, this is not generally true; indeed, it is valid when the equation of state of the background satisfies special conditions.
In particular, the evolution equations for $E_{ab}$ and $H_{ab}$ are not of the same order, the $E_{ab}$ wave equation is of third order, whereas $H_{ab}$ has a second-order wave equation. However, both of them reduce to  second order if we suppose $(\rho +p)=0$, i.e. in vacuum (also de Sitter spacetime), as well as for the equation of state $p=-{{\frac{1}{3}}}\rho $. Thus, under those special conditions, both $E_{ab}$ and $H_{ab}$
provide the second-order wave solutions \cite{Dunsby1997a}, which also have the SO(2) electric-magnetic invariance \cite{Danehkar2019}.

\subsection{Distortions of the gravitoelectromagnetic fields}

The distortion of the curvature is a part of its derivative that is dissociated from the Bianchi identities and is not locally connected to the matter. The divergences of the locally free fields $E_{ab} $ and $H_{ab}$ are pointwise connected to the dynamic quantities of the matter, such as the energy density gradient. 
The curl of $E_{ab} $ (or $H_{ab}$) are pointwise associated with the time derivative of $H_{ab}$ (or $E_{ab}$) and the matter terms. Additionally, the distortion of $E_{ab}$ (or $H_{ab}$), similar to the curl, covariantly characterizes a locally free part of the spacetime gradient of the gravitational field. It was shown by \cite{Maartens1997c} that the spatial Laplacian ($\mathrm{D}^2 \equiv \mathrm{D}^{a}\mathrm{D}_{a}$), which is essential for describing a wave solution, can be produced by the distortion. Thus, the non-vanishing distortions of $E_{ab} $ and $H_{ab}$, along with the non-zero curls, is another condition for the propagation of gravitational waves.

Considering the decomposition (\ref{eq:e_2_16}) of the covariant distortion of rank-2 tensors, the covariant and spatial distortions of $E_{ab}$ are split into various terms enclosed by the kinematic quantities  \cite{Maartens1997c}: 
\begin{align}
\nabla _{c}E_{ab} =&-u_{c}\left\{ {\dot{E}_{\left\langle {ab}\right\rangle
}+2u_{(a}E_{b)d}\dot{u}^{d}}\right\}  \notag \\
&+2u_{(a}\left\{ {E_{b)}{}^{d}(\sigma _{cd}+\varepsilon _{cde}\omega ^{e})+{%
{\frac{1}{3}}}\Theta E_{b)c}}\right\} +\mathrm{D}_{a}E_{bc},
\label{eq:e_4_50} \\
\mathrm{D}_{a}E_{bc}=&{{\frac{3}{5}}}(\mathrm{div}E)_{\left\langle a\right.
}h_{\left. b\right\rangle c}-{{\frac{2}{3}}}\varepsilon _{dc(a}\mathrm{curl}%
E_{b)}{}^{d}+\hat{E}_{cab},  \label{eq:e_4_51}
\end{align}%
where $\hat{E}_{cab}$ is the symmetric, traceless gravitoelectric distortion defined by $\hat{E}_{cab} = \hat{E}_{(cab)}\equiv \mathrm{D}_{\left\langle c\right.
}E_{\left. {ab}\right\rangle }$, and $\dot{E}_{\left\langle {ab}\right\rangle }\equiv
h_{(a}{}^{c}h_{b)}{}^{d}\dot{E}_{ab}-{{\frac{1}{3}}}h_{cd}\dot{E}^{cd}h_{ab}$
is the time derivative of the spatially projected gravitoelectric tensor.\footnote{Note that the projected time derivative of $E_{ab}$ satisfies $h_{a}{}^{d}h_{b}{}^{e}\dot{E}_{de}=h_{\left\langle a\right. }{}^{d}h_{\left.
b\right\rangle }{}^{e}\dot{E}_{de}$.}

Similarly, $\nabla _{c}H_{ab}$ can also be decomposed. Comparison of Eq. (\ref{eq:e_4_51}) with Eq. (\ref{eq:e_2_19}) indicates that the gravitoelectric and gravitomagnetic distortions, $\hat{E}_{cab}$ and $\hat{H}_{cab}$, which are the divergence-less and curl-less spatial variation of the Weyl tensor, could be analogous to the shear $\sigma _{ab}$ of the fluid flow. We may also apply the decomposition (\ref{eq:e_4_50})--(\ref{eq:e_4_51}) to any PSTF tensors such as the shear $\sigma _{ab}$:
\begin{equation}
\mathrm{D}_{a}\sigma _{bc}={{\frac{3}{5}}}(\mathrm{div}\sigma
)_{\left\langle a\right. }h_{\left. b\right\rangle c}-{{\frac{2}{3}}}%
\varepsilon _{dc(a}\mathrm{curl}\sigma _{b)}{}^{d}+\hat{\sigma}_{cab}.
\label{eq:e_4_53}
\end{equation}%
Although the divergence and curl of $\sigma _{ab}$ are previously seen in the Ricci identities (\ref{eq:e_4_22}) and (\ref{eq:e_4_24}), respectively, its distortion does not appear in the Ricci and Bianchi equations. If we take a spatial distortion from (\ref{eq:e_4_21}), it is noticed that the evolution of $\hat{\sigma}_{cab}$ is controlled by $\hat{E}_{cab}$.

The Laplacian equation, which plays a key role in the wave solution, also emerges from the distortion. Let us consider the gravitomagnetic distortion: 
\begin{equation}
\hat{H}_{cab}\equiv\mathrm{D}_{(c}H_{ab)}-{{\frac{2}{5}}}h_{(ab}\mathrm{D}%
^{d}H_{c)d},  \label{eq:e_4_54}
\end{equation}%
which has the following divergence equation:
\begin{equation}
\mathrm{D}^{c}\hat{H}_{cab}={{\frac{1}{3}}}\mathrm{D}^{2}H_{ab}-{{\frac{4}{{%
15}}}}\mathrm{D}_{(a}\mathrm{D}^{c}H_{b)c}+{{\frac{2}{3}}}\mathrm{D}^{c}%
\mathrm{D}_{(a}H_{b)c}.  \label{eq:e_4_55}
\end{equation}%
Now if we use the commutation (\ref{eq:e_3_20}), the last term on
the right of Eq. (\ref{eq:e_4_55}) can be transferred into a divergence term and curvature
correction terms \cite{Maartens1997c}: 
\begin{equation}
\mathrm{D}^{2}H_{ab}\mathrm{\ =3D}^{c}\hat{H}_{cab}-2\left( {\rho -{{\frac{1%
}{3}}}\Theta ^{2}}\right) H_{ab}-{{\frac{6}{5}}}\mathrm{D}^{c}\mathrm{D}%
_{(a}H_{b)c}+\mathcal{N}[H]_{ab},  \label{eq:e_4_56}
\end{equation}%
where $\mathcal{N}[H]_{ab}$ is a non-linear term defined by \cite{Maartens1997c}, 
\begin{align}
\mathcal{N}[H]_{ab} \equiv&4\omega ^{c}\omega _{c}H_{ab}+2\Theta \sigma
^{c}{}_{\left\langle a\right. }H_{\left. b\right\rangle c}-6\omega
^{c}\omega _{\left\langle a\right. }H_{\left. b\right\rangle c}  \notag \\
&+\sigma _{c\left\langle a\right. }\sigma _{\left. b\right\rangle
d}H^{cd}-2\sigma ^{cd}H_{cd}\sigma _{ab}+2\Theta \omega ^{c}\varepsilon
_{cd\left\langle a\right. }H_{\left. b\right\rangle }{}^{d}  \notag \\
&-\sigma ^{cd}\sigma _{c\left\langle a\right. }H_{\left. b\right\rangle
d}-6E^{c}{}_{\left\langle a\right. }H_{\left. b\right\rangle c}+2\omega
^{c}\varepsilon _{cde}\sigma ^{d}{}_{\left\langle a\right. }H_{\left.
b\right\rangle }{}^{e}  \notag \\
&+2\omega ^{c}\varepsilon _{cd\left\langle a\right. }\sigma _{\left.
b\right\rangle e}H^{de}+2\omega _{c}\varepsilon ^{cd}{}_{a}\dot{H}%
_{\left\langle {bd}\right\rangle }+2\omega _{c}\varepsilon ^{cd}{}_{b}\dot{H}%
_{\left\langle {ad}\right\rangle }.
\end{align}%
In an almost-FLRW model, these covariant terms in $\mathcal{N}[H]_{ab}$ are very small \cite{Ellis1989}, so this non-linear term could be negligible. As the non-linear term $\mathcal{N}[H]_{ab}$ vanishes in the linearization mode, while the divergence is also zero for gravitational waves, we get 
\begin{equation}
\mathrm{D}^{2}H_{ab}\mathrm{\ =3D}^{c}\hat{H}_{cab}-2\left( {\rho -{{\frac{1%
}{3}}}\Theta ^{2}}\right) H_{ab}.  \label{eq:e_4_57}
\end{equation}%
If the gravitomagnetic distortion $\hat{H}_{cab}$ vanishes, a spatial Helmholtz equation is satisfied by $H_{ab}$, which as argued by \cite{Maartens1997c} excludes general gravitational waves (see also \cite{Hawking1966}). A similar result follows if we start with zero distortion of $E_{ab}$.
Hence, a necessary condition for gravitational waves is non-vanishing distortions of the gravitoelectric and gravitomagnetic fields, i.e.,
\begin{equation}
\mathrm{D}_{\left\langle c\right. }E_{\left. {ab}\right\rangle }\neq 0\neq 
\mathrm{D}_{\left\langle c\right. }H_{\left. {ab}\right\rangle }.
\label{eq:e_4_58}
\end{equation}%
Therefore, the non-zero curls and distortions are essential for the propagation of
gravitational waves. Additionally, the distortions of the gravitoelectric and gravitomagnetic fields can also broaden our understanding of cosmological models, which are typically characterized by the divergences and curls of $E_{ab}$ and $H_{ab}$ (see e.g. \cite{Ellis1971,vanElst1996,vanElst1997a,Maartens1997b,Maartens1997}).

\section{Purely Gravitoelectric/-magnetic Models}

\label{sec6}

The gravitoelectric field $E_{ab}$ is a general-relativistic generalization of Newtonian tidal forces, while there is no analogy for the gravitomagnetic field ($H_{ab}$) in Newtonian theory \cite{Ellis1971,Ellis1997}. 
However, a model with $H_{ab} = 0$ cannot support gravitational waves as shown in dust spacetimes \cite{Hogan1997,Dunsby1997a}, so called \textit{silent universe} owing to the absence of propagating signals \cite{Matarrese1994,Croudace1994,Bruni1995}. The silent universe ($H_{ab} = 0$) is sometimes called \textit{Newtonian-like} due to the presence of a purely Newtonian counterpart, i.e. the gravitoelectric field.
The Newtonian-like model with only $E_{ab}$ corresponds to the general-relativistic generalization of \textit{Newtonian theory}. However, a consequence of vanishing $H_{ab}$ in post-Newtonian models is that
the locally free characteristics of $E_{ab}$ cannot restore in the
Newtonian limit, implying that it is generally incorrect to suppose $H_{ab} = 0$. The purely gravitoelectric model ($H_{ab} = 0$) has limited applications in cosmological models, particularly in studies of gravitational instabilities \cite{Kofman1995,Hui1996,vanElst1997a}, so it is essential to have the gravitomagnetic field in physically realistic inhomogeneous models of the late universe. Models with $H_{ab} \ne 0$ cannot be Newtonian-like, but they consistently satisfy the locally free fields in general \cite{Ellis1997}.

A purely gravitomagnetic model with $E_{ab} = 0 \ne H_{ab} $ is sometimes called \textit{anti-Newtonian} \cite{Maartens1998} due to the presence of only a field with no Newtonian counterpart.
A purely gravitomagnetic model ($E_{ab} = 0$) was first studied in \cite{Truemper1965} in which either the shear or the vorticity are shown to be non-vanishing. We note that $E_{ab} = 0 = H_{ab}$ is associated with an FLRW model (see \S\,\ref{sec4_1_2}). A linearized irrotational dust model with $E_{ab} = 0$ was also found to be exactly FLRW, resulting in $H_{ab} = 0$ \cite{Maartens1998}, so linearized anti-Newtonian irrotational dust models are inconsistent and cannot generally exist.

\subsection{Purely Gravitoelectric Dust Spacetimes}
\label{sec6_1}

Here we discuss an irrotational dust Newtonian-like model ($H_{ab} = 0$), so-called silent \cite{Bruni1995}, where $E_{ab} $ acts as Newtonian tidal forces  \cite{Ellis1997,Kofman1995}. We know that the gravitomagnetic tensor $H_{ab}$ has no Newtonian analogue. The silent models are known to be generally inconsistent and not likely to surpass the spatially homogeneous spaces \cite{Ellis1997,Maartens1998}.

Consider the evolution of the constraints ($\dot{\mathcal{C}}^{A}$) in irrotational dust spacetimes, namely Eqs. (\ref{eq:e_5_15})--(\ref{eq:e_5_19}). 
Although placing $H_{ab}=0$ modifies the constraints $\mathcal{C}^{A}$, they are still
consistent, since the $\dot{\mathcal{C}}^{A}$-equations are still valid.
Eq. (\ref{eq:e_5_4}) offers an extra constraint for the Newtonian-like model: 
\begin{equation}
\mathcal{C}^{5}{}_{ab}\equiv \mathrm{curl}E_{ab}=0,  \label{eq:e_7_1}
\end{equation}%
which should be satisfied, along with the propagation. The spatial divergence and the time derivative of $\mathcal{C}^{5}{}_{ab}$ are calculated by applying the algebraic identities (\ref{eq:e_3_22})--(\ref{eq:e_3_23}) as follows \cite{Maartens1998}: 
\begin{align}
\mathrm{D}^{b}\mathcal{C}^{5}{}_{ab}=&{{\frac{1}{2}}}\mathrm{curl}\mathcal{C}%
^{3}{}_{a}-{{\frac{1}{3}}}\Theta \mathcal{C}^{4}{}_{a}-\sigma _{a}{}^{b}%
\mathcal{C}^{4}{}_{b}.  \label{eq:e_7_2} \\
\dot{\mathcal{C}}^{5}{}_{ab} =&-{{\frac{4}{3}}}\Theta \mathcal{C}%
^{5}{}_{ab}-{{\frac{3}{2}}}\varepsilon ^{cd}{}_{(a}E_{b)c}\mathcal{C}%
^{1}{}_{d}  \notag \\
&-{{\frac{1}{2}}}\rho \mathcal{C}^{2}{}_{ab}-{{\frac{3}{2}}}\varepsilon
^{cd}{}_{(a}\sigma _{b)c}\mathcal{C}^{3}{}_{d}+{{\frac{3}{2}}}\mathcal{H}%
_{ab},  \label{eq:e_7_3}
\end{align}%
where $\mathcal{H}_{ab}$ is defined by \cite{Maartens1998}
\begin{align}
\mathcal{H}_{ab} \equiv \varepsilon _{cd(a}\Big\{ & \mathrm{D}^{e}\left[ {%
E_{b)}{}^{c}\sigma ^{d}{}_{e}}\right] +2\mathrm{D}^{c}\left[ {\sigma
_{b)e}E^{de}}\right]   \notag \\
&+ \sigma _{b)}{}^{c}\mathrm{D}^{e}E^{d}{}_{e}+{{\frac{1}{3}}}\sigma
^{c}{}_{|e|}\mathrm{D}^{e}E_{b)}{}^{d}\Big\} .  \label{eq:e_7_4}
\end{align}%
The consistency of the evolution of the constraints in the Newtonian-like model is satisfied if $\mathcal{H}_{ab}=0$. Eq. (\ref{eq:e_7_4}) implies the condition $\mathcal{H}_{ab}=0$ and its evolution is equally fulfilled by linearization around an FLRW background, which is also characterized by $\mathrm{D}_{a}\rho =\mathrm{D}_{a}\Theta =0$ and $\sigma_{ab}=E_{ab}=H_{ab}=0$ \cite{Ellis1989}.

The purely gravitoelectric models generally have inconsistent solutions that are contingent on linearization instabilities in such a way that their linearized solutions are not constrained by any consistent solutions of the exact (nonlinear) equations. According to \cite{vanElst1997a}, it is unlikely that silent solutions could be a physically realistic model for the late universe or gravitational instabilities.
Consistent, realistic models require the gravitomagnetic field, as independently confirmed by \cite{Kofman1995}. As shown by  \cite{Maartens1998}, the Newtonian-like silent models have a few limited applications, so general relativity does not straightforwardly correspond to Newtonian theory.

\subsubsection{Newtonian Limit}

A Newtonian-like universe with only the Poisson equation for gravitational potential is impractical without some extensions to Newtonian theory as pointed out by Heckmann and Sch\"{u}cking \cite{Heckmann1955,Heckmann1956,Heckmann1959}.  Newtonian theory should be extended in a way to satisfy the essence of the non-locality, so that local physical laws cannot be detached from instantaneous boundary conditions at infinity. The singular limit of general relativity, where gravitational interaction is at an infinite speed, reduces to the Newtonian theory. Obviously, this Newtonian limit of general relativity excludes the gravitomagnetic field.

Here we discuss a covariant Newtonian approximation of general relativity, known as the Heckmann-Sch\"{u}cking limit, having boundary conditions for spatially homogeneous spacetimes. 
Under certain boundary conditions, the gravitoelectric tensor can covariantly be approximated in the Newtonian limit by 
\begin{equation}
E_{ab}^{\mathrm{(N)}}\equiv \mathrm{D}_{\left\langle a\right. }\mathrm{D}%
_{\left. b\right\rangle }\Phi =\mathrm{D}_{a}\mathrm{D}_{b}\Phi -{{\frac{1}{3%
}}}h_{ab}\mathrm{D}^{2}\Phi ,  \label{eq:e_7_7}
\end{equation}%
where $E_{ab}^{\mathrm{(N)}}$ is the gravitoelectric tensor equivalent to tidal forces in the Newtonian limit \cite{Kofman1995,Ellis1997}, and $\Phi$ is the Newtonian potential.\footnote{Note that the spatial derivative $\mathrm{D}_{a}$ satisfies $\mathrm{D}_{a}h_{bc}=0$ and $\left[ {%
\mathrm{D}_{a},\mathrm{D}_{b}}\right] =0$.}

In irrotational spacetimes, Eq. (\ref{eq:e_4_20}) reduces to  $\mathrm{curl}\dot u_a = 0$, which implies the presence of an acceleration scalar potential $\hat{\Phi}$  defined by \cite{Ehlers1961,Ehlers1993,York1979}: 
\begin{equation}  \label{eq:e_7_22}
\dot u_a \equiv \mathrm{D}_a \hat{\Phi}.
\end{equation}
It can be seen that the acceleration vector is covariantly proportional to the spatial gradient of a scalar.
According to the equivalence principle in general relativity, it is indeed the Newtonian potential ($\hat{\Phi}={\Phi}$). Eq.~(\ref{eq:e_7_22}) can also be solved for the velocity field $u_a $ defined by the normalization of the stationary Killing vector $\xi_a = \xi u_a $. Accordingly, Killing's equations result in $\Theta =
\sigma _{ab} = 0$ \cite{Ehlers1961,Ehlers1993,Stephani2003}, so Eq. (\ref{eq:e_2_19}) reduces to $\mathrm{D}_b u_a = - \frac{1}{3} \varepsilon _{abc} \omega ^c $. As $\Phi $ is invariant under $\xi _a $, we should have $\dot \Phi = 0$ of the Newtonian limit, and Eq. (\ref{eq:e_7_22}) satisfies $\mathrm{curl}\dot u_a = 0$. Thus, as covariantly described by Eq. (\ref{eq:e_7_22}), the acceleration vector in irrotational spacetimes strictly corresponds to the gradient of the acceleration scalar potential $\Phi $.

The Newtonian limit of $E_{ab}$ in the Heckmann-Sch\"{u}cking approach satisfies the condition, $\lim_{c\rightarrow \infty }E_{ab}=E_{ab}(t)|_{\infty }$, in a way that the gravitoelectric field (\ref{eq:e_7_7}) prescribed as a function of time arbitrarily propagates to any point with infinite speed ($c\rightarrow \infty $), although is constrained by Eq.(\ref{eq:e_4_30}), resulting in the Poisson equation of Newtonian gravity:
\begin{equation}
\mathrm{D}^{2}\Phi ={{\frac{1}{2}}}\rho .  \label{eq:e_7_10}
\end{equation}%
However, we have no analogue of $H_{ab}$ in the Newtonian limit \cite{Ellis1971}, as shown by strictly mapping general relativity onto Newtonian theory in the singular limit \cite{Ehlers2009}. In Newtonian theory, we do not have $\dot{\Phi}$ , as well as $\dot{E}_{ab}$ seen in the propagation (\ref{eq:e_4_32}). 
We may also approximate the gravitoelectric field to first order as $E_{ab}=E_{ab}^{\mathrm{(N)}}+E_{ab}^{\mathrm{(PN)}}$, where $E_{ab}^{\mathrm{(N)}}$ is the Newtonian term given by Eq. (\ref{eq:e_7_7}), and $E_{ab}^{\mathrm{(PN)}}$ is the first post-Newtonian term. In the Newtonian limit, the gravitoelectric field $E_{ab}$ simply reduces to $E_{ab}^{\mathrm{(N)}}$, which is tidal forces.

\subsubsection{Quasi-Newtonian Gravitational Attraction}

The Raychaudhuri equation (\ref{eq:e_4_19}) \cite{Raychaudhuri1955,Raychaudhuri1957}, which is the basic equation for gravitational attraction \cite{Ehlers1961,Ehlers1993,Ellis1971,Ellis1973} and the singularity theorem \cite{Ellis1973,Ellis1999a}, can also be employed to provide a simpler interpretation of the Newtonian limit \cite{Heckmann1961,Ellis1971}: 
\begin{equation}
\dot{\Theta}+{{\frac{1}{3}}}\Theta ^{2}-2(\omega ^{2}-\sigma ^{2})-\mathrm{D}%
_{a}A^{a} - A^{a}A_{a}+{{\frac{1}{2}}}(\rho +3p)=0,  \label{eq:e_7_8}
\end{equation}%
where $A_{a}\equiv\dot{u}_{a} = \mathrm{D}_a \Phi $ is the \textit{acceleration vector} in the Newtonian limit.  
In Eq. (\ref{eq:e_7_8}), the $(\rho+3p)$ acts as the gravitational source. In the static case, where the expansion of the timelike congruence vanishes ($\Theta =0$), and the Raychaudhuri equation represents the general-relativistic generalization of the Poisson equation of gravity.
In a \textit{quasi-Newtonian model} \cite{vanElst1998},
where a congruence of worldlines is irrotational and shearless ($\omega _{a}=0=\sigma _{ab}$), it reduces to the well-known static gravitational attraction: 
\begin{equation}
\mathrm{D}_{a}\dot{u}^{a}+\dot{u}^{a}\dot{u}_{a}={{\frac{1}{2}}}(\rho +3p).
\label{eq:e_7_29}
\end{equation}%
The above equation covariantly generalizates the Poisson equation of Newtonian gravity, correlating the acceleration $\dot{u}_{a}$ with the active gravitational mass $(\rho +3p)$. In quasi-Newtonian dust spacetimes ($p=0$), Eqs. (\ref{eq:e_7_10}) and (\ref{eq:e_7_29}) imply $\mathrm{D}_a \dot u^a + \dot u^a \dot u_a = \mathrm{D}^2 \Phi$, which is the Laplace equation of the Newtonian potential. In quasi-Newtonian dust models, the Raychaudhuri equation determines the acceleration divergence. However, in expanding spacetimes, it includes the evolution of the expansion, along with the divergence of the acceleration.

\subsection{Purely Gravitomagnetic Dust Spacetimes}
\label{sec6_2}

We here consider a case of purely gravitomagnetic dust models, known as anti-Newtonian universes \cite{Maartens1998}. Substituting $E_{ab}=0$ and $\mathrm{curl}E_{ab}=0$ into Eqs. (\ref{eq:e_5_4}) and (\ref{eq:e_5_5}) yields the following two constraints: 
\begin{align}
-\dot{\sigma}_{\left\langle {ab}\right\rangle }-\sigma _{c\left\langle
a\right. }\sigma _{\left. b\right\rangle }{}^{c}-{{\frac{2}{3}}}\sigma
_{ab}\Theta = &0,  \label{eq:e_7_11} \\
-\dot{H}_{\left\langle {ab}\right\rangle }-\Theta H_{ab}+3\sigma
_{c\left\langle a\right. }H_{\left. b\right\rangle }{}^{c}=&0.
\label{eq:e_7_12}
\end{align}%
The former equation determines the evolution of the shear $\sigma _{ab}$, whereas the latter equation 
is associated with the propagation of $H_{ab}$ without influencing the other quantity evolution. 
It is seen that the the matter evolution is entirely detached from the gravitomagnetic fields, i.e. no coupling between the Weyl tensorial fields and the matter source in the geodesic deviation equation \cite{Ellis1999}. 
The evolution of the dynamic and kinematic quantities is controlled by Eqs. (\ref{eq:e_5_8}) and (\ref{eq:e_7_11}), which is consequently supplied to the propagation of the gravitomagnetic field via Eq. (\ref{eq:e_7_12}). 

In the case of irrotational dust anti-Newtonian spacetimes (\S\,\ref{sec4_1_2}), the constraint (\ref{eq:e_5_14}) implies $\mathrm{D}^{b}H_{ab}=0$, which still maintains Eq. (\ref{eq:e_5_19}) with $E_{ab}=0$. However, the propagation is no longer provided by Eq. (\ref{eq:e_5_3}) with vanishing the gravitoelectric terms, which turns into a new constraint analogous to the Newtonian-like equation (\ref{eq:e_7_1}): 
\begin{equation}
\mathcal{C}^{6}{}_{ab}\equiv \mathrm{curl}H_{ab}-{{\frac{1}{2}}}\rho \sigma
_{ab}=0.  \label{eq:e_7_13}
\end{equation}%
The spatial divergence and the time derivative of $\mathcal{C}%
^{6}{}_{ab}$ are calculated as follows \cite{Maartens1998}: 
\begin{align}
\mathrm{D}^{b}\mathcal{C}^{6}{}_{ab}=&-{{\frac{1}{2}}}\rho \mathcal{C}%
^{1}{}_{a}+{{\frac{1}{3}}}\Theta \mathcal{C}^{3}{}_{a}+{{\frac{1}{2}}}%
\mathrm{curl}\mathcal{C}^{4}{}_{a}+{{\frac{1}{9}}}\mathcal{J}_{a},
\label{eq:e_7_14} \\
\dot{\mathcal{C}}^{6}{}_{ab}=&-{{\frac{4}{3}}}\Theta \mathcal{C}^{6}{}_{ab}-{{%
\frac{3}{2}}}\varepsilon ^{cd}{}_{(a}H_{b)c}\mathcal{C}^{1}{}_{d}+\mathcal{E}%
_{ab},  \label{eq:e_7_15}
\end{align}%
where $\mathcal{J}_{a}$ and $\mathcal{E}_{ab}$ are defined by \cite{Maartens1998}
\begin{align}
\mathcal{J}_{a}\equiv &\Theta \mathrm{D}_{a}\rho -3\rho \mathrm{D}_{a}\Theta -{{%
\frac{3}{2}}}\sigma _{a}{}^{b}\mathrm{D}_{b}\rho ,  \label{eq:e_7_16} \\
\mathcal{E}_{ab} \equiv &{{\frac{1}{6}}}\rho \Theta \sigma _{ab}+{{\frac{1}{2}}}%
\rho \sigma _{c\left\langle a\right. }\sigma _{\left. b\right\rangle
}{}^{c}+3H_{c\left\langle a\right. }H_{\left. b\right\rangle }{}^{c}+3%
\mathrm{curl}\left[ {\sigma ^{c}{}_{(a}H_{b)c}}\right]  \notag \\
&+{{\frac{3}{2}}}\varepsilon _{cd(a}H_{b)}{}^{c}\mathrm{D}^{e}\sigma
^{d}{}_{e}-\sigma _{e}{}^{c}\varepsilon _{cd(a}\mathrm{D}^{e}H_{b)}{}^{d}.
\label{eq:e_7_17}
\end{align}%
The evolution of the constraints is consistent in the anti-Newtonian model if
the two conditions $\mathcal{J}_{a}=0$ and $\mathcal{E}_{ab}=0$ are satisfied.
Thus, the consistency of the constraints on an initial surface is covariantly fulfilled by the following condition, 
\begin{equation}
\rho \mathrm{D}_{a}\Theta ={{\frac{1}{3}}}\Theta \mathrm{D}_{a}\rho -{{\frac{%
1}{2}}}\sigma _{a}{}^{b}\mathrm{D}_{b}\rho .  \label{eq:e_7_20}
\end{equation}%
We see that an algebraic relation between the spatial gradients of the energy density and the expansion scalar appears as a main integrability condition for the consistency. There is at least one special
situation where this condition is identically satisfied (\ref{eq:e_7_20}).
Nevertheless, only specific initial conditions $\left\{ {\rho ,\Theta ,\sigma _{ab}}\right\} $ on the initial surface can be incorporated into the covariant condition (\ref{eq:e_7_20}), which lead to inconsistencies of the evolution equations (\ref{eq:e_5_8}) and (\ref{eq:e_7_11}) in general cases. Thus, anti-Newtonian spacetimes are generally inconsistent.

Linearization about an FLRW background under the consistency condition $\mathcal{J}_{a}=0$ implies that, unlike the Newtonian-like model, linearized integrabilities in the anti-Newtonian case are not insignificant \cite{Ellis1989}. Indeed, all the linearized anti-Newtonian solutions appear to be inconsistent, since
the linearized integrability conditions are satisfied only with $H_{ab} = 0$. 
As we already impose $E_{ab} = 0$ in the anti-Newtonian case, $H_{ab} = 0$ shall make an FLRW spacetime. 
Linearization about an FLRW space under the condition $\mathcal{E}_{ab}=0$ implies $\Theta \sigma _{ab} = 0$ that requires either $\Theta = 0$ or $\sigma _{ab} = 0$. Moreover, the linearized form of the covariant condition $\mathcal{J}_{a}=0$ is $3\rho\mathrm{D}_a \Theta - \Theta \mathrm{D}_a \rho = 0$ that holds if one imposes $\Theta = 0$. If we additionally have $\sigma _{ab} = 0$, there is an FLRW model, which is is not anti-Newtonian. We also see that vanishing the expansion scalar results in $\rho = 0$ according to the linearized  form of Eq.(\ref{eq:e_5_8}a), subsequently $\sigma _{ab} = 0$ based on Eq.(\ref{eq:e_5_8}b). 
Hence, we must rule out any possibilities of linearized anti-Newtonian models (see \cite{Maartens1998} for full discussion).

\subsubsection{Anti-Newtonian Limit}

In the Newtonian limit, the interaction propagates to any point at an infinite
speed, and the gravitoelectric field to first order is roughly written as $E_{ab}=E_{ab}^{\mathrm{(N)}%
}+E_{ab}^{\mathrm{(PN)}}$, where the Newtonian term $E_{ab}^{\mathrm{(N)}}$ is 
associated with tidal forces. Similarly, for the
gravitomagnetic field we may also assume $H_{ab}=H_{ab}^{\mathrm{(AN)}}+H_{ab}^{%
\mathrm{(NN)}}$, where $H_{ab}^{\mathrm{(AN)}}$ is the anti-Newtonian limit of gravitomagnetic field, and $H_{ab}^{\mathrm{(NN)}}$ the first-order non-Newtonian term. Under the SO(2) electric-magnetic duality transformation $E_{ab} \rightarrow H_{ab}$ (see e.g. \cite{Danehkar2019})), the gravitomagnetic tensor may also be written similar to Eq. (\ref{eq:e_7_7}) in the anti-Newtonian limit as follows:
\begin{equation}
H_{ab}^{\mathrm{(AN)}}\equiv \mathrm{D}_{\left\langle a\right. }\mathrm{D}%
_{\left. b\right\rangle }\Psi =\mathrm{D}_{a}\mathrm{D}_{b}\Psi -{{\frac{1}{3%
}}}h_{ab}\mathrm{D}^{2}\Psi ,  \label{eq:e_7_25}
\end{equation}%
where $\Psi $ is the anti-Newtonian potential. The above equation is based on the gravitoelectric/-magnetic duality invariance (see \cite{Maartens1998b,Dadhich2000,Danehkar2019}). The electric-magnetic duality has important implications in quantum field theory.

Considering (\ref{eq:e_4_22}) and (\ref{eq:e_4_23}) in an anti-Newtonian irrotational shear-free model ($\Theta = \sigma _{ab} = 0$), along with the velocity field $u_a $ defined by the
normalization of the stationary Killing vector $\xi _a = \xi u_a $,  
the constrain (\ref{eq:e_4_22}) then yields $\mathrm{curl}\omega _a = - 2[\dot u, \omega]_a $ \cite{Ehlers1961,Ehlers1993}. It can be seen that the vorticity curl corresponds  to the product of vorticity and acceleration, which implies the presence of a vorticity scalar potential $\hat{\Psi}$ defined by 
\begin{equation}  \label{eq:e_7_27}
\omega _a \equiv \mathrm{D}_a \hat{\Psi}.
\end{equation}
The vorticity vector is then covariantly characterized by the vorticity scalar.

Suppose the equivalence between the anti-Newtonian potential and the vorticity scalar (${\Psi}=\hat{\Psi}$), by analogy with the equivalence principle (${\Phi}=\hat{\Phi}$), the ani-Newtonian limit (\ref{eq:e_7_25}) of $H_{ab}$ constrained by Eq.(\ref{eq:e_4_30}b), results in the Helmholtz equation \cite{Danehkar2009}:
\begin{equation}
\mathrm{D}^{2}\Psi +{{\frac{3}{2}}}(\rho +p)\Psi =0.  \label{e_7_28}
\end{equation}%
The Bianchi equations relate the gravitomagnetic tensor to the source $(\rho +p)\omega _{a}$, in another word, the angular momentum density \cite{Ellis1971}. The gravitomagnetic tensor $H_{ab}$, which has no Newtonian analogue, is associated with either gravitational radiation \cite{Ellis1997a} or cosmic inflation \cite{Tsagas2005}. In \S ~\ref{sec5}, we saw that the spatial curls of both the gravitoelectric and gravitomagnetic fields characterize gravitational waves \cite{Maartens1997c}. In post-Newtonian models, the
non-locality -- in a way that local physics cannot be decoupled from boundary conditions at infinity -- of  Newtonian tidal forces induced by the gravitoelectric field cannot be restored without the gravitomagnetic field.

\section{Multi-fluid Models}

\label{sec7}

There are many situations in cosmology where it is necessary to describe a system containing multiple fluids of various matter species rather than a single fluid, which is made possible by studies of multi-component systems (see the review by \cite{Kodama1984}). 
There are many examples of multi-fluid systems such as CMB inhomogeneities, including radiation, baryonic matter and neutrinos \cite{Challinor1999}, and incorporating peculiar
velocities into nonlinear gravitational collapse \cite{Ellis2002}. In either multi-component systems or analyses of peculiar velocities, it is required to include the velocity tilt between the matter species and the fundamental observers (see \cite{King1973,Kodama1984,Bruni1992,Ellis2002,Giovannini2005,Tsagas2008}).

In a multi-fluid system,  the 4-velocity $u_{a}$ of the fundamental
observers and the 4-velocities $u_{a}^{(i)}$ of $i$-th species satisfy $u_{a}u^{a}=u_{a}^{(i)}u_{(i)}^{a}=-1$, so the projector tensors orthogonal to $u_{a}$ and $u_{a}^{(i)}$ are defined, respectively, as follows: 
\begin{equation}
\begin{array}{cc}
{h_{ab}=g_{ab}+u_{a}u_{b},} & {h_{ab}^{(i)}=g_{ab}+u_{a}^{(i)}u_{b}^{(i)}.}%
\end{array}
\label{eq:e_3_1_1}
\end{equation}%
The velocity $u_{a}$ is related to $u_{a}^{(i)}$ via the following \textit{Lorentz boost}, ($v_{a}^{(i)}u^{a}=0$):
\begin{equation}
u_{a}^{(i)}=\gamma ^{(i)}(u_{a}+v_{a}^{(i)}),  \label{eq:e_3_1_2}
\end{equation}%
where $v_{a}^{(i)}$ is the peculiar velocity of the $i$-th species with respect to $u_{a}$, and $\gamma_{(i)}=(1-v_{a}^{(i)}v_{(i)}^{a})^{-1/2}$ is the Lorentz-boost factor, so $\gamma_{(i)}=1$ in non-relativistic peculiar motion.

The boost relation is a reaction to the hyperbolic angle of tilt
$\beta ^{(i)}$ \cite{King1973} between $u_{a}$ and $u_{a}^{(i)}$ in a way that satisfies
\begin{equation}
\begin{array}{ccc}
{\cosh \beta ^{(i)}=-u_{a}^{(i)}u^{a}=\gamma ^{(i)},} & {\sinh \beta
^{(i)}e_{a}=\gamma ^{(i)}v_{a}^{(i)}=h_{a}^{b}u_{b}^{(i)},} & {%
v_{a}^{(i)}=v^{(i)}e_{a}},
\end{array}%
\label{eq:e_3_1_3}
\end{equation}%
which turns Eq. (\ref{eq:e_3_1_2}) to \cite{King1973} 
\begin{equation}
u_{a}^{(i)}=\cosh \beta ^{(i)}u_{a}+\sinh \beta ^{(i)}e_{a}.
\label{eq:e_3_1_4}
\end{equation}%
It can easily be seen that $v_{(i)}=\tanh \beta _{(i)}$, so $v_{(i)}\simeq \beta _{(i)}$ for small tilt angles ($\beta _{(i)}\ll 1$) in non-relativistic peculiar motion.

The dynamic quantities in a multi-fluid system should be defined to include all dynamically significant species:
\begin{align}  
T_{ab} =  &\sum\limits_i {T_{ab}^{(i)} } = \rho u_a u_b + ph_{ab} + 2q_{(a}
u_{b)} + \pi _{ab} , \label{eq:e_3_1_14} \\
T_{ab}^{(i)} = & \rho ^{(i)} u_a^{(i)} u_b^{(i)} + p^{(i)} h_{ab}^{(i)} +
2q_{(a}^{(i)} u_{b)}^{(i)} + \pi _{ab}^{(i)} .  \label{eq:e_3_1_15}
\end{align}
where $\rho ^{(i)}$, $p^{(i)}$, $q_{(a}^{(i)}$, and $\pi _{ab}^{(i)}$ are the energy density, pressure, energy flux, and anisotropic stress measured in the frames of $i$-th species with $u_{a}^{(i)}$ defined by Eq. (\ref{eq:e_3_1_2}). As an example, a multi-fluid system may include the electromagnetic radiation ($i = R$), baryonic matter ($i = B$) modeled by a perfect fluid, cold dark matter ($i = C$) described by a dust model over the era of CMB anisotropies, neutrinos ($i = N$), and dark energy modeled by a cosmological constant ($i = V$) (see \cite{Maartens1999} for further details).

The inverse form of Eq. (\ref{eq:e_3_1_2}) can be expressed as (where $\hat{v}_{a}^{(i)}u_{(i)}^{a}=0$ and $\hat{v}_{a}^{(i)}\hat{v}_{(i)}^{a}=v_{a}^{(i)}v_{(i)}^{a}$): 
\begin{equation}
\begin{array}{cc}
{u_{a}=\gamma ^{(i)}(u_{a}^{(i)}+\hat{v}_{a}^{(i)}),} & {\hat{v}%
_{a}^{(i)}=\gamma ^{(i)}(v_{a}^{(i)}+v_{a}^{(i)}v_{(i)}^{a}u_{a}).}%
\end{array}
\label{eq:e_3_1_18}
\end{equation}%
The above equations directly lead to the following nonlinear energy-momentum tensors of $i$-th species measured in the fundamental $u_{a}$-frame: 
\begin{equation}
T_{ab}^{(i)}=\hat{\rho}^{(i)}u_{a}u_{b}+\hat{p}^{(i)}h_{ab}+2u_{(a}\hat{q}%
_{b)}^{(i)}+\hat{\pi}_{ab}^{(i)},  \label{eq:e_3_1_19}
\end{equation}%
where the nonlinear dynamic quantities of $i$-th species are ($v^{2} \equiv v_{a}v^{a}$) \cite{Maartens1998a,Maartens1999}: 
\begin{align}
\hat{\rho}^{(i)}= & \rho ^{(i)}+\left\{ {\gamma _{(i)}^{2}v_{(i)}^{2}\left( {%
\rho ^{(i)}+p^{(i)}}\right) +2\gamma _{(i)}q_{(i)}^{a}v_{a}^{(i)}+\pi
_{(i)}^{ab}v_{a}^{(i)}v_{b}^{(i)}}\right\} ,  \label{eq:e_3_1_20} \\
\hat{p}^{(i)}=& p^{(i)}+{{\frac{1}{3}}}\left\{ {\gamma
_{(i)}^{2}v_{(i)}^{2}\left( {\rho ^{(i)}+p^{(i)}}\right) +2\gamma
_{(i)}q_{(i)}^{a}v_{a}^{(i)}+\pi _{(i)}^{ab}v_{a}^{(i)}v_{b}^{(i)}}\right\} ,
\label{eq:e_3_1_21} \\
\hat{q}_{a}^{(i)} =&q_{a}^{(i)}+\left( {\rho ^{(i)}+p^{(i)}}\right)
v_{a}^{(i)}+\left\{ {\left( {\gamma _{(i)}^{2}-1}\right) q_{a}^{(i)}-\gamma
_{(i)}q_{(i)}^{b}v_{b}^{(i)}u_{a}}\right.  \notag \\
&\left. {\ +\gamma _{(i)}^{2}v_{(i)}^{2}\left( {\rho ^{(i)}+p^{(i)}}\right)
v_{a}^{(i)}+\pi _{ab}^{(i)}v_{(i)}^{b}-\pi
_{bc}^{(i)}v_{(i)}^{b}v_{(i)}^{c}u_{a}}\right\} ,  \label{eq:e_3_1_22} \\
\hat{\pi}_{ab}^{(i)} =&\pi _{ab}^{(i)}+\left\{ {\ -2u_{(a}\pi
_{b)c}^{(i)}v_{(i)}^{c}+\pi _{bc}^{(i)}v_{(i)}^{b}v_{(i)}^{c}u_{a}u_{b}}%
\right\}  \notag \\
&+\left\{ {\ -{{\frac{1}{3}}}\pi
_{cd}^{(i)}v_{(i)}^{c}v_{(i)}^{d}h_{ab}+\gamma _{(i)}^{2}\left( {\rho
^{(i)}+p^{(i)}}\right) v_{\left\langle a\right. }^{(i)}v_{\left.
b\right\rangle }^{(i)}+2\gamma _{(i)}v_{\left\langle a\right.
}^{(i)}q_{\left. b\right\rangle }^{(i)}}\right\} .  \label{eq:e_3_1_23}
\end{align}%
These nonlinear dynamic quantities generalize the well-known linearized results (see \cite{Israel1979,Bruni1992}). The total dynamic quantities are simply obtained using 
\begin{equation}
\begin{array}{cccc}
{\rho =\sum\limits_{i}{\hat{\rho}^{(i)}},} & {p=\sum\limits_{i}{\hat{p}^{(i)}%
},} & {q_{a}=\sum\limits_{i}{\hat{q}_{a}^{(i)}},} & {\pi
_{ab}=\sum\limits_{i}{\hat{\pi}_{ab}^{(i)}}.}%
\end{array}
\label{eq:e_3_1_24}
\end{equation}%
To linear order, the dynamic quantities measured in the $i$-th frame are roughly equal to those in the fundamental frame, expect for the energy flux $q_{a}$ that needs a velocity correction: 
\begin{equation}
\begin{array}{cccc}
{\hat{\rho}^{(i)}\approx \rho ^{(i)},} & {\hat{p}^{(i)}\approx p^{(i)},} & {%
\hat{q}_{a}^{(i)}\approx q_{a}^{(i)}+\left( {\rho ^{(i)}+p^{(i)}}\right)
v_{a}^{(i)},} & {\hat{\pi}_{ab}^{(i)}\approx \pi _{ab}^{(i)}.}%
\end{array}
\label{eq:e_3_1_25}
\end{equation}%
In the nonlinear situation, as seen in Eqs. (\ref{eq:e_3_1_20})--(\ref{eq:e_3_1_23}), the dynamic quantities observed in the $i$-th frame and the fundamental frame are not generally identical.

\subsection{Multi-Perfect Fluids}

\label{sec7_1}

Here we consider a multi-component system consisting of $i$-th perfect-fluid species defined by
energy density $\rho _{(i)}$ and pressure $p_{(i)}$ moving along the timelike 4-velocity $u_{a}^{(i)}$. 
The energy-momentum tensor of each $i$-th fluid with respect to $u_{a}^{(i)}$ is given by 
\begin{equation}
T_{ab}^{(i)}=\rho ^{(i)}u_{a}^{(i)}u_{b}^{(i)}+p^{(i)}h_{ab}^{(i)},
\label{eq:e_3_1_5}
\end{equation}%
where $h_{ab}^{(i)}$ defined by Eq. (\ref{eq:e_3_1_1}). With respect to the fundamental frame $u_{a}$, Eq. (\ref{eq:e_3_1_5}) turns into an imperfect fluid as follows
\begin{equation}
T_{ab}^{(i)}=\hat{\rho}^{(i)}u_{a}u_{b}+\hat{p}^{(i)}h_{ab}+2u_{(a}\hat{q}%
_{b)}^{(i)}+\hat{\pi}_{ab}^{(i)},  \label{eq:e_3_1_6}
\end{equation}%
where the dynamic quantities $\hat{\rho}^{(i)}$, $\hat{p}^{(i)}$, $\hat{q}_{a}^{(i)}$, and $\hat{\pi}_{ab}^{(i)}$ are given by \cite{Bruni1992} 
\begin{align}
\hat{\rho}^{(i)}= & \gamma _{(i)}^{2}\left( {\rho ^{(i)}+p^{(i)}}\right)
-p^{(i)},  \label{eq:e_3_1_7} \\
\hat{p}^{(i)}= & p^{(i)}+{{\frac{1}{3}}}\left( {\gamma _{(i)}^{2}-1}\right)
\left( {\rho ^{(i)}+p^{(i)}}\right) ,  \label{eq:e_3_1_8} \\
\hat{q}_{a}^{(i)}= & \gamma _{(i)}^{2}\left( {\rho ^{(i)}+p^{(i)}}\right)
v_{a}^{(i)},  \label{eq:e_3_1_9} \\
\hat{\pi}_{ab}^{(i)}=& \gamma _{(i)}^{2}\left( {\rho ^{(i)}+p^{(i)}}\right)
\left( {v_{a}^{(i)}v_{b}^{(i)}-{{\frac{1}{3}}}v_{c}^{(i)}v_{(i)}^{c}h_{ab}}%
\right) .  \label{eq:e_3_1_10}
\end{align}

In non-relativistic peculiar motion ($\beta _{(i)}\ll 1$), quadratic terms in $v_{(i)}$ are negligible, resulting in $\gamma _{(i)}\simeq 1$, so Eqs. (\ref{eq:e_3_1_7})--(\ref{eq:e_3_1_10}) reduce to 
\begin{equation}
\begin{array}{cccc}
{\hat{\rho}^{(i)}=\rho ^{(i)},} & {\hat{p}^{(i)}=p^{(i)},} & {\hat{q}%
_{a}^{(i)}=\left( {\rho ^{(i)}+p^{(i)}}\right) v_{a}^{(i)},} & {\hat{\pi}%
_{ab}^{(i)}=0.}%
\end{array}
\label{eq:e_3_1_11}
\end{equation}%
The energy-momentum tensors of $i$-th species for a combination of interacting and non-comoving perfect fluids are therefore written as 
\begin{equation}
T_{ab}^{(i)}=\rho ^{(i)}u_{a}u_{b}+p^{(i)}h_{ab}+2u_{(a}v_{b)}^{(i)}\left( {%
\rho ^{(i)}+p^{(i)}}\right) ,  \label{eq:e_3_1_12}
\end{equation}%
where $v_{a}^{(i)}$ are the peculiar velocities of $i$-th species. The energy-momentum tensors of $i$-th species hold the conservation law, $\nabla ^{b}T_{ab}^{(i)}=I_{a}^{(i)}$, where $I_{a}^{(i)}$ are the interaction terms of $i$-th species, which must satisfy $\sum\nolimits_{i}{I_{a}^{(i)}}=0$ owing to the conservation law of $T_{ab}=\sum\nolimits_{i}{T_{ab}^{(i)}}$.

\subsection{Tilted Frames}

\label{sec7_2}

Let us assume $u_a $ and $\tilde u_a $ to be two timelike, ($u_a u^a = \tilde u_a \tilde u^a = - 1$) 4-velocities of two observers $O$ and $\tilde O$ satisfying the following projection
tensors ($h_a {}^b u_b = 0 = \tilde h_a {}^b \tilde u_b$):
\begin{equation}
\begin{array}{cc}
{\tilde h_{ab} = g_{ab} + \tilde u_a \tilde u_b,} & {h_{ab} = g_{ab} + u_a u_b},%
\end{array}
\label{eq:e_3_2_4}
\end{equation}%
which represent the spatial parts of the local rest frames of the $O$ and $\tilde O$ observers. For each hypersurface-orthogonal, the projection tensor is a spatial metric in the surface. In other words, the projectors are the metrics in the subspaces of the tangent space orthogonal to the corresponding 4-velocities. 

Suppose that there is a relation between $u_{a}$ and $\tilde{u}_{a}$ determined by the
hyperbolic angle of tilt $\beta $ \cite{King1973}: 
\begin{equation}
\begin{array}{cc}
{u^{a}\tilde{u}_{a}=-\cosh \beta ,} & {\beta \geq 0,}%
\end{array}
\label{eq:e_3_2_6}
\end{equation}%
also constrained by the direction of tilt, either given by the direction $%
\tilde{c}^{a}$ of the $O$ observer motion (i.e. projection of $u^{a}$) in the $\tilde{O}$ local rest frame:
\begin{equation}
\tilde{h}^{a}{}_{b}u^{b}=\sinh \beta \tilde{c}^{a}\Rightarrow \tilde{c}_{a}%
\tilde{u}^{a}=0,\tilde{c}^{a}\tilde{c}_{a}=1,  \label{eq:e_3_2_7}
\end{equation}%
or specified by the $c^{a}$ direction of the $\tilde{O}$ motion (i.e. projection of $%
\tilde{u}^{a}$) in the $O$ local rest frame: 
\begin{equation}
h^{a}{}_{b}\tilde{u}^{b}=\sinh \beta c^{a}\Rightarrow
c_{a}u^{a}=0,c^{a}c_{a}=1.  \label{eq:e_3_2_8}
\end{equation}%
We thus have:
\begin{equation}
\begin{array}{cc}
{u^{a}=\cosh \beta \tilde{u}^{a}+\sinh \beta \tilde{c}^{a},} & {\tilde{u}%
^{a}=\cosh \beta u^{a}-\sinh \beta c^{a}.}%
\end{array}
\label{eq:e_3_2_9}
\end{equation}%
In the case, where $u_{a}$ and $\tilde{u}_{a}$ are orthogonal to the homogenous surface everywhere ($u_{a}=\tilde{u}_{a}$), the hyperbolic angle vanishes ($\beta =0$) \cite{Ellis1969,MacCallum1970,MacCallum1971}. However, in the case of tilted models \cite{King1973}, $c^{a}$ and $\tilde{c}^{a}$ are uniquely expressed by (\ref{eq:e_3_2_7}), (\ref{eq:e_3_2_8}) and (\ref{eq:e_3_2_9}) in a way that $u_{a}$ is tilted with respect to the homogenous surface ($u^{a}\neq \tilde{u}^{a}$), so $\beta >0$. 

It is also useful to consider $\gamma \equiv \cosh \beta $ rather than $\beta $, where $\gamma $ is the contraction factor for the relativistic peculiar velocity of the fluid relative to the homogeneous
surface \cite{King1973}:
\begin{equation}
u_{a}=\gamma (\tilde{u}_{a}+v\tilde{c}_{a}).  \label{eq:e_3_2_11}
\end{equation}%
We note that $\sinh \beta =(\gamma ^{2}-1)^{1/2}$. The 3-velocity $v_{\left\langle a\right\rangle }=vc_{\left\langle a\right\rangle }$ is related to $\gamma $ by \cite{King1973} 
\begin{equation}
\begin{array}{cc}
{\gamma \equiv \cosh \beta =(1-v^{2})^{-1/2},} & {v=\tanh \beta .}%
\end{array}%
\label{eq:e_3_2_12}
\end{equation}%
From Eq. (\ref{eq:e_3_2_9}), it follows some algebraic relations: 
\begin{align}
& \begin{array}{cc}
{c^{a}=\sinh \beta \tilde{u}^{a}+\cosh \beta \tilde{c}^{a},} & {\tilde{c}%
^{a}=-\sinh \beta u^{a}+\cosh \beta c^{a},}%
\end{array}
\label{eq:e_3_2_13} \\
& \begin{array}{cc}
{\tilde{c}_{a}u^{a}=\sinh \beta =-c_{a}\tilde{u}^{a},} & {c_{a}\tilde{c}%
^{a}=\cosh \beta .}%
\end{array}
\label{eq:e_3_2_14}
\end{align}%
There is a spacelike difference vector $d^{a}$ that holds the following relations \cite{King1973}: 
\begin{align}
& \begin{array}{cc}
{d^{a}\equiv u^{a}-\tilde{u}^{a},} & {d^{a}d_{a}=2(\cosh \beta -1)\geq 0,}%
\end{array}
\label{eq:e_3_2_15} \\
& d^{a}=(\cosh \beta -1)\tilde{u}^{a}+\sinh \beta \tilde{c}^{a}=(1-\cosh \beta
)u^{a}+\sinh \beta c^{a},  \label{eq:e_3_2_16} \\
&\begin{array}{cc}
{d^{a}c_{a}=d^{a}\tilde{c}_{a}=\sinh \beta ,} & {d^{a}\tilde{u}%
_{a}=-d^{a}u_{a}=1-\cosh \beta ,}%
\end{array}
\label{eq:e_3_2_17} \\
& \begin{array}{cc}
{h^{a}{}_{b}d^{b}=-h^{a}{}_{b}\tilde{u}^{b}=-V^{a},} & {\tilde{h}%
_{b}^{a}d^{b}=-\tilde{h}^{a}{}_{b}u^{b}=\tilde{V}^{a}.}%
\end{array}
\label{eq:e_3_2_18}
\end{align}%
The projection tensor $\tilde{h}_{ab}$ is also related to ${h}_{ab}$ as follows: \cite%
{Bruni1992}: 
\begin{align}
& \begin{array}{cc}
{\tilde{h}_{ab}=h_{ab}-2d_{(a}u_{b)}+d_{a}d_{b},} & {h_{ab}=\tilde{h}%
_{ab}+2d_{(a}\tilde{u}_{b)}+d_{a}d_{b},}%
\end{array}
\label{eq:e_3_2_19} \\
& h_{a}{}^{b}\tilde{h}_{bc}=h_{ac}-V_{a}\tilde{u}_{c}=\tilde{h}_{ac}+\tilde{V}%
_{c}u_{a},  \label{eq:e_3_2_20} \\
& \begin{array}{cc}
{h_{a}{}^{c}h_{b}{}^{d}\tilde{h}_{cd}=h_{ab}-V_{a}V_{b},} & {\tilde{h}%
_{a}{}^{c}\tilde{h}_{b}{}^{d}h_{cd}=\tilde{h}_{ab}+\tilde{V}_{a}\tilde{V}%
_{b}.}%
\end{array}
\label{eq:e_3_2_21}
\end{align}%
Linearization of the above relations under the condition $\beta \ll 1$, where the motion of $\tilde{O}$ with respect to $O$ is non-relativistic, yields: 
\begin{equation}
\begin{array}{cc}
{d^{a}\equiv u^{a}-\tilde{u}^{a}\simeq \beta \tilde{c}^{a}\equiv \tilde{V}%
^{a}\simeq \beta c^{a}\equiv -V^{a},} & {\tilde{h}_{ab}\simeq
h_{ab}+2u_{(a}V_{b)}.}%
\end{array}
\label{eq:e_3_2_22}
\end{equation}%
The change between the frames $O$ and $\tilde{O}$ with a small relative velocity is referred to as a first-order change in $\beta $.

For a set of timelike worldlines tangent to the 4-velocity $u_a $, normalized such that $u_a u^a = - 1$, we consider the matter field moving with $\tilde u_a $ \cite{Ellis2002,Ellis2001}:
\begin{equation}  \label{eq:e_3_2_1}
\tilde u_a = \gamma (u_a + v_a ),
\end{equation}
where $u_a $ is the 4-velocity with respect to observers comoving with the matter (i.e. $v_a u^a = 0$), and $v_a $ is the peculiar velocity of the matter relative to $u_a $. In an FLRW background, as $u_a $ and $\tilde u_a $ are chosen parallel to the canonical time direction, the peculiar velocity vanishes ($v_a v^a \ll 1$), meaning that $\gamma \simeq 1$ to ensure $\tilde u_a \tilde u^a \simeq - 1$. For two 
independent frames $u_a $ and $\tilde u_a $, the 4-velocity field $\tilde u_a $ correspond to an timelike direction and an associated projection tensor given by Eq. (\ref{eq:e_3_2_4}). As $v_a $ is
not orthogonal to $\tilde u_a $, even for $\gamma = 1$, Eq. (\ref{eq:e_3_2_1}) implies that $\tilde u_a v^a = v^2\ne 0$, and then Eq. (\ref{eq:e_3_2_4}) ensures that $\tilde h_{ab} v^b = v_a + v^2 \tilde u_a \ne v_a $.

\subsection{Tilted Frame Transformations}

\label{sec7_3}

For tilted ${O}$ and $\tilde{O}$ frames described by 4-velocity ${u}_{a}$ and $\tilde{u}_{a}$, respectively, the kinematic and dynamic quantities observed by $O$ undergo some transformations as measured by $\tilde{O}$.
Here, we summarize the exact forms of these transformations.

Transformation of $u_{a}$ to the comoving frame $\tilde{O}$ performed by Eq. (\ref{eq:e_3_2_1}), have the following algebraic identities: 
\begin{align}
g_{ab}=&h_{ab}-u_{a}u_{b}=\tilde{h}_{ab}-\tilde{u}_{a}\tilde{u}_{b},
\label{eq:e_3_2_23}\\
\tilde{h}_{ab}=&h_{ab}+\gamma ^{2}\left( {%
v_{c}v^{c}u_{a}u_{b}+2u_{(a}u_{b)}+v_{a}v_{b}}\right) ,  \label{eq:e_3_2_24} \\
\eta _{abcd}= & 2\varepsilon _{ab[c}u_{d]}-2u_{[a}\varepsilon _{b]cd}=2\tilde{%
\varepsilon}_{ab[c}\tilde{u}_{d]}-2\tilde{u}_{[a}\tilde{\varepsilon}_{b]cd},
\label{eq:e_3_2_25} \\
\tilde{\varepsilon}_{abc}=&\gamma \varepsilon _{abc}+\gamma \left\{ {%
2u_{[a}\varepsilon _{b]cd}+u_{c}\varepsilon _{abd}}\right\} v^{d}.
\label{eq:e_3_2_26}
\end{align}%
The above relations, together with the decomposition (\ref{eq:e_2_15}), and 
$\nabla _{a}\gamma =\gamma ^{3}v^{b}\nabla _{a}v_{b}$, result in
the following nonlinear transformation of the kinematic quantities \cite{Maartens1998a} 
\begin{align}
\tilde{\Theta}=&\gamma \Theta +\gamma \left( {\mathrm{D}_{a}v^{a}+\dot{u}%
_{a}v^{a}}\right) +\gamma ^{3}\left( {\dot{v}_{a}v^{a}+{{\frac{1}{3}}}%
v_{b}v^{b}\mathrm{D}_{a}v^{a}+v^{a}v^{b}\mathrm{D}_{\left\langle a\right.
}v_{\left. b\right\rangle }}\right) ,  \label{eq:e_3_2_28} \\
\dot{\tilde{u}}_{a} =&\gamma ^{2}\dot{u}_{a}+\gamma ^{2}\left\{ {\dot{v}%
_{\left\langle a\right\rangle }+{{\frac{1}{3}}}\Theta v_{a}+\sigma
_{ab}v^{b}-[\omega ,v]_{a}}\right. +\left( {{{\frac{1}{3}}}\Theta v_{b}v^{b}+%
\dot{u}_{b}v^{b}+\sigma _{bc}v^{b}v^{c}}\right) u_{a}  \notag \\
&\left. {\ +{{\frac{1}{3}}}\left( {\mathrm{D}_{a}v^{a}}\right) v_{a}+{{%
\frac{1}{2}}}[v,\mathrm{curl}v]_{a}+v^{b}\mathrm{D}_{\left\langle b\right.
}v_{\left. a\right\rangle }}\right\}  \notag \\
&+\gamma ^{4}\left( {\dot{v}_{c}v^{c}+{{\frac{1}{3}}}v_{b}v^{b}\mathrm{D}%
_{c}v^{c}+v^{b}v^{c}\mathrm{D}_{\left\langle b\right. }v_{\left.
c\right\rangle }}\right) \left( {u_{a}+v_{a}}\right) ,  \label{eq:e_3_2_29} \\
\tilde{\omega}_{a} =&\gamma ^{2}\left\{ {\left( {1-{{\frac{1}{2}}}v_{b}v^{b}%
}\right) \omega _{a}-}\right. {{\frac{1}{2}}}\mathrm{curl}v_{a}+{{\frac{1}{2}%
}}v_{b}\left( {2\omega ^{b}-\mathrm{curl}v^{b}}\right) u_{a}  \notag \\
&\left. {\ +{{\frac{1}{2}}}v_{b}\omega ^{b}v_{a}+{{\frac{1}{2}}}[\dot{u}%
,v]_{a}+{{\frac{1}{2}}}[\dot{v},v]_{a}+{{\frac{1}{2}}}\varepsilon
_{abc}\sigma ^{bd}v^{c}v{}_{d}}\right\} ,  \label{eq:e_3_2_30}
\end{align}%
\begin{align}
\tilde{\sigma}_{ab} =&\gamma \sigma _{ab}+\gamma \left( {1+\gamma ^{2}}%
\right) u_{(a}\sigma _{b)c}v^{c}+\gamma ^{2}\dot{u}_{(a}\left( {%
v_{b)}+u_{b)}v_{c}v^{c}}\right)  \notag \\
&+\gamma \mathrm{D}_{\left\langle a\right. }v_{\left. b\right\rangle }-{{%
\frac{1}{3}}}h_{ab}\left\{ {\dot{u}_{c}v^{c}+\gamma ^{2}\left( {{{\frac{1}{3}%
}}v_{c}v^{c}\mathrm{D}_{d}v^{d}+v^{c}v^{d}\mathrm{D}_{\left\langle c\right.
}v_{\left. d\right\rangle }}\right) }\right\}  \notag \\
&+\gamma ^{3}u_{a}u_{b}\left( {\sigma _{cd}v^{c}v^{d}+{{\frac{2}{3}}}%
v_{d}v^{d}\dot{u}_{c}v^{c}-v^{c}v^{d}\mathrm{D}_{\left\langle c\right.
}v_{\left. d\right\rangle }}\right)  \notag \\
&+\gamma ^{3}u_{a}u_{b}\left( {\gamma ^{4}-{{\frac{1}{3}}}\gamma
^{2}v_{c}v^{c}-1}\right) \left( {\dot{v}_{c}v^{c}+{{\frac{1}{3}}}v_{c}v^{c}%
\mathrm{D}_{d}v^{d}+v^{c}v^{d}\mathrm{D}_{\left\langle c\right. }v_{\left.
d\right\rangle }}\right)  \notag \\
&+\gamma ^{3}u_{(a}u_{b)}\left\{ {\dot{u}_{c}v^{c}+\sigma _{cd}v^{c}v^{d}-%
\dot{u}_{c}v^{c}+2\gamma ^{2}\left( {\gamma ^{2}-{{\frac{1}{3}}}}\right)
\left( {\dot{v}_{c}v^{c}+{{\frac{1}{3}}}v_{c}v^{c}\mathrm{D}%
_{d}v^{d}+v^{c}v^{d}\mathrm{D}_{\left\langle c\right. }v_{\left.
d\right\rangle }}\right) }\right\}  \notag \\
&+{{\frac{1}{3}}}\gamma ^{3}v_{a}v_{b}\left\{ {\mathrm{D}_{c}v^{c}-\dot{v}%
_{c}v^{c}+3\gamma ^{2}\left( {\gamma ^{2}-{{\frac{1}{3}}}}\right) \left( {%
\dot{v}_{c}v^{c}+{{\frac{1}{3}}}v_{c}v^{c}\mathrm{D}_{d}v^{d}+v^{c}v^{d}%
\mathrm{D}_{\left\langle c\right. }v_{\left. d\right\rangle }}\right) }%
\right\}  \notag \\
&+\gamma ^{3}v_{\left\langle a\right. }\dot{v}_{\left. b\right\rangle
}+\gamma ^{3}\dot{v}_{c}v^{c}u_{(a}\dot{v}_{\left\langle b\right\rangle
)}+\gamma ^{3}v^{c}v_{(a}\sigma _{b)c}-\gamma ^{3}[\omega ,v]_{(a}\left\{ {%
v_{b)}+u_{b)}v_{c}v^{c}}\right\}  \notag \\
&+2\gamma ^{3}v^{c}\mathrm{D}_{\left\langle c\right. }v_{\left. {(a}%
\right\rangle }\left\{ {v_{b)}+u_{b)}}\right\} .  \label{eq:e_3_2_31}
\end{align}%
Moreover, the dynamic quantities are nonlinearly transformed as follows \cite{Maartens1998a} (also compare with \cite{Maartens1994}) 
\begin{align}
\tilde{\rho}= &\rho +\gamma ^{2}\left\{ {v_{b}v^{b}\left( {\rho +p}\right)
-2q_{a}v^{a}+\pi _{ab}v^{a}v^{b}}\right\} ,  \label{eq:e_3_2_32} \\
\tilde{p}=&p+{{\frac{1}{3}}}\gamma ^{2}\left\{ {v_{b}v^{b}\left( {\rho +p}%
\right) -2q_{a}v^{a}+\pi _{ab}v^{a}v^{b}}\right\} ,  \label{eq:e_3_2_33} \\
\tilde{q}_{a} =&\gamma q_{a}-\gamma \pi _{ab}v^{b}-\gamma ^{3}\left\{ {%
\left( {\rho +p}\right) -2q_{b}v^{b}+\pi _{bc}v^{b}v^{c}}\right\} v_{a} 
\notag \\
&-\gamma ^{3}\left\{ {v_{b}v^{b}-(1+v_{c}v^{c})q_{b}v^{b}+\pi
_{bc}v^{b}v^{c}}\right\} u_{a},  \label{eq:e_3_2_34} \\
\tilde{\pi}_{ab} =&\pi _{ab}+2\gamma ^{2}v_{d}v^{d}v^{c}\pi _{c(a}\left\{ {%
u_{b)}+v_{b)}}\right\} -2\gamma ^{2}v_{c}v^{c}q_{(a}u_{b)}-2\gamma
^{2}q_{(a}v_{b)}  \notag \\
&-{{\frac{1}{3}}}\gamma ^{2}\left\{ {v_{c}v^{c}\left( {\rho +p}\right) +\pi
_{cd}v^{c}v^{d}}\right\} h_{ab}  \notag \\
&+{{\frac{1}{3}}}\gamma ^{4}u_{a}u_{b}\left\{ {2v_{c}v^{c}v_{d}v^{d}\left( {%
\rho +p}\right) -4v_{c}v^{c}q_{d}v^{d}+\left( {3-v_{e}v^{e}}\right) \pi
_{cd}v^{c}v^{d}}\right\}  \notag \\
&+{{\frac{2}{3}}}\gamma ^{4}u_{(a}u_{b)}\left\{ {2v_{c}v^{c}\left( {\rho +p}%
\right) -\left( {1+3v_{c}v^{c}}\right) q_{d}v^{d}+2\pi _{cd}v^{c}v^{d}}%
\right\}  \notag \\
&+{{\frac{1}{3}}}\gamma ^{4}v_{a}v_{b}\left\{ {\left( {3-v_{c}v^{c}}\right)
\left( {\rho +p}\right) -4q_{c}v^{c}+2\pi _{cd}v^{c}v^{d}}\right\} .
\label{eq:e_3_2_35}
\end{align}%
Eqs. (\ref{eq:e_3_2_28})--(\ref{eq:e_3_2_35}) to linear order reduce to: 
\begin{align}
& \begin{array}{cc}
{\tilde{\Theta}\approx \Theta +\mathrm{D}_{a}v^{a},} & {\dot{\tilde{u}}%
_{a}\approx \dot{u}_{a}+\dot{v}_{a}+{{\frac{1}{3}}}\Theta v_{a},}%
\end{array}
\label{eq:e_3_2_36} \\
& \begin{array}{cc}
{\tilde{\omega}_{a}\approx \omega _{a}-\mathrm{curl}v_{a},} & {\tilde{\sigma}%
_{ab}\approx \sigma _{ab}+\mathrm{D}_{\left\langle a\right. }v_{\left.
b\right\rangle },}%
\end{array}
\label{eq:e_3_2_37} \\
& \begin{array}{cccc}
{\tilde{\rho}\approx \rho ,} & {\tilde{p}\approx p,} & {\tilde{q}_{a}\approx
q_{a}-\left( {\rho +p}\right) v_{a},} & {\tilde{\pi}_{ab}\approx \pi _{ab}.}%
\end{array}
\label{eq:e_3_2_38}
\end{align}

Furthermore, the transformation of the Weyl tensor relative to the comoving frame $\tilde{O}$ yields: 
\begin{equation}
C_{ab}{}^{cd}=4\{\tilde{u}_{[a}\tilde{u}^{[c}+\tilde{h}_{[a}{}^{[c}\}\tilde{E%
}_{b]}{}^{d]}+2\tilde{\varepsilon}_{abe}\tilde{u}^{[c}\tilde{H}^{d]e}+2%
\tilde{\varepsilon}^{cde}\tilde{u}_{[a}\tilde{H}_{b]e}.  \label{eq:e_3_3_1}
\end{equation}%
The kinematic quantities in the above equation are measured in the frame of $\tilde{O}$, whose projection tensor and velocity are defined by (\ref{eq:e_3_2_4}) and (\ref{eq:e_3_2_1}) respectively, so the gravitoelectric and gravitomagnetic tensors are transformed to the frame $\tilde{O}$ as follows:
\begin{align}
\tilde{E}_{ab} =&\gamma ^{2}\left\{ {\left( {1+v_{a}v^{a}}\right)
E_{ab}+v^{c}\left( {2\varepsilon _{cd(a}H_{b)}{}^{d}}\right.
+2E{}_{c(a}u_{b)}}\right.  \notag \\
&\left. {\left. +{\left( {u_{a}u_{b}+h_{ab}}\right)
E_{cd}v^{d}-2E_{c(a}v_{b)}+2u_{(a}\varepsilon _{b)cd}H^{de}v_{e}}\right) }%
\right\} ,  \label{eq:e_3_3_2} \\
\tilde{H}_{ab} =&\gamma ^{2}\left\{ {\left( {1+v_{a}v^{a}}\right)
H_{ab}+v^{c}\left( -{2\varepsilon _{cd(a}E_{b)}{}^{d}}\right.
+2H{}_{c(a}u_{b)}}\right.  \notag \\
&\left. {\left. +{\left( {u_{a}u_{b}+h_{ab}}\right)
H_{cd}v^{d}-2H_{c(a}v_{b)}-2u_{(a}\varepsilon _{b)cd}E^{de}v_{e}}\right) }%
\right\} .  \label{eq:e_3_3_3}
\end{align}%
To linear order, we have ${\tilde{E}_{ab}\approx E_{ab}}$ and $\tilde{H}_{ab}\approx H_{ab}$. 

In a multi-fluid system, the same transformations can be employed to analyze the gravitoelectric/-magnetic fields of multiple matter species relative to the fundamental frame (see \cite{Maartens1998a} for details). In multi-component models, it is essential to consider the 4-velocities of various species with respect to the fundamental frame. Each matter component has a distinct 4-velocity, which needs to be analyzed in a fundamental frame. Each of the 4-velocities leads to slightly different covariant formulations of the gravitoelectromagnetic fields. These different variations may also be regarded as partial gauge-fixings that can be solved using a covariant method. However, in an FLRW spacetime, any differences among the 4-velocities of different matter species vanish, as covariantly demonstrated by the consistent and gauge-invariant linearization around an FLRW background \cite{Ellis1989,Challinor1998}.

\section{1 + 3 Tetrad Formalism}

\label{sec8}

In addition to the $1 + 3$ covariant formalism, there is the so-called $1 + 3$ tetrad approach that also allows us to express geometric interpretations in terms of the projected vectors and PSTF tensors coupled with the dynamic and kinematic quantities. In particular, it is also useful to consider the tetrad formulations along with the covariant formalism that represents covariant relations of quantities with geometrical and/or physical meanings. In the covariant formalism, we do not have a full set of equations that demonstrates how the metric and connection are linked to each other. A tetrad description allows us to formulate a complete set of equations linking the metric and connection that are equivalent to the covariant equations. The development of the tetrad formulations was started by Pirani \cite{Pirani1956}, and followed by Newman and Penrose \cite{Newman1962},  Ellis \cite{Ellis1967}, Stewart and Ellis \cite{Stewart1968}, and MacCallum \cite{MacCallum1973,MacCallum1998} (see reviews by \cite{Papapetrou1970,Papapetrou1971a,Papapetrou1971b,Edgar1977,Edgar1980,Wald1984,Wainwright1997}).
This approach has been extensively employed by a number of authors \cite{deFelice1992,Lesame1995,vanElst1996a,Velden1997,vanElst1997,vanElst1996,Ellis1999a,vanElst1998a}.  
In this section, we briefly introduce the tetrad formalism and summarize its key identities, which are useful for analyzing the gravitoelectric/-magnetic tensorial fields. Rewriting the Bianchi equations using the tetrad approach puts the constraint and evolution equations of gravitoelectromagnetism on another identical footing.

To establish the tetrad method, a vector basis is chosen for an orthonormal tetrad $\left\{ {\mathrm{e}_{a}%
}\right\} $ with the timelike vector $\mathrm{e}_{0}$ as follows:
\begin{equation}
e_{0}{}^{i}=u^{i}.  \label{eq:e_8_46}
\end{equation}%
A local coordinate system is built by $\left\{ {x^{i}}\right\} $ and a
tetrad by $\left\{ {\mathrm{e}_{a}}\right\} $ in a way that  \cite{Ellis1967} 
\begin{equation}
{\mathrm{e}_{a}=e_{a}{}^{i}\left( \partial /\partial x^{i}\right) }\equiv
e_{a}{}^{i}\partial _{i},
\label{eq:e_8_1}
\end{equation}%
where $e_{a}{}^{i}$ are the functions ($\det |{e_{a}{}^{i}}%
| \neq 0$) including components of the tetrad
vectors $\mathrm{e}_{a}$ relative to the basis $\partial /\partial x^{i}$ and directional derivatives of the coordinate functions $x^{i}$ as $e_{a}{}^{i}=\partial _{a}(x^{i})$ \cite{Ellis1967} .

The local tetrad transformation, $\mathrm{e}_{a'}=\Lambda _{a'}^{a}\mathrm{e}_{a}$, where $\Lambda _{a'}^{a}$ is a position-dependent Lorentz matrix, along with 
the coordinate transformation, $x^{i'}=x^{i'}(x^{j}) \rightarrow \partial /\partial x^{j'}=({\partial x^{i}}/{\partial x^{j'}})\partial /\partial x^{i} $, results in changes of the functions $e_{a}{}^{i}$ \cite{Ellis1967}. The components $e^{a}{}_{i}$ of $\partial /\partial x^{i}$ are typically referred to as the basis $\left\{ {%
\mathrm{e}_{a}}\right\} $ specified by \cite{Stewart1968} 
\begin{equation}
\partial _{i}\equiv \partial /\partial x^{i}=e^{a}{}_{i}\mathrm{e}_{a}.
\label{eq:e_8_5}
\end{equation}%
As the tetrad is orthonormal, the tetrad components of the spacetime metric are written by 
\begin{equation}
g_{ab}=g_{ij}e_{a}{}^{i}e_{b}{}^{j}=\mathrm{e}_{a}\cdot \mathrm{e}_{b}=\eta
_{ab},  \label{eq:e_8_6}
\end{equation}%
where $\eta _{ab}=\mathrm{diag}(-1,1,1,1)$ is the Minkowski metric, so the basis unit vectors ${e}_{a}$ and ${e}_{b}$
are orthogonal to each other. As the metric tensor $g^{ab}$ are numerically equivalent to $g_{ab}$, tetrad indices can be raised and lowered by useing $g_{ab}=\eta _{ab}$, and vice versa.

In the tetrad method, a set of four linearly independent 1-forms $\left\{ {%
\omega ^a } \right\}$ may be chosen at each point of the spacetime manifold in such a way that
the line elements can locally be written by $ds^2 = \eta _{ab} \omega ^a \omega ^b$, so the vector fields $\left\{ {\mathrm{e}_a }
\right\}$ are dual to the 1-form fields $\left\{ {\omega ^a } \right\}$ such that $\left\langle {\omega ^a ,e_b } \right\rangle = \delta^a{}_b $ \cite{vanElst1996a}.
Suppose the components of the metric tensor expressed by $\delta _a^b $, from Eq. (\ref%
{eq:e_8_6}) it follows that $\delta _a^b = e_a{}^i e_i{}^b = \mathrm{e}_a \cdot 
\mathrm{e}^b $, so $e_a{}^i $ and $e_j{}^b $ are inverse matrices \cite%
{Stewart1968}. Thus, we may write $\delta _j^i = e_j{}^b e_b{}^i$ that implies
\begin{equation}  \label{eq:e_8_8}
g_{ij} = e_i{}^a e_j{}^b g_{ab} .
\end{equation}
Accordingly, Eqs. (\ref{eq:e_8_5}) and (\ref{eq:e_8_8}) are inverse to Eqs. (\ref{eq:e_8_1}) and (\ref{eq:e_8_6}), respectively.

We may obtain components of any vectors $V^a$  or tensors $T^{ab} {}_{cd}$ with respect to the basis $\partial /\partial x^i $ or $\mathrm{e}_a $ (see e.g. \cite%
{Schouten1954,Heckmann1962,Ellis1964,Ellis1999a}):
\begin{align}  
V^a =& e^a {}_i V^i , \label{eq:e_8_9} \\ 
T^{ab} {}_{cd} =& e^a {}_i e^a {}_j e_c{}^k e_d{}^l T^{ij} {}_{kl}.\label{eq:e_8_10}
\end{align}

The Ricci rotation coefficients are typically regarded as tetrad components of the Christoffel symbols \cite{Ellis1967} or connection components for the tetrad \cite%
{Ellis1999a}:
\begin{equation}
\Gamma ^{c}{}_{ab}\equiv \mathrm{e}_{a}\cdot \nabla _{b}\mathrm{e}%
^{c}=e_{i}^{c}e_{b}^{j}\nabla _{j}e_{a}^{i},  \label{eq:e_8_13}
\end{equation}%
where $\Gamma ^{c}{}_{ab}$ calculates the $c$-component of the covariant derivative in the $\mathrm{e}_{b}$-direction of the basic vector $\mathrm{e}_{a}$. 

Using the rotation coefficients, the \textit{covariant derivatives} of any vectors $V^a$  or tensors $T_{bc}$ can be expressed in terms of tetrad components as follows \cite{Schouten1954,Ellis1964}: 
\begin{align}
&\begin{array}{cc}
{\nabla _{b}V^{a}=\mathrm{e}_{b}\left( {V^{a}}\right) +\Gamma
^{a}{}_{cb}V^{c},} & {\nabla _{b}V_{a}=\mathrm{e}_{b}\left( {V_{a}}\right)
-\Gamma ^{c}{}_{ab}V_{c},}%
\end{array}%
\label{eq:e_8_14} \\
& \nabla _{a}T_{bc}=\mathrm{e}_{a}\left( {T_{bc}}\right) -\Gamma
^{d}{}_{ba}T_{dc}-\Gamma ^{d}{}_{ca}T_{bd}.  \label{eq:e_8_15}
\end{align}%
But, for any scalars $f$, we shall calculate the derivative of $f$ in the $\mathrm{e}_{a}$-direction by
\begin{equation}
\mathrm{e}_{a}(f)=e_{a}{}^{i}\partial _{i}f.  \label{eq:e_8_16}
\end{equation}%

We know that $\mathrm{e}_{a}(g_{bc})=\nabla _{a}g_{bc}=0$, so the rotation coefficients are symmetric on the two indices
\begin{equation}
\Gamma _{(ab)c}=\Gamma _{abc}+\Gamma _{bac}=0,  \label{eq:e_8_17}
\end{equation}%
where they are only raised and lowered.

\subsection{Commutators}

The Lie derivative of $\mathrm{e}_{b}$ relative to $\mathrm{e}_{a}$ is expressed by
the $\mathrm{[e}_{a},\mathrm{e}_{b}]$ commutator calculated by 
\begin{equation}
\mathrm{[e}_{a},\mathrm{e}_{b}]f=\partial _{a}\left( {\partial _{b}f}\right)
-\partial _{b}\left( {\partial _{a}f}\right) .  \label{eq:e_8_18}
\end{equation}%
Following \cite{Ellis1967}, the $\mathrm{[e}_{a},\mathrm{e}_{b}]$ commutator is connected to a basic vector $\mathrm{e}_{c}$ via commutation functions $\gamma_{ab}^{c}$: 
\begin{equation}
\begin{array}{cc}
{\mathrm{[e}_{a},\mathrm{e}_{b}]\equiv \gamma _{ab}^{c}\mathrm{e}_{c},} & {%
\gamma _{ab}^{c}}=\gamma _{\lbrack ab]}^{c},%
\end{array}
\label{eq:e_8_19}
\end{equation}%
where the commutation functions $\gamma_{ab}^{c}$ satisfy
\begin{equation}
\gamma _{ab}^{c}=2\Gamma _{\lbrack ab]}^{c}=\left( \Gamma _{ab}^{c}-\Gamma
_{ba}^{c}\right) .  \label{eq:e_8_20}
\end{equation}%
Eq. (\ref{eq:e_8_17}) leads to the inverse of Eq. (\ref{eq:e_8_20}), the so-called the Christoffel relation: 
\begin{equation}
\Gamma _{abc}={\textstyle{\frac{1}{2}}}\left( {\gamma _{abc}+\gamma
_{cab}-\gamma _{bca}}\right) .  \label{eq:e_8_21}
\end{equation}%
It can be seen that the rotation coeﬃcients $\Gamma _{abc}$ are linearly connected to the commutation functions $\gamma _{abc}$, and vice versa.

Suppose that the timelike direction of the orthonormal frame $\left\{ {\mathrm{e}%
_{a}}\right\} $ is aligned with the preferred timelike reference congruence $\mathrm{e}%
_{0}=\mathrm{u}$ ($u^{a}=\delta ^{a}{}_{0}$, $u_{a}=-\delta ^{0}{}_{a}$),
the commutation functions with one or two indices set to zero can be
written in term of the frame components of the kinematic quantities
(\ref{eq:e_2_18}) and (\ref{eq:e_2_19}) of the timelike congruence, so we have \cite{Estabrook1964,Wahlquist1966} 
\begin{equation}
\Omega ^{a}\equiv -{{\frac{1}{2}}}\varepsilon ^{abc}\mathrm{e}_{b}\cdot 
\mathrm{\dot{e}}_{c},  \label{eq:e_8_47}
\end{equation}%
where $\Omega^{a}$ is the local angular velocity in the rest-frame of an observer with
4-velocity $u^{a}$, and $\mathrm{\dot{e}}_{a}\equiv u^{b}\nabla _{b}\mathrm{e}_{a}$ is a Fermi-propagation axes relative to  $\left\{ {\mathrm{e}_{a}}\right\} $. For the Fermi-derivatives $\mathrm{e}_{a}\cdot \mathrm{\dot{e}}_{b}\equiv e_{a}^{i}u^{j}\nabla_{j}e_{bi}$, we get $\mathrm{e}_{a}\cdot \mathrm{\dot{e}}_{b}=-\mathrm{e}_{b}\cdot \mathrm{\dot{e}}_{a}$. 

The spatial functions $\gamma ^{a}{}_{bc}$ are split into a symmetric tensor $\mathrm{n}^{ab}=\mathrm{n}^{(ab)}$ and an antisymmetric term specified by a vector $\mathrm{a}^{a}$:\footnote{This notation was first introduced by Engelbert Sch\"{u}cking and Wolfgang Kundt in the Hamburg Relativity Seminars organized by Pascual Jordan in 1962--1963.}
\begin{equation}
\gamma ^{a}{}_{bc}=\varepsilon _{bce}\mathrm{n}^{ae}+2\delta ^{a}{}_{[c}%
\mathrm{a}_{b]}, \label{eq:e_8_48_1}
\end{equation}%
where $\mathrm{n}^{ab}$ and $\mathrm{a}_{b}$ are defined by 
\begin{equation}
\begin{array}{cc}
{\mathrm{n}^{ab}\equiv {{\frac{1}{2}}}\gamma ^{(a}{}_{cd}\varepsilon ^{b)cd},} & {\mathrm{a}_{b}\equiv {{\frac{1}{2}}}\gamma ^{a}{}_{ba},} 
\end{array} \label{eq:e_8_48}
\end{equation}%
where $\varepsilon _{abc}$ is the spatial permutation tensor with 
$\varepsilon _{123}=1=\varepsilon ^{123}$.

From Eq. (\ref{eq:e_8_21}), the rotation coefficients have the following components 
\cite{Ellis1967,Ellis1969,vanElst1998a}:
\begin{align}
\Gamma _{a00}=&\dot{u}_{a},  \label{eq:e_8_28} \\
\Gamma _{a0b}=&{{\frac{1}{3}}}\Theta \delta _{ab}+\sigma _{ab}-\varepsilon
_{abc}\omega ^{c},  \label{eq:e_8_29} \\
\Gamma _{ab0}=&\varepsilon _{abc}\Omega ^{c},  \label{eq:e_8_30} \\
\Gamma _{abc}=&2\mathrm{a}_{[a}\delta _{b]c}+\varepsilon _{ce[a}\mathrm{n}%
^{e}{}_{b]}+{{\frac{1}{2}}}\varepsilon _{abe}\mathrm{n}^{e}{}_{c}.
\label{eq:e_8_31}
\end{align}%
The first two equations contain the kinematic quantities, whereas the latter two equations include the rotation rate of the spatial frame $\left\{ {\mathrm{e}_{a}}\right\} $ relative to a Fermi-propagation basis and the quantities ($\mathrm{a}_{b}$ and $\mathrm{n}_{bc}$) determining the nine rotation coefficients, respectively.

Considering the commutators (\ref{eq:e_8_19}) and the variables introduced in Eq. (\ref{eq:e_8_48_1}), we have
\begin{align}
\lbrack \mathrm{e}_{0},\mathrm{e}_{a}]= & \dot{u}_{a}\mathrm{e}_{0}-\left( {{{%
\frac{1}{3}}}\Theta \delta ^{b}{}_{a}+\sigma ^{b}{}_{a}+\varepsilon
^{b}{}_{ac}(\omega ^{c}+\Omega ^{c})}\right) \mathrm{e}_{b},
\label{eq:e_8_51} \\
\lbrack \mathrm{e}_{a},\mathrm{e}_{b}]= & 2\varepsilon _{abc}\omega ^{c}\mathrm{%
e}_{0}+\left( {2\mathrm{a}_{[a}\delta ^{c}{}_{b]}+\varepsilon _{abd}n^{dc}}%
\right) \mathrm{e}_{c}.  \label{eq:e_8_52}
\end{align}%
Under the conditions, where a spatial frame vector $\mathrm{e}_{a}$ is hypersurface orthogonal (see e.g. \cite{Wald1984}), we get \cite{vanElst1997} 
\begin{equation}
\begin{array}{cc}
{0=\sigma _{ab}+\varepsilon _{abc}(\omega ^{c}+\Omega ^{c})} & {a\neq b\neq
c,} \\ 
{0=n^{a}{}_{a}} & {\mathrm{no~summation}\mathrm{.}}%
\end{array}
\label{eq:e_8_53}
\end{equation}

\subsection{Tetrad Curvature}

Let us consider $u^{a}$ as the basis vector $\mathrm{e}_{b}$ (i.e. $\mathrm{e}_{b}\cdot
u^{a}=\delta _{b}^{a}$). Applying Eq. (\ref{eq:e_8_15}) to the Ricci identities (\ref{eq:e_4_7}) results in the \textit{tetrad Riemann curvature} given by
\begin{equation}
R^{a}{}_{bcd}=\mathrm{e}_{c}\left( {\Gamma ^{a}{}_{bd}}\right) -\mathrm{e}%
_{d}\left( {\Gamma ^{a}{}_{bc}}\right) +\Gamma ^{a}{}_{ec}\Gamma
^{e}{}_{bd}-\Gamma ^{a}{}_{ed}\Gamma ^{e}{}_{bc}-\Gamma ^{a}{}_{be}\gamma
^{e}{}_{cd}.  \label{eq:e_8_24}
\end{equation}%
Contracting the tetrad Riemann curvature leads to the \textit{tetrad Ricci curvature}: 
\begin{equation}
R_{bd}=\mathrm{e}_{a}\left( {\Gamma ^{a}{}_{bd}}\right) -\mathrm{e}%
_{d}\left( {\Gamma ^{a}{}_{ba}}\right) +\Gamma ^{a}{}_{ea}\Gamma
^{e}{}_{bd}-\Gamma ^{a}{}_{de}\Gamma ^{e}{}_{ba}=T_{bd}-{{\frac{1}{2}}}%
Tg_{bd}.  \label{eq:e_8_25}
\end{equation}

The antisymmetry properties of the Riemann curvature ($R^a {}_{[bcd]} = 0$) correspond to the \textit{Jacobi identity} 
\begin{equation}  \label{eq:e_8_26}
[e_b ,[e_c ,e_d ]] + [e_c ,[e_d ,e_b ]] + [e_d ,[e_b ,e_c ]] = 0,
\end{equation}
which can be simplified as
\begin{equation}  \label{eq:e_8_27}
\mathrm{e}_{[a} \left( {\gamma ^d {}_{bc]} } \right) + \gamma ^e {}_{[ab}
\gamma ^d {}_{c]e} = 0.
\end{equation}
It represents the integrability conditions, where $\gamma _{ab}^c $ are the
commutation functions for a set of basic vectors $\mathrm{e}_a $.

Substituting the tetrad forms (\ref{eq:e_8_24}) and (\ref{eq:e_8_25}) into Eq. (\ref{eq:e_3_1}), we 
derive the Einstein equations, along with the 16 Jacobi identities and constraint and evolution equations for $E_{ab} $ and $H_{ab} $ in terms of the tetrad derivatives ($\mathrm{e}_a $), the tetrad variables ($\Omega_a$, $\mathrm{a}_a$, and $\mathrm{n}_{ab}$), and the kinematic and dynamical quantities (see \cite{vanElst1996a,vanElst1997,Ellis1999a,vanElst1996,vanElst1998a} for full details).

\subsection{Correspondence between Covariant and Tetrad Formulations}

Here we summarize how the 1 + 3 covariant notations correspond to the 1+3 tetrad analogues. 
To obtain the equivalent tetrad formulations, we can use the following conversion rules \cite{vanElst1996a,Ellis1983a,Ellis1983b}: 
\begin{align}
\dot{f}\rightarrow & \mathrm{e}_{0}\left( f\right) ,  \label{eq:e_8_32} \\
\dot{V}_{\left\langle a\right\rangle }\rightarrow & \mathrm{e}_{0}\left( {V_{a}%
}\right) -\varepsilon _{abc}\Omega ^{b}V^{c}=\mathrm{e}_{0}\left( {V_{a}}%
\right) -[\Omega ,V]_{a},  \label{eq:e_8_33} \\
\dot{S}_{\left\langle {ab}\right\rangle }\rightarrow &\mathrm{e}_{0}\left( {%
S_{ab}}\right) -2\varepsilon _{cd\left\langle a\right. }\Omega ^{c}S_{\left.
b\right\rangle }{}^{d}=\mathrm{e}_{0}\left( {S_{ab}}\right) -2[\Omega
,S]_{\left\langle {ab}\right\rangle },  \label{eq:e_8_34} \\
\mathrm{D}_{a}f\rightarrow & \delta _{ab}\mathrm{e}^{b}\left( f\right) ,
\label{eq:e_8_35} \\
\mathrm{D}^{a}V_{a}\rightarrow & \left( {\mathrm{e}^{a}-2\mathrm{a}^{a}}%
\right) \left( {V_{a}}\right) ,  \label{eq:e_8_36} \\
\mathrm{D}^{b}S_{ab}\rightarrow & \left( {\mathrm{e}^{b}-3\mathrm{a}^{b}}%
\right) \left( {S_{ab}}\right) -\varepsilon _{abc}\mathrm{n}%
^{bd}S_{d}{}^{c}=\left( {\mathrm{e}^{b}-3\mathrm{a}^{b}}\right) \left( {%
S_{ab}}\right) -[\mathrm{n},S]_{a},  \label{eq:e_8_37} \\
\mathrm{curl}\left( {V_{a}}\right) \rightarrow & \varepsilon _{abc}\left( {%
\mathrm{e}^{b}-\mathrm{a}^{b}}\right) V^{c}-\mathrm{n}_{a}{}^{b}V_{b}=[%
\mathrm{e}-\mathrm{a},V]_{a}-\mathrm{n}_{a}{}^{b}V_{b},  \label{eq:e_8_38} \\
\mathrm{curl}(S_{ab}) \rightarrow &\varepsilon _{cd\left\langle a\right.
}\left( {\mathrm{e}^{c}-\mathrm{a}^{c}}\right) \left( {S_{\left.
b\right\rangle }{}^{d}}\right) -3\mathrm{n}_{\left\langle a\right.
}^{c}S_{\left. b\right\rangle c}+{{\frac{1}{2}}}\mathrm{n}^{c}{}_{c}S_{ab} 
\notag \\
&=[\mathrm{e}-\mathrm{a},S]_{\left\langle {ab}\right\rangle }-3\mathrm{n}%
_{\left\langle a\right. }{}^{c}S_{\left. b\right\rangle c}+{{\frac{1}{2}}}%
\mathrm{n}^{c}{}_{c}S_{ab},  \label{eq:e_8_39} \\
\mathrm{D}_{(a}V_{b)}\rightarrow & \left( {\mathrm{e}_{(a}+\mathrm{a}_{(a}}%
\right) \left( {V_{b)}}\right) -\delta _{ab}\mathrm{a}^{c}V_{c}-\varepsilon
_{cd(a}\mathrm{n}_{b)}{}^{c}V^{d},  \label{eq:e_8_40} \\
\mathrm{D}_{\left\langle a\right. }V_{\left. b\right\rangle }\rightarrow & 
\left( {\mathrm{e}_{\left\langle a\right. }+\mathrm{a}_{\left\langle
a\right. }}\right) \left( {V_{\left. b\right\rangle }}\right) -\varepsilon
_{cd\left\langle a\right. }\mathrm{n}_{\left. b\right\rangle
}{}^{c}V^{d}=\left( {\mathrm{e}_{\left\langle a\right. }+\mathrm{a}%
_{\left\langle a\right. }}\right) \left( {V_{\left. b\right\rangle }}\right)
-[\mathrm{n},V]_{\left\langle {ab}\right\rangle }.  \label{eq:e_8_41}
\end{align}

Applying the above rules to the Bianchi equations (\ref{eq:e_4_12})--(\ref{eq:e_4_15}) with a perfect-fluid matter yields
\begin{align}
\mathrm{e}^{b}E_{ab}-3\mathrm{a}^{b}E_{ab}-[\mathrm{n},E]_{a}&={{\frac{1}{3}}}%
\delta _{ab}\mathrm{e}^{a}\left( \rho \right) +3\omega ^{b}H_{ab}+[\sigma
,H]_{a},  \label{eq:e_8_42} \\
\mathrm{e}^{b}H_{ab}-3\mathrm{a}^{b}H_{ab}-[\mathrm{n},H]_{a}&=-\omega
_{a}(\rho +p)-3\omega ^{b}E_{ab}-[\sigma ,E]_{a},  \label{eq:e_8_43} 
\end{align}%
\begin{align}
\lbrack \mathrm{e},H]_{\left\langle {ab}\right\rangle }-[\mathrm{a}-2\dot{u}%
,H]_{\left\langle {ab}\right\rangle } &\notag \\
-3\mathrm{n}_{\left\langle a\right.
}{}^{c}H_{\left. b\right\rangle c}+{{\frac{1}{2}}}\mathrm{n}^{c}{}_{c}H_{ab}
 &= \mathrm{e}_{0}\left( {E_{ab}}\right)
-[\omega +2\Omega ,E]_{\left\langle {ab}\right\rangle }
\notag \\
&-3\sigma
_{c\left\langle a\right. }E_{\left. b\right\rangle }{}^{c}+\Theta E_{ab}
+{{\frac{1}{2}}}\sigma _{ab}(\rho +p)
,%
\label{eq:e_8_44} \\
\lbrack \mathrm{e},E]_{\left\langle {ab}\right\rangle }-[\mathrm{a}-2\dot{u}%
,E]_{\left\langle {ab}\right\rangle } &\notag \\
-3\mathrm{n}_{\left\langle a\right.
}{}^{c}E_{\left. b\right\rangle c}+{{\frac{1}{2}}}\mathrm{n}^{c}{}_{c}E_{ab}
&=-\mathrm{e}_{0}\left( {H_{ab}}\right) +[\omega +2\Omega ,H]_{\left\langle {%
ab}\right\rangle }\notag \\
&+3\sigma _{c\left\langle a\right. }H_{\left.
b\right\rangle }{}^{c}-\Theta H_{ab}.%
\label{eq:e_8_45}
\end{align}%
The above equations can be expressed in a symmetric normal hyperbolic form that determines their hyperbolic characteristics \cite{vanElst1998a}.

\section{1 + 1 + 2 Semi-covariant Formalism}

\label{sec9}

A semi-covariant approach has been developed \cite{Clarkson2003,Clarkson2004,Betschart2004,Clarkson2007}, which keeps the timelike vector of the 1 + 3 method, but separates one spacelike vector from the 3-D space manifold (see also \cite{Greenberg1970,Tsamparlis1983,Mason1985,Tsamparlis1992} for similar formulations). The other two dimensions remain untouched, in contrast to the 3 spatial dimensions in the $1 + 3$ covariant approach. The $1 + 1 + 2$ semi-covariant formalism could be a halfway between the $1+3$ covariant and tetrad methods.

While the 1+3 covariant approach split spacetime into time and space using a timelike 4-velocity vector $u^{a}$ ($u^{a}u_{a}=-1$), the $1+1+2$ semi-covariant formalism additionally decomposes space into one dimension and 2D surface with the help of a unit spacelike vector $n^{a}$ (i.e. $n^{a}n_{a}=1$ and $u^{a}n_{a}=0$). 
The projection tensor $h_{ab}$ given by Eq. (\ref{eq:e_2_31}) can be split into the spacelike vector $n^{a}$ and a new projection tensor $N_{ab}$ as follows: 
\begin{equation}
N_{ab}\equiv h_{ab}-n_{a}n_{b}=g_{ab}+u_{a}u_{b}-n_{a}n_{b}.
\label{eq:e_9_1}
\end{equation}%
The tensor $N_{ab}$, which projects vectors orthogonal to $n^{a}$ and $u^{a}$ onto a 2-surface referred to as the \textit{sheets} \cite{Betschart2004}, has the following properties: 
\begin{equation}
\begin{array}{ccc}
{N_{ab}n^{b}=0=N_{ab}u^{b},} & {N_{a}{}^{c}N_{cb}=N_{ab},} & {N_{a}{}^{a}=2.}%
\end{array}
\label{eq:e_9_2}
\end{equation}%
The \textit{alternating Levi-Civita 2-tensor} $\varepsilon _{ab}$, is determined from the volume element of the observers' rest-spaces
\begin{equation}
\varepsilon _{ab}\equiv \varepsilon _{abc}n^{c}=u^{d}\eta _{dabc}n^{c},
\label{eq:e_9_7}
\end{equation}%
which possesses the following identities and contractions:
\begin{align}
&\begin{array}{cc}
{\varepsilon _{abc}=2n_{[a}\varepsilon _{b]c}+n_{c}\varepsilon _{ab},} & {%
\varepsilon _{ab}\varepsilon ^{cd}=2!N_{[a}{}^{c}N_{b]}{}^{d},} 
\end{array}\\
&\begin{array}{cccc}
{\varepsilon _{abc}u^{c}=0,} & {\varepsilon _{abc}=\varepsilon _{\lbrack abc]},} & {\varepsilon _{ac}\varepsilon ^{bc}=N_{a}{}^{b},} & {\varepsilon _{ac}\varepsilon ^{ac}=2.}%
\end{array}%
\end{align}%

The projection tensor $N_{ab}$ can be employed to irreducibly decompose any 3-vector $V_{a}$ into a scalar part $\mathcal{V}$ of the vector parallel to $n_{a}$, and a 2-vector $\mathcal{V}_{a}$ sitting on the sheet orthogonal to $n_{a}$: 
\begin{equation}
V_{a}=\mathcal{V}n_{a}+\mathcal{V}_{a},  \label{eq:e_9_3}
\end{equation}%
where $\mathcal{V}$ and $\mathcal{V}_{a}$ can be specified by
\begin{equation}
\begin{array}{cc}
{\mathcal{V}\equiv n_{a}V^{a},} & {\mathcal{V}_{a} \equiv V_{\{a\}} = N_{a}{}^{b}V_{b}.}%
\end{array}
\label{eq:e_9_4}
\end{equation}%
Similarly, a PSTF tensor $W_{ab}$ can be split into a scalar $\mathcal{W}$, a vector $\mathcal{W}_{a}$, and a 2-tensor $\mathcal{W}_{ab}$ as follows:
\begin{equation}
W_{ab}=\mathcal{W}(n_{a}n_{b}-{{\frac{1}{2}}}N_{ab})+2\mathcal{W}_{(a}n_{b)}+%
\mathcal{W}_{ab},  \label{eq:e_9_5}
\end{equation}%
where $\mathcal{W}$, $\mathcal{W}_{a}$, and $\mathcal{W}_{ab}$ can be obtained using
\begin{align}
\mathcal{W}=&n_{a}n_{b}W^{ab}, \\ 
\mathcal{W}_{a}=&n^{c}W_{\{b\}c}=N_{a}{}^{b}n^{c}W_{bc}, \\ 
\mathcal{W}_{ab}=&W_{\{ab\}}=\left( {N_{(a}{}^{c}N_{b)}{}^{d}-{{\frac{1}{2}}}%
N_{ab}N^{cd}}\right) W_{ab}.%
\label{eq:e_9_6}
\end{align}%
The curly brackets are utilized to denote the projection of a vector with $N_{ab}$
and the \textit{transverse-traceless} (TT) part of a tensor. This approach allows us to decompose any object into scalars, 2-vectors laying in the sheet, and TT 2-tensors again in the sheet.

Using $n^{a}$ and $N_{ab}$, one can also define the following two new derivatives of any tensors:
\begin{align}
\hat{T}_{a\cdots }=&n^{b}\mathrm{D}_{b}T_{a\cdots },  \label{eq:e_9_10} \\
\delta _{b}T_{a\cdots }=&N_{b}{}^{d}N_{a}{}^{c}\cdots \mathrm{D}%
_{d}T_{c\cdots },  \label{eq:e_9_11}
\end{align}%
where the hat ($\hat{~}$) denotes the derivative along $n^{a}$ in the
surfaces orthogonal to $u^{a}$, and $\delta $ denotes a projected derivative on the sheet,
with projection on every free index, which is analogous to the spatial derivative $\mathrm{D}$ in the $1+3$ covariant formalism

The above derivatives can be employed to provide the following algebraic identities \cite{Clarkson2007}: 
\begin{align}
&{\dot{N}_{ab}=2u_{(a}\dot{u}_{b)}-2n_{(a}\dot{n}_{b)}=2u_{(a}\mathcal{A}%
_{b)}-2n_{(a}\alpha _{b)},}\\
&{\dot{\varepsilon}_{ab}=-2u_{[a}\varepsilon_{b]c}\mathcal{A}^{c}+2n_{[a}\varepsilon _{b]c}\alpha ^{c},}
\label{eq:e_9_12}\\
&\begin{array}{cc}
{\hat{N}_{ab}=-2n_{(a}\hat{n}_{b)},} & {\delta _{c}N_{ab}=0,}%
\end{array} \label{eq:e_9_13} \\
& \begin{array}{cc}
{\hat{\varepsilon}_{ab}=2n_{[a}\varepsilon _{b]c}\hat{n}^{c},} & {\delta
_{c}\varepsilon _{ab}=0,}%
\end{array}
\label{eq:e_9_14}
\end{align}%
where we have
\begin{align}
&\begin{array}{ccc}
{\dot{n}_{a}=\mathcal{A}u_{a}+\alpha _{a},} & {\mathcal{A}=n_{a}\dot{u}^{a},}
& {\alpha _{a}=\dot{n}_{\{a\}},}%
\end{array}
\label{eq:e_9_17} \\
&\begin{array}{cc}
{\dot{u}_{a}=\mathcal{A}n_{a}+\mathcal{A}_{a},} & {\mathcal{A}_{a}=\dot{u}%
_{\{a\}}.}%
\end{array}
\label{eq:e_9_18}
\end{align}

The spatial derivatives $\mathrm{D}$ of any scalars, projected 2-vectors, and TT 2-tensors, their spatial derivatives can be split into the $1+1+2$ notations:
\begin{align}
\mathrm{D}_{a}\mathcal{F}=&\delta _{a}\mathcal{F} + n_{a}\hat{\mathcal{F}},
\label{eq:e_9_19} \\
\mathrm{D}_{a}\mathcal{V}_{b}=&\delta _{a}\mathcal{V}_{b} + n_{a}\left\{ {\hat{\mathcal{V}}_{\{b\}}-\hat{n}%
_{c}\mathcal{V}^{c}u_{b}}\right\} \notag \\
& -n_{b}\left\{ {{{\frac{1}{2}}}\phi 
\mathcal{V}_{a}+(\zeta _{ac}+\xi \varepsilon _{ac})\mathcal{V}^{c}}\right\},  \label{eq:e_9_20} \\
\mathrm{D}_{a}\mathcal{W}_{bc}=& \delta _{a}\mathcal{W}_{bc} + n_{a}\left\{ {\hat{\mathcal{W}}_{bc}-2n_{(b}%
\mathcal{W}_{c)d}\hat{n}^{d}}\right\} \notag \\
& -2n_{(b}\left\{ {{{\frac{1}{2}}}\phi 
\mathcal{W}_{c)a}+\mathcal{W}_{c)}{}^{d}(\zeta _{ad}+\xi \varepsilon _{ad})}%
\right\} .  \label{eq:e_9_21}
\end{align}

By analogy with Eqs. (\ref{eq:e_2_18}) and (\ref{eq:e_2_19}), the spatial derivative $\mathrm{D}$  of $n^{a}$ orthogonal to $u^{a}$ is split into the following irreducible kinematic quantities: 
\begin{align}
\mathrm{D}_{b}n_{a}=&\delta _{b}n_{a}+\hat{n}_{a}n_{b},  \label{eq:e_9_15} \\
\delta _{b}u_{a}=&{{\frac{1}{2}}}\phi N_{ab}+\zeta _{ab}+\xi \varepsilon
_{ab},  \label{eq:e_9_16}
\end{align}%
where $\hat{n}_{a}=n^{b}\mathrm{D}_{b}n_{a}$ is the acceleration of the sheet moving along $n^{a}$, $\phi =\delta ^{a}n_{a}$ is the expansion of the sheet, $%
\zeta _{ab}=\delta _{\{a}n_{b\}}$ represents the distortion of the sheet calculated by the shear of $%
n^{a}$, and $\xi =-{{\frac{1}{2}}}\varepsilon _{ab}\delta ^{a}n^{b}$ is the twist of the sheet described by the rotation of $n^{a}$. These kinematic quantities of $n^{a}$ in the $1 + 1 + 2$ approach can be treated in the same way as the kinematical variables of $u^{a}$ in the $1 + 3$ method.

To formulate the $1 + 1 + 2$ approach, we should also decompose the kinematical quantities defined by Eqs. (\ref{eq:e_2_18}) and (\ref{eq:e_2_19}), the Weyl fields, and the energy-momentum tensor. The acceleration is provided by (\ref{eq:e_9_18}), but for the vorticity and shear, we have the following decomposition:
\begin{align}
\omega _{a}= &\Omega n_{a}+\Omega _{a},  \label{eq:e_9_22} \\
\sigma _{ab}= &\Sigma (n_{a}n_{b}-{{\frac{1}{2}}}N_{ab})+2\Sigma
_{(a}n_{b)}+\Sigma _{ab}. \label{eq:e_9_23}
\end{align}%
The gravitoelectric tensor is splits into $\mathcal{E}$, $\mathcal{E}_{a}$, and $\mathcal{E}_{ab}$, similarly the gravitomagnetic tensor, as follows  
\begin{align}
E_{ab}=&\mathcal{E}(n_{a}n_{b}-{{\frac{1}{2}}}N_{ab})+2\mathcal{E}_{(a}n_{b)}+%
\mathcal{E}_{ab},  \label{eq:e_9_24} \\
H_{ab}=&\mathcal{H}(n_{a}n_{b}-{{\frac{1}{2}}}N_{ab})+2\mathcal{H}_{(a}n_{b)}+%
\mathcal{H}_{ab}.  \label{eq:e_9_25}
\end{align}%
The shear scalar $\sigma $ is calculated by
\begin{equation}
\sigma ^{2}={{\frac{1}{2}}}\sigma _{ab}\sigma ^{ab}={{\frac{3}{4}}}\Sigma
^{2}+\Sigma _{a}\Sigma ^{a}+{{\frac{1}{2}}}\Sigma _{ab}\Sigma ^{ab}.
\label{eq:e_9_26}
\end{equation}

Finally, the energy flux $q^{a}$ and the anisotropic stress $\pi _{ab}$ of
the energy-momentum tensor (\ref{eq:e_2_17}) are split into 
\begin{align}
q_{a}=&\mathcal{Q}n_{a}+\mathcal{Q}_{a},  \label{eq:e_9_27} \\
\pi _{ab}=&\Pi (n_{a}n_{b}-{{\frac{1}{2}}}N_{ab})+2\Pi _{(a}n_{b)}+\Pi _{ab}.
\label{eq:e_9_28}
\end{align}%
We see how the $1 + 3$ variables are split to formulate their $1 + 1 + 2$ counterparts, which produce new variables associated with the irreducible parts of $\nabla _{a}n_{b}$. Accordingly, the $1 + 1 + 2$ Bianchi and Ricci equations can be constructed for the timelike vector $u^{a}$ and the spacelike vector $n^{a}$ in terms of all the quantities introduced above (see e.g. \cite{Betschart2004,Clarkson2004,Clarkson2007}).

\subsection{Correspondence between Covariant and Semi-covariant Formulations}

The $1 + 3$ covariant formulations can simply be mapped into their $1 + 1 + 2$ semi-covariant analogues using the following rules \cite%
{Clarkson2007}: 
\begin{align}
\dot{V}_{\langle a\rangle }\rightarrow &\dot{\mathcal{V}}_{\{a\}}+(\dot{%
\mathcal{V}}-\mathcal{V}_{b}\dot{n}^{\{b\}})n_{a}+\mathcal{V}\dot{n}_{\{a\}},
\label{eq:e_9_29} \\
\dot{W}_{\langle ab\rangle } \rightarrow &\dot{\mathcal{W}}_{\{ab\}}+(\dot{%
\mathcal{W}}-2\mathcal{W}_{c}\dot{n}^{\{c\}})n_{a}n_{b}-{{\frac{1}{2}}}\dot{%
\mathcal{W}}N_{ab}  \notag \\
&+\left\{ {3\mathcal{W}\dot{n}_{(\{a\}}+2\mathcal{W}_{(\{a\}}-2\dot{n}%
^{\{c\}}\mathcal{W}_{c(a}}\right\} n_{b)}+2\mathcal{W}_{(a}\dot{n}_{\{b\})},
\label{eq:e_9_30} \\
\mathrm{D}_{a}V^{a}\rightarrow & (\delta _{a}-\hat{n}_{a})\mathcal{V}^{a}+\hat{%
\mathcal{V}}+\phi \mathcal{V},  \label{eq:e_9_31} \\
\mathrm{curl}V_{a}\rightarrow & \varepsilon _{bc}\delta ^{b}\mathcal{V}%
^{c}n_{a}-\varepsilon _{ab}\hat{\mathcal{V}}^{b}+\xi (\mathcal{V}_{a}+2%
\mathcal{V}n_{a})+\varepsilon _{ab}\left\{ {(\delta ^{b}-\hat{n}^{b})%
\mathcal{V}-\zeta ^{ab}\mathcal{V}_{c}-{{\frac{1}{2}}}\phi \mathcal{V}^{b}}%
\right\} ,  \label{eq:e_9_32} \\
\mathrm{D}_{\langle a}V_{b\rangle } \rightarrow &\delta _{\{a}\mathcal{V}%
_{b\}}+\mathcal{V}\zeta _{ab}+\left\{ {\mathcal{V}\hat{n}_{(a}+\mathcal{V}%
\delta _{(a}+\hat{\mathcal{V}}_{(\{a\}}-{{\frac{1}{2}}}\phi \mathcal{V}_{(a}+%
\mathcal{V}^{c}(\xi \varepsilon _{c(a}-\zeta _{c(a})}\right\} n_{b)}  \notag
\\
&+{{\frac{1}{3}}}\left\{ {2\hat{\mathcal{V}}-\phi \mathcal{V}-\mathcal{V}%
_{c}(\delta ^{c}+2\hat{n}^{c})}\right\} (n_{a}n_{b}-{{\frac{1}{2}}}N_{ab}),
\label{eq:e_9_33} \\
\mathrm{D}^{b}W_{ab} \rightarrow &(\delta ^{b}-\hat{n}^{b})\mathcal{W}%
_{ab}-\zeta ^{ab}\mathcal{W}_{ab}n_{a}+(\delta _{b}-2\hat{n}_{b})\mathcal{W}%
^{b}n_{a}+\hat{\mathcal{W}}_{\{a\}}+(\zeta _{ab}-\xi \varepsilon _{ab})%
\mathcal{W}^{b}  \notag \\
&+{{\frac{3}{2}}}\phi \mathcal{W}_{a}-{{\frac{1}{2}}}(\delta _{a}-3\hat{n}%
_{a})\mathcal{W}+\hat{\mathcal{W}}n_{a}+{{\frac{3}{2}}}\phi \mathcal{W}n_{a},
\label{eq:e_9_34} \\
\mathrm{(curl}W)_{\langle ab\rangle } \rightarrow &\varepsilon _{c\{a}\hat{%
\mathcal{W}}_{b\}}{}^{c}+\varepsilon ^{cd}\delta _{c}\mathcal{W}%
_{d(a}n_{b)}+\xi \mathcal{W}_{ab}-{{\frac{3}{2}}}\mathcal{W}\varepsilon
_{c\{a}\zeta _{b\}}{}^{c}+{{\frac{1}{2}}}\phi \varepsilon _{c\{a}\mathcal{W}%
_{b\}}{}^{c}  \notag \\
&+2\varepsilon _{c\{a}\hat{n}^{c}\mathcal{W}_{b\}}-\varepsilon
_{c\{a}\delta ^{c}\mathcal{W}_{b\}}+\varepsilon ^{cd}\mathcal{W}_{d}\zeta
_{c(a}n_{b)}+5\xi \mathcal{W}_{(a}n_{b)}  \notag \\
&+\varepsilon _{c(a}\left\{ {\ -{{\frac{3}{2}}}(\delta ^{c}-\hat{n}^{c})%
\mathcal{W}+\hat{\mathcal{W}}^{c}+{{\frac{1}{2}}}\phi {\mathcal{W}}^{c}+2{%
\mathcal{W}}_{d}\zeta ^{cd}}\right\} n_{b)}  \notag \\
&+\left( {3\xi \mathcal{W}+\varepsilon _{cd}\delta ^{c}{\mathcal{W}}%
^{d}-\varepsilon _{cd}{\mathcal{W}}^{de}\zeta ^{c}{}_{e}}\right) \left( {%
n_{a}n_{b}-{{\frac{1}{2}}}N_{ab}}\right) ,  \label{eq:e_9_35}
\end{align}%
The above relations can be put into the covariant formulas to get
their equivalences in the semi-covariant formalism.

Applying the above rules to the perfect-fluid formulas (\ref{eq:e_4_30})--(\ref{eq:e_4_33}) in nonperturbative shear-free spacetimes, we obtain:
\begin{align}
(\delta ^{b}-\hat{n}^{b})\mathcal{E}_{ab}+(\delta _{b}-2\hat{n}_{b})\mathcal{%
E}^{b}n_{a}+\hat{\mathcal{E}}_{\{a\}} &  \notag \\
-{{\frac{1}{2}}}(\delta _{a}-3\hat{n}_{a})\mathcal{E}+\hat{\mathcal{E}}%
n_{a}&={{\frac{1}{3}}}\delta _{a}\rho +{{\frac{1}{3}}}n_{a}\hat{\rho},
\label{eq:e_9_36} \\
(\delta ^{b}-\hat{n}^{b})\mathcal{H}_{ab}+(\delta _{b}-2\hat{n}_{b})\mathcal{%
H}^{b}n_{a}+\hat{\mathcal{H}}_{\{a\}} & \notag \\
-{{\frac{1}{2}}}(\delta _{a}-3\hat{n}_{a})\mathcal{H}+\hat{\mathcal{H}}%
n_{a}&=-\Omega _{a}(\rho +p)-\Omega n_{a}(\rho +p),  \label{eq:e_9_37}
\end{align}%
\begin{align}
\varepsilon ^{cd}\delta _{c}\mathcal{H}_{d(a}n_{b)}+&\varepsilon _{c\{a}\hat{%
\mathcal{H}}_{b\}}{}^{c}+2\varepsilon _{c\{a}\hat{n}^{c}\mathcal{H}%
_{b\}}-\varepsilon _{c\{a}\delta ^{c}\mathcal{H}_{b\}} \notag \\ 
+&\varepsilon _{cd}\delta ^{c}\mathcal{H}^{d}\left( {n_{a}n_{b}-{{\frac{1}{2}}%
}N_{ab}}\right) +\varepsilon _{c(a}\hat{\mathcal{H}}^{c}n_{b)}-{{\frac{3}{2}}%
}\mathcal{H}\varepsilon _{c(a}(\delta ^{c}-\hat{n}^{c})n_{b)} \notag \\ 
=&\dot{\mathcal{E}}_{\{ab\}}-2\dot{n}^{\{c\}}\mathcal{E}_{c(a}n_{b)}+\Theta 
\mathcal{E}_{ab}-2\mathcal{E}_{c}\dot{n}^{\{c\}}n_{a}n_{b}+2\mathcal{E}_{(a}%
\dot{n}_{\{b\})} \notag \\
&+2\mathcal{E}_{(\{a\}}n_{b)}+2\Theta \mathcal{E}_{(a}n_{b)}+\dot{\mathcal{E}}%
\left( {n_{a}n_{b}-{{\frac{1}{2}}}N_{ab}}\right) +3\mathcal{E}\dot{n}%
_{(\{a\}}n_{b)} \notag \\
&+\Theta \mathcal{E}(n_{a}n_{b}-{{\frac{1}{2}}}N_{ab})+\Sigma _{ab}(\rho
+p)+2\Sigma _{(a}n_{b)}(\rho +p) \notag \\
&+{{\frac{1}{2}}}\Sigma (n_{a}n_{b}-{{\frac{1}{2}}}N_{ab})(\rho +p),%
\label{eq:e_9_38}
\end{align}%
\begin{align}
\varepsilon ^{cd}\delta _{c}\mathcal{E}_{d(a}n_{b)}+&\varepsilon _{c\{a}\hat{%
\mathcal{E}}_{b\}}{}^{c}+2\varepsilon _{c\{a}\hat{n}^{c}\mathcal{E}%
_{b\}}-\varepsilon _{c\{a}\delta ^{c}\mathcal{E}_{b\}} \notag \\ 
+&\varepsilon _{cd}\delta ^{c}\mathcal{E}^{d}\left( {n_{a}n_{b}-{{\frac{1}{2}}%
}N_{ab}}\right) +\varepsilon _{c(a}\hat{\mathcal{E}}^{c}n_{b)}-{{\frac{3}{2}}%
}\mathcal{E}\varepsilon _{c(a}(\delta ^{c}-\hat{n}^{c})n_{b)} \notag \\ 
=&-\dot{\mathcal{H}}_{\{ab\}}+2\dot{n}^{\{c\}}\mathcal{H}_{c(a}n_{b)}-\Theta 
\mathcal{H}_{ab}+2\mathcal{H}_{c}\dot{n}^{\{c\}}n_{a}n_{b}-2\mathcal{H}_{(a}%
\dot{n}_{\{b\})} \notag \\ 
&-2\mathcal{H}_{(\{a\}}n_{b)}-2\Theta \mathcal{H}_{(a}n_{b)}-\dot{\mathcal{H}}%
\left( {n_{a}n_{b}-{{\frac{1}{2}}}N_{ab}}\right) -3\mathcal{H}\dot{n}%
_{(\{a\}}n_{b)} \notag \\ 
&-\Theta \mathcal{H}(n_{a}n_{b}-{{\frac{1}{2}}}N_{ab}).%
\label{eq:e_9_39}
\end{align}%
The above linear equations describe the evolution of the Weyl fields in a perfect-fluid model according to in the $1 + 1 + 2$ approach (see \cite{Clarkson2007} for a generic matter field). We see that the $1 + 1 + 2$ formalism offers an alternative description of the gravitoelectric/-magnetic fields in
terms of quantities with clear physical or geometrical meanings, namely scalar, 2-vector, and TT 2-tensors. 
This instrument has been used for a number of studies, such as electromagnetic perturbations on
locally rotationally symmetric spacetimes \cite{Betschart2004}, electromagnetic radiation made by gravitational waves interacting with a strong magnetic field of vibrating massive compact objects  \cite{Clarkson2004}, and
perturbations in rotationally symmetric spacetimes \cite{Clarkson2007}.

\section{Observational Effects}

\label{sec10}

Astrophysical observations are performed by detecting electromagnetic waves (light) and gravitational waves that have traveled through the past lightcone. We know from where the light originates and how it reaches us. 
Although astronomical observations allow us to collect information about the universe, there is a strict limit on the information due to constraints imposed by the past lightcone \cite{Ellis1985}. 
In addition to astrophysics, observations also enable us to collect information about the kinematic properties of large-scale structures such as the CMB anisotropies, where are described by global properties of $u^a$ e.g. the shear (see \cite{Ellis1983a,Ellis1983b,Stoeger1995,Maartens1995}). The propagation of electromagnetic waves and gravitational waves is affected by the spacetime metric, in other words the kinematic quantities, so the description of it in terms of the kinematic quantities will help us to better interpret the observed effects of the gravitoelectric/-magnetic field on small scales near compact objects such as around black holes, as well as large-scale structures (e.g. CMB).

Covariant handling of observations was first implemented by Kristian and Sachs \cite{Kristian1966}, MacCallum and Ellis \cite{MacCallum1970}. They tried to link different distance measurements to redshift based on the source's intrinsic and observed brightness.
The expansion of the universe results in an energy loss of the electromagnetic waves reaching us.  
The energy loss of each photon is described by the redshift $z$, i.e. ${\nu _{\mathrm{obsv}} } = (1 + z)^{-1} {\nu _{\mathrm{emit}} } $, where $\nu_{\mathrm{emit}}$ is the emitted frequency and $\nu_{\mathrm{obsv}}$ is the observed frequency. Cosmological redshift is one of the important measurable quantities associated with the stretching effect of space expansion in cosmology.

The redshift of electromagnetic waves can be investigated in a more careful way via solving the Maxwell
equations, $\nabla ^{a}F_{ab}=J_{b}$ (where $F_{ab}=2\nabla _{\lbrack a}A_{b]} $) on a pseudo-Riemannian manifold for current and current-free assumptions, respectively,  
\begin{equation}
\begin{array}{cc}
{\nabla _{b}\nabla ^{b}A^{a}-R^{a}{}_{b}A^{b}=J^{a},} & {\nabla _{b}\nabla
^{b}A^{a}-R^{a}{}_{b}A^{b}=0.}%
\end{array}
\label{eq:e_11_2}
\end{equation}%
Suppose that the light wavelengths are small relative to the curvature, light is then taken to be tangent to null surfaces of the constant phase $\phi $, so travels on null geodesics. If the velocity of a photon is
specified by a null (geodesic) vector $k^{a}\equiv \nabla ^{a}\phi$, we then have $\nabla _{\lbrack a}k_{b]}=0$, and
\begin{equation}
\begin{array}{cc}
{\nabla _{a}\phi \nabla ^{a}\phi =k_{a}k^{a}=0,} & {k^{a}\nabla _{a}k^{b}=0.}%
\end{array}
\label{eq:e_11_4}
\end{equation}%
The former is an eikonal equation \cite{Ehlers2000}. The angle between the
vector $k^{a}$ and the traveling velocity $u^{a}$ of an observer is associated with the relative angular
frequency $\omega =2\pi \nu =-u^{a}k_{a}$ of electromagnetic waves measured by the observer \cite{Jordan1961,Jordan2013}. A photon traveling between two points is then redshifted based on the relative
frequency difference between the two points, i.e., 
$1+z = {\left( {u^{a}k_{a}}\right) _{\mathrm{emit}}}/{\left( {u^{b}k_{b}}%
\right)_{\mathrm{obsv}}}$. This redshift effect can be explicitly described by the change rate of frequency
along the photon path: 
\begin{align}  
k^{a}\nabla _{a}\omega =&-k^{a}k^{b}\nabla _{a}u_{b}  
=k^{a}k^{b}\left( {u_{a}\dot{u}_{b}-{%
{\frac{1}{3}}}\Theta h_{ab}-\sigma _{ab}-\omega _{ab}}\right) .  \label{eq:e_11_7}
\end{align}%
As $k^{a}$ is a null vector, it may be expressed as a linear
mixture of a parallel term and a term orthogonal to $u^{a}$: 
\begin{equation}
\begin{array}{ccc}
{k^{a}=-u_{b}k^{b}(u^{a}+e^{a}),} & {e^{a}e_{a}=1,} & {u_{a}e^{a}=0,}%
\end{array}
\label{eq:e_11_8}
\end{equation}%
where $e^{a}$ is a normalized spacelike vector orthogonal to $u^{a}$ that represents
the propagation direction of the electromagnetic waves with respect to $u^{a}$. 

Substituting the kinematic quantities into the above equation yields 
\begin{equation}
k^{a}\nabla _{a}\omega =-\omega ^{2}\left( {{{\frac{1}{3}}}\Theta +e^{a}\dot{%
u}_{a}+e^{a}e^{b}\sigma _{ab}-\varepsilon _{abc}e^{a}e^{b}\omega ^{c}}%
\right) ,  \label{eq:e_11_9}
\end{equation}%
which results in
\begin{align}
\frac{{d\nu }}{\nu }=\lambda d\left( {\frac{1}{\lambda }}\right) &=-\left( {{{%
\frac{1}{3}}}\Theta +e^{a}\dot{u}_{a}+e^{a}e^{b}\sigma _{ab}-\varepsilon
_{abc}e^{a}e^{b}\omega ^{c}}\right) \delta l \notag \\  
&=-H\delta l-\dot{u}_{a}\delta _{\bot }x^{a}-\delta
_{\bot }x^{a}e^{b}\sigma _{ab}+\delta _{\bot }x^{a}\varepsilon
_{abc}e^{b}\omega ^{c},
\label{eq:e_11_10}
\end{align}%
where $\delta l^{2}=h_{ab}dx^{a}dx^{b}=(u_{a}k^{a})dv^{2}=-\omega dv^{2}.$
The relative direction of electromagnetic waves is affected as their frequency changes, which can be explained by the kinematic quantities. 
The expansion $\Theta$ reduces the energy and frequency of photons and increases the wavelength, which causes a red shift. However, as seen  in Eq. (\ref{eq:e_11_10}), the changes in frequency caused by the shear $\sigma_{ab}$, vorticity $\omega_{a}$, and acceleration $\dot{u}_{a}$ are direction-dependent. 
Depending on the direction of the incoming photon, relative to the direction of vorticity and acceleration, $\omega_{a}$ and $\dot{u}_{a}$ can make either an increase or a decrease in the frequency. Similarly, the shear $\sigma_{ab}$ also modifies the light frequency according to the difference between its stress direction and the incoming light direction \cite{Clarkson2000,Kristian1966,Ehlers1961,Ehlers1993,MacCallum1970}.

Furthermore, the Ricci equations (\ref{eq:e_4_10}) and (\ref{eq:e_4_18}) imply that the
gravitoelectric and gravitomagnetic fields, which are governed by the Bianchi equations (\ref{eq:e_4_1_1})--(\ref{eq:e_4_1_4}), induce the shear and vorticity. In this way, the gravitoelectric/-magnetic fields, along with the kinematical quantities, indirectly have some observational consequences \cite{Kristian1966}.
The effects of the Weyl conformal tensor occur in the geodesic deviation equation, where
the distortions affect bundles of null geodesics, leading to strong gravitational lensing, multiple images and Einstein rings, as well as weak gravitational lensing and image distortions (see e.g. \cite{Schneider1992,Holz1998}).\footnote{Also see \cite{Iorio2001,Mashhoon2001,Ruggiero2002,Iorio2002,Iorio2005,Lichtenegger2006} for the gravitomagnetic clock effect, and \cite{Braginskii1977,Peng1983,Mashhoon1989,Peng1990,Li1991,Tartaglia2004,Ummarino2017,Ummarino2020} for proposed laboratory experiments.}  
These effects have profound implications for cosmology, where observations allow us to determine the large-scale structure of the universe in the past null cone \cite{Ellis1985}.

\section{Conclusions}

\label{sec11}

In this paper, we have reviewed the 1 + 3 covariant formalism \cite%
{Heckmann1955,Heckmann1956,Ehlers1961,Ehlers1993,Ellis1971} (in addition to
alternative methods in \S ~\ref{sec8} and \S ~\ref{sec9}), which can be employed
to covariantly describe the locally free fields of the (traceless) Weyl conformal tensor in general relativity.
In this approach, we have the projected vectors and PSTF tensors instead of the spacetime metric, together with the kinematic quantities of the fluid flow, and the dynamic quantities of the matter field. This formalism leads to the definitions of spatial divergences, curls, and distortions of the projected vectors and PSTF tensors as introduced in \S ~\ref{sec2}. Moreover, we have the kinematic quantities that describe the relative motion of comoving observers \cite{Ehlers1957,Ellis1964}, and the dynamic quantities, including the energy density and pressure in a perfect fluid, in addition to the energy flux and anisotropic pressure in an imperfect fluid.
These quantities are contingent upon either a single-type of matter or a multiple-type of matter species \cite{Maartens1999,King1973,Maartens1998a,Bruni1992,Tsagas2008} (as seen in \S\,\ref{sec7}), 
such as implications of inhomogeneities for CMB \cite{Challinor1999} and exact (nonlinear) solutions of gravitational collapse \cite{Ellis2002}.

In the theory of general relativity, the Riemann curvature, which completely describes 
the curvature of spacetime, is decomposed into the Ricci curvature and the Weyl
conformal tensor \cite{Jordan1960,Jordan2009}. 
The matter field has local gravitational effects on the surrounding  Ricci curvature that are explained by the Einstein field equations \cite{Einstein1918}. However, the Weyl tensor \cite{Weyl1918}, which is the tracefree part of the Riemann curvature, offers the locally-free long-range fields, enabling distant interactions, namely tidal forces \cite{Ellis1971}, frame-dragging effects  \cite{Owen2011,Nichols2011,Zhang2012,Nichols2012,Danehkar2020}, and gravitational waves \cite{Hawking1966,Dunsby1997a,Danehkar2009}). 
Furthermore, the Weyl tensor is decomposed into the gravitoelectric and gravitomagnetic tensorial ﬁelds \cite{Pirani1957,Bel1958a,Bel1958b,Bel1962,Penrose1960,Matte1953,Pirani1962a,Pirani1962b}, owing to some analogies with the electric and magnetic fields in the Maxwell theory.
The Bianchi and Ricci identities \cite{Kundt1962,Kundt2016} provide some constraint and evolution equations, which covariantly govern couplings of the gravitoelectric/-magnetic fields with the dynamic and kinematic quantities \cite{Truemper1964,Ehlers1961,Ehlers1993,Ellis1971,Ellis1973}. 

As seen in Section~\ref{sec4}, the Bianchi identities are split into constraint and propagation equations, which covariantly characterize the roles played by the gravitoelectric and gravitomagnetic fields. These equations includes the spatial divergences, spatial curls, and time derivatives of the gravitoelectric and gravitomagnetic tensors \cite{Ellis1971,Ellis1973,Ellis1999a}. 
The gravitoelectric divergence equation can simply be interpreted as tidal forces in Newtonian theory (see \S\,\ref{sec6}). The Newtonian analogy occurs due to the presence of variables in the gravitoelectric dynamics that have obvious Newtonian counterparts. In this way, the gravitoelectric field that originates from the energy density gradient acts as a general-relativistic generalization of Newtonian tidal forces
\cite{Ellis1997}. In \S ~\ref{sec6_1}, a Newtonian-like model was built from the gravitoelectric field  using the Heckmann-Sch\"{u}cking approach \cite{Heckmann1955,Heckmann1956,Heckmann1959} under the assumption of instantaneous gravitational interaction that produces the Poisson equation of Newtonian theory \cite{Ellis1997,Kofman1995}. However, in particular, the gravitomagnetic field, which orginates from the angular momentum density \cite{Ellis1971}, has no analogy in Newtonian theory.
In \S ~\ref{sec6}, a purely gravitoelectric (Newtonian-like) model and a purely gravitomagnetic (anti-Newtonian \cite{Maartens1998}) model were shown to be generally inconsistent without their counterpart field.

The decomposition of the Bianchi identities also generates the gravitoelectric evolution ($\dot E_{ab}$), which has no analogy in Newtonian gravity, as well as the gravitomagnetic evolution ($\dot H_{ab}$).
The absence of $\dot E_{ab}$ from Newtonian theory is obvious since the instantaneous interaction prevents any
possibility for wave solutions. We also notice a close analogy between the evolution equations of $E_{ab} $ and $H_{ab} $ and the Maxwell equations of $E_{a} $ and $H_{a} $ in expanding spacetimes \cite{Maartens1998b}. As discussed in Section~\ref{sec5}, wave solutions of $E_{ab} $ and $H_{ab} $ appear under some certain conditions, where their curls \cite{Maartens1997b} and distortions \cite{Maartens1997c} support gravitational waves linearly characterized in free space by $\ddot{E}_{ab}-\mathrm{D}^{2}E_{ab}=0$ and $\ddot{H}_{ab}-\mathrm{D}^{2}H_{ab}=0$ via the propagation equations \cite{Ellis1989,Hawking1966,Hogan1997,Dunsby1997a}.

The geodesic deviation equation can be used to perceive how gravitoelectric and gravitomagnetic fields affect astrophysical and cosmological observations of electromagnetic waves, such as strong and weak lensings \cite{Kristian1966}. Although the effects were covariantly explained in terms of the kinematic quantities in \S ~\ref{sec10}, gravitoelectric /-magnetic fields can also induce shear and vorticity, which subsequently modify the frequency and relative direction of light. The expansion of the universe reduces the photon energy (frequency), resulting in cosmological redshift \cite{Ehlers1961,Ehlers1993,MacCallum1970}. 
Nevertheless, the observational effects on light caused by either shear or vorticity are direction-dependent.
The light frequency can be changed based on its direction relative to the shear or vorticity direction.

\appendixtitles{no} 
\appendixstart
\appendix
\section{Higher-rank PSTF Tensor Formulism}
\label{app_1}

Higher-rank PSTF tensors are useful for describing the spherical harmonic quantities with multiple directions specified by $e^{a}$ at a position $x^{i}$, the so-called \textit{multipoles}.  Rank-$\ell $ PSTF tensors can be employed to represent $\ell $ multipoles of radiation anisotropies such as CMB. 

Following \cite{Thorne1980}, for simplicity of style, we may abbreviate a series of $\ell$ indices on a tensor: 
\begin{equation}
\begin{array}{cc}
{T_{A^{\ell }}\equiv T_{a_{1}a_{2}\cdots a_{\ell }},} & {T^{A^{\ell }}\equiv
T^{a_{1}a_{2}\cdots a_{\ell }}.}%
\end{array}
\label{eq:e_10_1}
\end{equation}%
We may also associate $\ell =0,1,2$ with scalars, vectors, and rank-2 tensors, respectively: 
\begin{equation}
\begin{array}{ccc}
{T\equiv T_{A^{0}},} & {T_{a}=T_{A^{1}},} & {T_{ab}=T_{A^{2}}.}%
\end{array}
\label{eq:e_10_6}
\end{equation}%
Similarly, the tensor product of vectors may be abbreviated by a similar manner, along with a ${}^{\times }$ denoting the product \footnote{In \cite{Ellis1983a}, the tensor product is shown by a tilde as $\tilde{T}_{A_{\ell
}}\equiv T_{a_{1}}T_{a_{2}}\cdots T_{a_{\ell }}$.} 
\begin{align}
&\begin{array}{cc}
{T_{A^{1\times \ell }}\equiv T_{a_{1}}T_{a_{2}}\cdots T_{a_{\ell }},} & {%
T^{A^{1\times \ell }}\equiv T^{a_{1}}T^{a_{2}}\cdots T^{a_{\ell }},}%
\end{array}
\label{eq:e_10_5}\\
& {T_{A^{2\times \ell }}\equiv T_{a_{1}a_{2}}T_{a_{2}a_{3}}\cdots T_{a_{2\ell
-1}a_{2\ell }},} %
\\
& T_{A^{n\times \ell }}\equiv T_{a_{1}\cdots
a_{n}}T_{a_{n+1}\cdots a_{2n}}\cdots T_{a_{n(\ell -1)+1}\cdots a_{n\ell }}.
\label{eq:e_10_19}
\end{align}%
It is convenient to use the following notations for the symmetric and antisymmetric indices of tensors $T_{A^{\ell }}$:
\begin{align}
T_{(A^{\ell })}\equiv & T_{(a_{1}a_{2}\cdots a_{\ell })}=\frac{1}{{\ell !}}%
\sum\limits_{\pi }{T_{a_{\pi (1)}\cdots a_{\pi (\ell )}}},  \label{eq:e_10_3} \\
T_{[A^{\ell }]}\equiv & T_{[a_{1}a_{2}\cdots a_{\ell }]}=\frac{1}{{\ell !}}%
\sum\limits_{\pi }{\delta _{\pi }T_{a_{\pi (1)}\cdots a_{\pi (\ell )}}},
\label{eq:e_10_4}
\end{align}%
where $\Sigma$ is a sum over $\ell !$ permutations $\pi $ of $1,\ldots
,\ell $, $\delta _{\pi }$ is +1 for even permutations and -1 for odd
permutations, and for integer $\ell $, we define ${\ell !\equiv \ell (\ell -1)\cdots 2\cdot 1}$ and $\ell !!\equiv \ell (\ell -2)(\ell -4)\cdots (2\mathrm{~or~}1)$.

Following \cite{Pirani1965,Sachs1961}, the symmetric tracefree (STF) part of a tensor $T_{(A^{\ell })}$ can be obtained by (i) constructing the symmetric part: 
\begin{equation}  
S_{A^\ell } \equiv \left[ {T_{A^\ell } } \right]^{\mathrm{S}} = \frac{1}{{%
\ell !}}\sum\limits_\pi {T_{a_{\pi (1)} \cdots a_{\pi (\ell )} } } , \label{eq:e_10_7}
\end{equation}
and (ii) removing all traces: 
\begin{equation}  
\left[ {T_{A^\ell } } \right]^{\mathrm{STF}} = \sum\limits_{n = 0}^{[\ell
/2]} {B^{\ell n} g _{(a_1 a_2 } \cdots g _{a_{2n - 1} a_{2n} }
S^{b_1 b_1 \cdots b_{n}b_{n} } }{}_{a_{2n + 1} \cdots a_\ell ),} \label{eq:e_10_8}
\end{equation}
where $[\ell /2]$ denotes the largest integer less than or equal to $\ell /2$, and $B^{\ell n}$ is calculated by
\begin{equation}  
B^{\ell n} = \frac{{( - 1)^n \ell !(2\ell - 2n - 1)!!}}{{(\ell - 2n)!(2\ell
- 1)!!(2n)!!}}. \label{eq:e_10_9}
\end{equation}
Similarly, the STF part of a spatial tensor is determined using the spatial metric $h_{ab}$ instead of the spacetime metric $g_{ab}$ \cite{Pirani1965}.

Following \cite{Ellis1983a,Ellis1983b}, the PSTF part of a tensor is denoted by
\begin{equation}  \label{eq:e_10_11}
T_{\langle A^\ell \rangle } \equiv \left[ {T_{A^\ell } } \right]^{\mathrm{%
PSTF}} = \left[ {\left[ {T_{A^\ell } } \right]^{\mathrm{STF}} } \right]^{%
\mathrm{P}} ,
\end{equation}
which can recursively be built using a vector basis $e^a$ (see \cite{Gebbie2000}; $e^{a}e_{a}=1$, $e^{a}u_{a}=0$): 
\begin{equation}
T^{A^{\ell }}=e^{\langle A^{1\times \ell }\rangle }=e^{\langle
a_{1}}e^{a_{2}}e^{a_{3}}\cdots e^{a_{\ell -1}}e^{a_{\ell }\rangle }.
\label{eq:e_10_12}
\end{equation}%

Replacing $g_{ab}$ with $h_{ab}$ in Eq. (\ref{eq:e_10_8}), the spatial multipole $T^{A^{\ell }}$ is defined by 
\begin{equation}
T^{A^{\ell }}=e^{\langle A^{1\times \ell }\rangle }=\sum\limits_{n=0}^{[\ell
/2]}{B_{\ell n}h^{(A^{2\times n}}e^{A^{\ell -2\times n})}.}
\label{eq:e_10_21}
\end{equation}%
As the spatial multipole is a higher-rank PSTF tensor, it is irreducible, i.e.,
\begin{equation}
\begin{array}{cccc}
{T^{A^{\ell }}=T^{\langle A^{\ell }\rangle }} & \Leftrightarrow & {%
T^{A^{\ell }}=T^{(A^{\ell })},} & {T^{A^{\ell }}u_{A^{1}}=0=T^{A^{\ell
}}h_{A^{2}}.}%
\end{array}
\label{eq:e_10_13}
\end{equation}%
Some recursion relations for $\ell$-rank PSTF tensors are established by \cite{Ellis1983a}: 
\begin{align}
T^{\langle A^{\ell }}e^{a_{\ell +1}\rangle }= & T^{(A^{\ell }}e^{a_{\ell +1})}-%
\frac{\ell }{{2\ell +1}}e^{b}h_{bc}T^{c(A^{\ell -1}}h^{a_{\ell }a_{\ell
+1})},  \label{eq:e_10_14} \\
e_{a_{1}}T^{A^{\ell }}=& \frac{\ell }{{2\ell -1}}T^{A^{\ell -1}},
\label{eq:e_10_15}
\end{align}%
which relate a $(\ell +1)$-rank PSTF tensor to a $\ell $-rank and $(\ell -1)$-rank PSTF tensor:
\begin{equation}
T^{A^{\ell +1}}=e^{(a_{\ell +1}}T^{A^{\ell })}-\frac{{\ell ^{2}}}{{(2\ell
+1)(2\ell -1)}}h^{(a_{\ell +1}a_{\ell }}T^{A^{\ell -1})}.  \label{eq:e_10_16}
\end{equation}%
Using $\ell =0$ and $1$ turns Eq. (\ref{eq:e_10_14}) into Eq.  (\ref{eq:e_2_4}) of projected vectors
and PSTF tensors, respectively. 

Using a $\ell$-rank PSTF tensor $T^{A^{\ell }}$ and a projected vector $V^{a}$, a $(\ell+1)$-rank PSTF tensor can be made by \cite{Ellis1983a} 
\begin{equation}
T^{\langle A^{\ell }}V^{a\rangle }=T^{(A^{\ell }}V^{a)}-\frac{\ell }{{2\ell
+1}}V^{b}h_{bc}T^{c(A^{\ell -1}}h^{a_{\ell }a)}.  \label{eq:e_10_22}
\end{equation}%
Similarly, a $(\ell+2)$-rank PSTF tensor can be generated using a $\ell$-rank PSTF tensor $T^{A^{\ell }}$ and a rank-2 PSTF tensor $S^{ab}$  \cite{Ellis1983a}:
\begin{align}
T^{\langle A^{\ell }}S^{ab\rangle }= &T^{(A^{\ell }}S^{ab)}-\frac{1}{{3+2\ell }%
}S^{cd}h_{cd}T^{c(A^{\ell -1}}h^{ab)} \notag \\ 
&-\frac{{2\ell }}{{3+2\ell }}T^{c(A^{\ell -1}}S^{\left\vert d\right\vert
a_{\ell }}h^{ab)}h_{cd}\notag \\ 
& + \frac{{\ell (\ell -1)}}{{(2\ell +1)(2\ell +3)}}%
h_{eg}h_{fh}S^{gh}T^{ef(A^{\ell -2}}h^{a_{\ell -1}a_{\ell }}h^{ab)}.
\end{align}
Contraction of $h^{( A^{2 \times \ell } ) } $ also leads to
\begin{equation}  
h_{( A^{2 \times \ell } ) } h^{( A^{2 \times \ell } ) } = 2\ell + 1. \label{eq:e_10_20}
\end{equation}

Following \cite{Thorne1980}, the integrals of the multipoles over the direction $e^a$ are \cite{Ellis1983a,Maartens1999}: 
\begin{align}  
\int {e^{A^{1 \times (2\ell + 1)} } \mathrm{d}\Omega } & = 0, \label{eq:e_10_17} \\
\int {e^{A^{1 \times 2\ell } } \mathrm{d}\Omega } & = \left( {\frac{{4\pi }}{{%
2\ell + 1}}} \right)h^{(A^{2 \times \ell } )} ,\label{eq:e_10_18}
\end{align}
where $\mathrm{d}\Omega = \mathrm{d}^2 e$ is the volume element covered by
two independent $\mathrm{d} e^a$.

From \cite{Ellis1983a}, we also have: 
\begin{align}  
\int {e^{A^{1 \times 2\ell } } e^{B^{1 \times 2m} } \mathrm{d}\Omega } =&
\left( {\frac{{4\pi }}{{2(\ell + m) + 1}}} \right)h^{(A^{2 \times \ell } }
h^{B^{2 \times m} )} , \label{eq:e_10_24} \\
\int {e^{A^{1 \times (2\ell + 1)} } e^{B^{1 \times 2m} } \mathrm{d}\Omega }
= &\int {e^{A^{2\ell } } e^{B^{2m + 1} } \mathrm{d}\Omega } = 0, \label{eq:e_10_25} \\
\int {e^{A^{1 \times (2\ell + 1)} } e^{B^{1 \times (2m + 1)} } \mathrm{d}%
\Omega } = &\left( {\frac{{4\pi }}{{2(\ell + m) + 3}}} \right)h^{(A^{2 \times
\ell } } h^{B^{2 \times m} } h^{A^1 B^1 )}  \label{eq:e_10_26} .
\end{align}

From Eqs. (\ref{eq:e_10_24}), (\ref{eq:e_10_21}), and (\ref{eq:e_10_8}), for $T^{A^\ell } = e^{\langle A^{1 \times \ell } \rangle } $ it also follows \cite{Gebbie2000} 
\begin{equation}  \label{eq:e_10_33}
\int {T^{A^\ell } T_{B^m } \mathrm{d}\Omega } = \delta ^\ell {}_m \Delta
_\ell h^{\langle A^{1 \times \ell } \rangle } {}_{\langle B^{1 \times \ell }
\rangle } ,
\end{equation}
where $h^{\langle A^{1 \times \ell } \rangle } {}_{\langle B^{1 \times \ell }
\rangle } \equiv h^{\langle a_1 } {}_{\langle b_1 } \cdots h^{a_{\ell } \rangle }
{}_{b_{\ell } \rangle } $, and $\Delta_\ell$ is given by
\begin{align} 
\Delta_\ell \equiv & \frac{{4\pi }}{{(2\ell + 1)}}\frac{{2^\ell (\ell !)^2 }}{%
{(2\ell )!}}.  \label{eq:e_10_34}
\end{align}
Therefore, we have
\begin{equation}  \label{eq:e_10_36}
e^{B^{1 \times n} } h^{\langle A^{1 \times \ell } \rangle } {}_{\langle B^{1
\times \ell } } h_{B^{1 \times (n - \ell )} \rangle } = e^{\langle A^{1
\times \ell } \rangle } ( + 1)^{n - \ell } = e^{\langle A^{1 \times \ell }
\rangle } .
\end{equation}
From (\ref{eq:e_10_33}), we find (compare to Eq. \ref{eq:e_10_20}),
\begin{equation}  
h^{\langle A^{1 \times \ell } \rangle } {}_{\langle A^{1 \times \ell }
\rangle } = 2\ell + 1. \label{eq:e_10_37}
\end{equation}

Consider the relation between the directions $e^{a}$ and $\tilde{e}^{a}$ at $x_{i} $ specified by \cite{Gebbie2000} 
\begin{equation}
e^{a}\tilde{e}_{a}=\cos (\beta )\equiv X.  \label{eq:e_10_30}
\end{equation}%
where $\beta $ is the angular distance $\beta $ separating $e^{a}$ and $\tilde{e}^{a}$. For $e^{A^{1\times \ell }}$ and $\tilde{e}^{A^{1\times \ell }}$, the generalized relation is given by
\begin{equation}
\begin{array}{ccc}
{e^{A^{1\times \ell }}\tilde{e}_{A^{1\times \ell }}=X^{\ell }} & \Rightarrow
& {\sum\limits_{\ell =0}^{\infty }{e^{A^{1\times \ell }}\tilde{e}%
_{A^{1\times \ell }}} =\sum\limits_{\ell =0}^{\infty }{X^{\ell }}=\frac{{%
\cos \beta }}{{1-\cos \beta }}. }%
\end{array}
\label{eq:e_10_31}
\end{equation}%
Using Eqs. (\ref{eq:e_10_24})--(\ref{eq:e_10_26}), we can also calculate the integral over $e^{a}$ with fixed $\tilde{e}^{a}$:
\begin{equation}
\int {e^{A^{1\times \ell }}\tilde{e}_{A^{1\times \ell }}\mathrm{d}\Omega }%
=\int {(e^{a}\tilde{e}_{a}\mathrm{)}^{\ell }\mathrm{d}\Omega }=2\pi
\int_{-1}^{+1}{X^{\ell }}dX=\left\{ {%
\begin{array}{cc}
0 & {\forall \mathrm{~}n\mathrm{~even}} \\ 
{\frac{{4\pi }}{{n+1}}} & {\forall \mathrm{~}n\mathrm{~even}}%
\end{array}%
}\right. .  \label{eq:e_10_32}
\end{equation}

Suppose the multipoles given by $T^{A^\ell } = e^{\langle A^{1 \times \ell } \rangle } $ and $\tilde
T^{A^\ell } = \tilde e^{\langle A^{1 \times \ell } \rangle } $, the following correlation between them exists: 
\begin{equation}  \label{eq:e_10_38}
T^{A^\ell } \tilde T_{A^\ell } = \sum\limits_{k = 0}^{[\ell /2]} {%
\sum\limits_{\tilde k = 0}^{[\ell /2]} {B_{\ell k} B_{\ell \tilde k}
h^{(A^{2 \times k} } h_{(2 \times \tilde k} e^{A^{1 \times \ell } )} \tilde
e_{A^{1 \times \ell } )} } } .
\end{equation}
Higher-rank PSTF tensors hold the following natural polynomial \cite{Gebbie2000}: 
\begin{equation}  \label{eq:e_10_39}
L_\ell (X) = T^{A^\ell } \tilde T_{A^\ell } = \sum\limits_{m = 0}^{[\ell
/2]} {B_{\ell m} X^{\ell - 2m} } .
\end{equation}

Extending the 1+3 tetrad formalism to higher-rank tensors  \cite{Ellis1983b,Treciokas1972}, the spatial divergences, time derivatives, and spatial distortions of higher-rank tensors may be calculated by
\begin{align}
\mathrm{D}^{b}T_{bA^{\ell -1}}\rightarrow & \left( {\mathrm{e}^{b}-(\ell+1)\mathrm{a}^{b}}\right) T_{bA^{(\ell-1) }}+(\ell-1) T^{b}{}_{c(A^{\ell -2}}\varepsilon
_{a_{\ell-1})bd}\mathrm{n}^{cd} \notag \\
& +(\ell-1) T^{b}{}_{b(A^{\ell -2}}\mathrm{a}%
_{a_{\ell-1})},  \label{eq:e_10_27} \\
\dot{T}_{\langle A^{\ell }\rangle }\rightarrow & \mathrm{e}_{0}\left( {%
T_{A^{\ell }}}\right) -\ell T^{d}{}_{\langle A^{\ell -1}}\varepsilon
_{da_{\ell }\rangle c}\Omega ^{c},  \label{eq:e_10_28} \\
\mathrm{D}_{(b}T_{A^{\ell})} \rightarrow &\left( {\mathrm{e}%
_{(b}+ \ell \mathrm{a}_{(b}}\right)
T_{A^{\ell})}-\ell\varepsilon _{dc(b}\mathrm{n}%
^{c}{}_{a_{\ell}}T_{A^{\ell -1})}{}^{d}  \notag \\
&-\ell \mathrm{a}^{d}T_{d(A^{\ell -1}}\delta _{a_{\ell}b)}.
\label{eq:e_10_29} 
\end{align}
With $\ell =1$ (projected vector) and $2$ (rank-2 PSTF tensor), Eq. (\ref{eq:e_10_27}) reduces to Eqs. (\ref{eq:e_8_36}) and (\ref{eq:e_8_37}), and Eq. (\ref{eq:e_10_28}) to Eqs. (\ref{eq:e_8_33}) and (\ref{eq:e_8_34}), respectively.

\vspace{6pt} 











\begin{adjustwidth}{-\extralength}{0cm}

\reftitle{References}

\end{adjustwidth}

\begin{thebibliography}{999}

\bibitem[{Einstein}(1918)]{Einstein1918}
{Einstein}, A.
\newblock {Prinzipielles zur allgemeinen Relativit{\"a}tstheorie}.
\newblock {\em Annalen der Physik} {\bf 1918}, {\em 360},~241--244.
\newblock
  doi:{\changeurlcolor{black}\href{https://doi.org/10.1002/andp.19183600402}{\detokenize{10.1002/andp.19183600402}}}.

\bibitem[{Weyl}(1918)]{Weyl1918}
{Weyl}, H.
\newblock {Reine Infinitesimalgeometrie}.
\newblock {\em Math. Z.} {\bf 1918}, {\em 2},~384--411.
\newblock
  doi:{\changeurlcolor{black}\href{https://doi.org/10.1007/BF01199420}{\detokenize{10.1007/BF01199420}}}.

\bibitem[{Jordan} \em{et~al.}(1960){Jordan}, {Ehlers}, and {Kundt}]{Jordan1960}
{Jordan}, P.; {Ehlers}, J.; {Kundt}, W.
\newblock {Strenge L\"{o}sungen der Feldgleichungen der Allgemeinen
  Relativit\"{a}tstheorie}.
\newblock {\em Akad. Wiss. Lit. Mainz, Abhandl. Math.-Nat. Kl.} {\bf 1960},
  {\em 2S},~21--105.

\bibitem[{Jordan} \em{et~al.}(2009){Jordan}, {Ehlers}, and {Kundt}]{Jordan2009}
{Jordan}, P.; {Ehlers}, J.; {Kundt}, W.
\newblock {Republication of: Exact solutions of the field equations of the
  general theory of relativity}.
\newblock {\em \gregr} {\bf 2009}, {\em 41},~2191--2280.
\newblock
  doi:{\changeurlcolor{black}\href{https://doi.org/10.1007/s10714-009-0869-8}{\detokenize{10.1007/s10714-009-0869-8}}}.

\bibitem[{Kundt} and {Tr\"{u}mper}(1962)]{Kundt1962}
{Kundt}, W.; {Tr\"{u}mper}, M.
\newblock {Beitr\"{a}ge zur Theorie der Gravitations-Strahlungsfelder. Strenge
  L\"{o}sungen der Feldgleichungen der Allgemeinen Relativit\"{a}tstheorie V}.
\newblock {\em Akad. Wiss. Lit. Mainz, Abhandl. Math.-Nat. Kl.} {\bf 1962},
  {\em 12}.

\bibitem[{Kundt} and {Tr{\"u}mper}(2016)]{Kundt2016}
{Kundt}, W.; {Tr{\"u}mper}, M.
\newblock {Republication of: Contributions to the theory of gravitational
  radiation fields. Exact solutions of the field equations of the general
  theory of relativity V}.
\newblock {\em \gregr} {\bf 2016}, {\em 48},~44.
\newblock
  doi:{\changeurlcolor{black}\href{https://doi.org/10.1007/s10714-015-2009-y}{\detokenize{10.1007/s10714-015-2009-y}}}.

\bibitem[{Tr{\"u}mper}(1964)]{Truemper1964}
{Tr{\"u}mper}, M.
\newblock {A Three Dimensional Formulation of the Bianchi Identities for Vacuum
  Gravitational Fields}. In {\em {Contributions to Actual Problems of General
  Relativity. With contributions from W. Beiglb\"{o}ck, K. Bichteler, W.
  Budich, W. Kundt, and M. Tr\"{u}mper}}; {Jordan, P.}., Ed.; US Air Force
  Report III (Contract AF 61 (052)-567),
  apps.dtic.mil/docs/citations/AD0455081,  1964; chapter~3, pp. 85--98.

\bibitem[{Tr{\"u}mper}(2021)]{Truemper2021}
{Tr{\"u}mper}, M.
\newblock {Republication of: A threedimensional formulation of the Bianchi
  identities for vacuum gravitational fields}.
\newblock {\em General Relativity and Gravitation} {\bf 2021}, {\em 53},~110.
\newblock
  doi:{\changeurlcolor{black}\href{https://doi.org/10.1007/s10714-021-02861-9}{\detokenize{10.1007/s10714-021-02861-9}}}.

\bibitem[{Pirani}(1962)]{Pirani1962a}
{Pirani}, F.A.E.
\newblock {Gravitational Radiation}. In {\em {Gravitation: An Introduction to
  Current Research}}; {Witten}, L., Ed.; New York: Wiley,  1962.

\bibitem[{Matte}(1953)]{Matte1953}
{Matte}, A.
\newblock {Sur De Nouvelles Solutions Oscillatoires Des Equations De La
  Gravitation}.
\newblock {\em Can. J. Math.} {\bf 1953}, {\em 5},~1--16.
\newblock
  doi:{\changeurlcolor{black}\href{https://doi.org/10.4153/CJM-1953-001-3}{\detokenize{10.4153/CJM-1953-001-3}}}.

\bibitem[{Pirani}(1957)]{Pirani1957}
{Pirani}, F.A.
\newblock {Invariant Formulation of Gravitational Radiation Theory}.
\newblock {\em \phrv} {\bf 1957}, {\em 105},~1089--1099.
\newblock
  doi:{\changeurlcolor{black}\href{https://doi.org/10.1103/PhysRev.105.1089}{\detokenize{10.1103/PhysRev.105.1089}}}.

\bibitem[{Bel}(1958{\natexlab{a}})]{Bel1958a}
{Bel}, L.
\newblock {Relativit\'{e} -- D\'{e}finition d'une densit\'{e} d'\'{e}nergie et
  d'un \'{e}tat de radiation totale g\'{e}n\'{e}ralis\'{e}e}.
\newblock {\em C.\,R.~Acad.~Sci.\,(Paris)} {\bf 1958}, {\em 246},~3015.

\bibitem[{Bel}(1958{\natexlab{b}})]{Bel1958b}
{Bel}, L.
\newblock {Relativit\'{e} -- Sur la Radiation gravitationnelle}.
\newblock {\em C.\,R.~Acad.~Sci.\,(Paris)} {\bf 1958}, {\em 247},~1094.

\bibitem[{Penrose}(1960)]{Penrose1960}
{Penrose}, R.
\newblock {A spinor approach to general relativity}.
\newblock {\em \anphy} {\bf 1960}, {\em 10},~171--201.
\newblock
  doi:{\changeurlcolor{black}\href{https://doi.org/10.1016/0003-4916(60)90021-X}{\detokenize{10.1016/0003-4916(60)90021-X}}}.

\bibitem[{Bel}(1962)]{Bel1962}
{Bel}, L.
\newblock {Radiation states and the problem of energy in general relativity}.
\newblock {\em Cahiers de Physique} {\bf 1962}, {\em 16},~59--80.

\bibitem[{Bel}(2000)]{Bel2000}
{Bel}, L.
\newblock {Radiation States and the Problem of Energy in General Relativity}.
\newblock {\em \gregr} {\bf 2000}, {\em 32},~2047--2078.
\newblock
  doi:{\changeurlcolor{black}\href{https://doi.org/10.1023/A:1001958805232}{\detokenize{10.1023/A:1001958805232}}}.

\bibitem[{Hawking}(1966)]{Hawking1966}
{Hawking}, S.W.
\newblock {Perturbations of an Expanding Universe}.
\newblock {\em \apj} {\bf 1966}, {\em 145},~544.
\newblock
  doi:{\changeurlcolor{black}\href{https://doi.org/10.1086/148793}{\detokenize{10.1086/148793}}}.

\bibitem[Ellis(1971)]{Ellis1971}
Ellis, G.F.R.
\newblock {Relativistic Cosmology}. In {\em {Proceedings of International
  School of Physics Enrico Fermi, Course 47:}}; {Sachs, R.K.}., Ed.; London:
  Academic Press,  1971; Vol.~47, pp. 104--182.

\bibitem[Ellis(1973)]{Ellis1973}
Ellis, G.F.R.
\newblock {Relativistic Cosmology}. In {\em Carg\'{e}se Lectures in Physics};
  {Schatzmann, E.}., Ed.; New York:Gordon and Breach,  1973.

\bibitem[{Ehlers}(1961)]{Ehlers1961}
{Ehlers}, J.
\newblock {Beitr\"{a}ge zur relativistischen Mechanik kontinuierlicher Medien}.
\newblock {\em Akad. Wiss. Lit. Mainz, Abhandl. Math.-Nat. Kl.} {\bf 1961},
  {\em 11},~792--837.

\bibitem[{Ehlers}(1993)]{Ehlers1993}
{Ehlers}, J.
\newblock {Contributions to the relativistic mechanics of continuous media}.
\newblock {\em \gregr} {\bf 1993}, {\em 25},~1225--1266.
\newblock
  doi:{\changeurlcolor{black}\href{https://doi.org/10.1007/BF00759031}{\detokenize{10.1007/BF00759031}}}.

\bibitem[{Heckmann} and {Sch{\"u}cking}(1955)]{Heckmann1955}
{Heckmann}, O.; {Sch{\"u}cking}, E.
\newblock {Bemerkungen zur Newtonschen Kosmologie. I. Mit 3 Textabbildungen in
  8 Einzeldarstellungen}.
\newblock {\em \zap} {\bf 1955}, {\em 38},~95.

\bibitem[{Heckmann} and {Sch{\"u}cking}(1956)]{Heckmann1956}
{Heckmann}, O.; {Sch{\"u}cking}, E.
\newblock {Bemerkungen zur Newtonschen Kosmologie. II}.
\newblock {\em \zap} {\bf 1956}, {\em 40},~81.

\bibitem[{Tr{\"u}mper}(1965)]{Truemper1965}
{Tr{\"u}mper}, M.
\newblock {On a Special Class of Type-I Gravitational Fields}.
\newblock {\em \jmp} {\bf 1965}, {\em 6},~584--589.
\newblock
  doi:{\changeurlcolor{black}\href{https://doi.org/10.1063/1.1704310}{\detokenize{10.1063/1.1704310}}}.

\bibitem[{Ehlers} \em{et~al.}(1968){Ehlers}, {Geren}, and {Sachs}]{Ehlers1968}
{Ehlers}, J.; {Geren}, P.; {Sachs}, R.K.
\newblock {Isotropic solutions of the Einstein-Liouville equations.}
\newblock {\em \jmp} {\bf 1968}, {\em 9},~1344--1349.
\newblock
  doi:{\changeurlcolor{black}\href{https://doi.org/10.1063/1.1664720}{\detokenize{10.1063/1.1664720}}}.

\bibitem[{Tr{\"u}mper}(1967)]{Truemper1967}
{Tr{\"u}mper}, M.
\newblock {Bemerkungen {\"u}ber scherungsfreie Str{\"o}mungen gravitierender
  Gase}.
\newblock {\em \zap} {\bf 1967}, {\em 66},~215.

\bibitem[{Ellis} and {Bruni}(1989)]{Ellis1989}
{Ellis}, G.F.R.; {Bruni}, M.
\newblock {Covariant and gauge-invariant approach to cosmological density
  fluctuations}.
\newblock {\em \prd} {\bf 1989}, {\em 40},~1804--1818.
\newblock
  doi:{\changeurlcolor{black}\href{https://doi.org/10.1103/PhysRevD.40.1804}{\detokenize{10.1103/PhysRevD.40.1804}}}.

\bibitem[{Maartens} \em{et~al.}(1999){Maartens}, {Gebbie}, and
  {Ellis}]{Maartens1999}
{Maartens}, R.; {Gebbie}, T.; {Ellis}, G.F.R.
\newblock {Cosmic microwave background anisotropies: Nonlinear dynamics}.
\newblock {\em \prd} {\bf 1999}, {\em 59},~083506,
  \href{https://arxiv.org/abs/astro-ph/9808163}{{\normalfont
  [arXiv:astro-ph/9808163]}}.
\newblock
  doi:{\changeurlcolor{black}\href{https://doi.org/10.1103/PhysRevD.59.083506}{\detokenize{10.1103/PhysRevD.59.083506}}}.

\bibitem[{Ellis} and {van Elst}(1999)]{Ellis1999a}
{Ellis}, G.F.R.; {van Elst}, H.
\newblock {Cosmological Models (Carg{\`e}se Lectures 1998)}.
\newblock  Theoretical and Observational Cosmology; {Lachi{\`e}ze-Rey}, M.,
  Ed.,  1999, Vol. 541, {\em NATO Advanced Study Institute (ASI) Series C}, pp.
  1--116,  \href{https://arxiv.org/abs/gr-qc/9812046}{{\normalfont
  [arXiv:gr-qc/9812046]}}.

\bibitem[{Ellis} and {Dunsby}(1997)]{Ellis1997}
{Ellis}, G.F.R.; {Dunsby}, P.K.S.
\newblock {Newtonian Evolution of the Weyl Tensor}.
\newblock {\em \apj} {\bf 1997}, {\em 479},~97--101,
  \href{https://arxiv.org/abs/astro-ph/9410001}{{\normalfont
  [arXiv:astro-ph/9410001]}}.
\newblock
  doi:{\changeurlcolor{black}\href{https://doi.org/10.1086/303839}{\detokenize{10.1086/303839}}}.

\bibitem[{Maartens} \em{et~al.}(1998){Maartens}, {Lesame}, and
  {Ellis}]{Maartens1998}
{Maartens}, R.; {Lesame}, W.M.; {Ellis}, G.F.R.
\newblock {Newtonian-like and anti-Newtonian universes}.
\newblock {\em \cqgra} {\bf 1998}, {\em 15},~1005--1017,
  \href{https://arxiv.org/abs/gr-qc/9802014}{{\normalfont
  [arXiv:gr-qc/9802014]}}.
\newblock
  doi:{\changeurlcolor{black}\href{https://doi.org/10.1088/0264-9381/15/4/021}{\detokenize{10.1088/0264-9381/15/4/021}}}.

\bibitem[{Maartens} and {Bassett}(1998)]{Maartens1998b}
{Maartens}, R.; {Bassett}, B.A.
\newblock {Gravito-electromagnetism}.
\newblock {\em \cqgra} {\bf 1998}, {\em 15},~705--717,
  \href{https://arxiv.org/abs/gr-qc/9704059}{{\normalfont
  [arXiv:gr-qc/9704059]}}.
\newblock
  doi:{\changeurlcolor{black}\href{https://doi.org/10.1088/0264-9381/15/3/018}{\detokenize{10.1088/0264-9381/15/3/018}}}.

\bibitem[{Maartens}(1997)]{Maartens1997b}
{Maartens}, R.
\newblock {Linearization instability of gravity waves?}
\newblock {\em \prd} {\bf 1997}, {\em 55},~463--467,
  \href{https://arxiv.org/abs/astro-ph/9609198}{{\normalfont
  [arXiv:astro-ph/9609198]}}.
\newblock
  doi:{\changeurlcolor{black}\href{https://doi.org/10.1103/PhysRevD.55.463}{\detokenize{10.1103/PhysRevD.55.463}}}.

\bibitem[{Raychaudhuri}(1957)]{Raychaudhuri1957}
{Raychaudhuri}, A.
\newblock {Relativistic and Newtonian Cosmology}.
\newblock {\em \zap} {\bf 1957}, {\em 43},~161.

\bibitem[{Ellis} \em{et~al.}(2012){Ellis}, {Maartens}, and
  {MacCallum}]{Ellis2012}
{Ellis}, G.F.R.; {Maartens}, R.; {MacCallum}, M.A.H.
\newblock {\em {Relativistic Cosmology}}; Cambridge, UK: Cambridge University
  Press,  2012.
\newblock
  doi:{\changeurlcolor{black}\href{https://doi.org/10.1017/CBO9781139014403}{\detokenize{10.1017/CBO9781139014403}}}.

\bibitem[{Lesame} \em{et~al.}(1996){Lesame}, {Ellis}, and {Dunsby}]{Lesame1996}
{Lesame}, W.M.; {Ellis}, G.F.R.; {Dunsby}, P.K.S.
\newblock {Irrotational dust with divH=0}.
\newblock {\em \prd} {\bf 1996}, {\em 53},~738--746,
  \href{https://arxiv.org/abs/gr-qc/9508049}{{\normalfont
  [arXiv:gr-qc/9508049]}}.
\newblock
  doi:{\changeurlcolor{black}\href{https://doi.org/10.1103/PhysRevD.53.738}{\detokenize{10.1103/PhysRevD.53.738}}}.

\bibitem[{Bruni} \em{et~al.}(1992){Bruni}, {Dunsby}, and {Ellis}]{Bruni1992}
{Bruni}, M.; {Dunsby}, P.K.S.; {Ellis}, G.F.R.
\newblock {Cosmological Perturbations and the Physical Meaning of
  Gauge-invariant Variables}.
\newblock {\em \apj} {\bf 1992}, {\em 395},~34.
\newblock
  doi:{\changeurlcolor{black}\href{https://doi.org/10.1086/171629}{\detokenize{10.1086/171629}}}.

\bibitem[{Maartens} \em{et~al.}(1997){Maartens}, {Ellis}, and
  {Siklos}]{Maartens1997c}
{Maartens}, R.; {Ellis}, G.F.R.; {Siklos}, S.T.C.
\newblock {Local freedom in the gravitational field}.
\newblock {\em \cqgra} {\bf 1997}, {\em 14},~1927--1936,
  \href{https://arxiv.org/abs/gr-qc/9611003}{{\normalfont
  [arXiv:gr-qc/9611003]}}.
\newblock
  doi:{\changeurlcolor{black}\href{https://doi.org/10.1088/0264-9381/14/7/025}{\detokenize{10.1088/0264-9381/14/7/025}}}.

\bibitem[{Ellis}(1967)]{Ellis1967}
{Ellis}, G.F.R.
\newblock {Dynamics of Pressure-Free Matter in General Relativity}.
\newblock {\em \jmp} {\bf 1967}, {\em 8},~1171--1194.
\newblock
  doi:{\changeurlcolor{black}\href{https://doi.org/10.1063/1.1705331}{\detokenize{10.1063/1.1705331}}}.

\bibitem[{Maartens} \em{et~al.}(1995){Maartens}, {Ellis}, and
  {Stoeger}]{Maartens1995}
{Maartens}, R.; {Ellis}, G.F.R.; {Stoeger}, W.R.
\newblock {Limits on anisotropy and inhomogeneity from the cosmic background
  radiation}.
\newblock {\em \prd} {\bf 1995}, {\em 51},~1525--1535,
  \href{https://arxiv.org/abs/astro-ph/9501016}{{\normalfont
  [arXiv:astro-ph/9501016]}}.
\newblock
  doi:{\changeurlcolor{black}\href{https://doi.org/10.1103/PhysRevD.51.1525}{\detokenize{10.1103/PhysRevD.51.1525}}}.

\bibitem[{Danehkar}(2009)]{Danehkar2009}
{Danehkar}, A.
\newblock {On the Significance of the Weyl Curvature in a Relativistic
  Cosmological Model}.
\newblock {\em \mpla} {\bf 2009}, {\em 24},~3113--3127,
  \href{https://arxiv.org/abs/0707.2987}{{\normalfont
  [arXiv:0707.2987]}}.
\newblock
  doi:{\changeurlcolor{black}\href{https://doi.org/10.1142/S0217732309032046}{\detokenize{10.1142/S0217732309032046}}}.

\bibitem[{Challinor} and {Lasenby}(1998)]{Challinor1998}
{Challinor}, A.; {Lasenby}, A.
\newblock {Covariant and gauge-invariant analysis of cosmic microwave
  background anisotropies from scalar perturbations}.
\newblock {\em \prd} {\bf 1998}, {\em 58},~023001,
  \href{https://arxiv.org/abs/astro-ph/9804150}{{\normalfont
  [arXiv:astro-ph/9804150]}}.
\newblock
  doi:{\changeurlcolor{black}\href{https://doi.org/10.1103/PhysRevD.58.023001}{\detokenize{10.1103/PhysRevD.58.023001}}}.

\bibitem[{Dunsby}(1997)]{Dunsby1997}
{Dunsby}, P.K.S.
\newblock {A fully covariant description of cosmic microwave background
  anisotropies}.
\newblock {\em \cqgra} {\bf 1997}, {\em 14},~3391--3405,
  \href{https://arxiv.org/abs/gr-qc/9707022}{{\normalfont
  [arXiv:gr-qc/9707022]}}.
\newblock
  doi:{\changeurlcolor{black}\href{https://doi.org/10.1088/0264-9381/14/12/021}{\detokenize{10.1088/0264-9381/14/12/021}}}.

\bibitem[{Maartens} and {Triginer}(1997)]{Maartens1997a}
{Maartens}, R.; {Triginer}, J.
\newblock {Density perturbations with relativistic thermodynamics}.
\newblock {\em \prd} {\bf 1997}, {\em 56},~4640--4650,
  \href{https://arxiv.org/abs/gr-qc/9707018}{{\normalfont
  [arXiv:gr-qc/9707018]}}.
\newblock
  doi:{\changeurlcolor{black}\href{https://doi.org/10.1103/PhysRevD.56.4640}{\detokenize{10.1103/PhysRevD.56.4640}}}.

\bibitem[{van Elst} and {Ellis}(1998)]{vanElst1998}
{van Elst}, H.; {Ellis}, G.F.R.
\newblock {Quasi-Newtonian dust cosmologies}.
\newblock {\em \cqgra} {\bf 1998}, {\em 15},~3545--3573,
  \href{https://arxiv.org/abs/gr-qc/9805087}{{\normalfont
  [arXiv:gr-qc/9805087]}}.
\newblock
  doi:{\changeurlcolor{black}\href{https://doi.org/10.1088/0264-9381/15/11/017}{\detokenize{10.1088/0264-9381/15/11/017}}}.

\bibitem[{van Elst}(1996)]{vanElst1996a}
{van Elst}, H.
\newblock {Extensions and applications of 1+3 decomposition methods in general
  relativistic cosmological modelling}.
\newblock PhD thesis, Astronomy Unit, Queen Mary and Westfield College,
  University of London,  1996.
\newblock
  doi:{\changeurlcolor{black}\href{https://doi.org/10.13140/RG.2.2.33294.77128}{\detokenize{10.13140/RG.2.2.33294.77128}}}.

\bibitem[{Ehlers}(1957)]{Ehlers1957}
{Ehlers}, J.
\newblock {Konstruktionen und Charakterisierungen von L\"{o}sungen der
  Einsteinschen Gravitationsfeldgleichungen}.
\newblock PhD thesis, Hamburg University,  1957.

\bibitem[{Ellis}(1964)]{Ellis1964}
{Ellis}, G.F.R.
\newblock {On general relativistic fluids and cosmological models}.
\newblock PhD thesis, St. John's College, University of Cambridge,  1964.

\bibitem[{Pirani}(1962)]{Pirani1962b}
{Pirani}, F.A.E.
\newblock {Survey of Gravitational Radiation Theory}. In {\em {Recent
  Developments in General Relativity}}; Oxford: Pergamon Press; Warsaw: Polish
  Scientific Publishers,  1962.

\bibitem[{Hawking} and {Ellis}(1973)]{Hawking1973}
{Hawking}, S.W.; {Ellis}, G.F.R.
\newblock {\em {The Large Scale Structure of Space-Time}}; Cambridge, UK:
  Cambridge University Press,  1973.

\bibitem[{Levi-Civita}(1927)]{Levi-Civita1927}
{Levi-Civita}, T.
\newblock {Sur l'\'{e}cart g\'{e}od\'{e}sique}.
\newblock {\em Math.~Ann.} {\bf 1927}, {\em 97},~291--320.
\newblock
  doi:{\changeurlcolor{black}\href{https://doi.org/10.1007/BF01447869}{\detokenize{10.1007/BF01447869}}}.

\bibitem[{Pirani}(1956)]{Pirani1956}
{Pirani}, F.A.E.
\newblock {On the Physical significance of the Riemann tensor}.
\newblock {\em Acta Physica Polonica} {\bf 1956}, {\em 15},~389--405.

\bibitem[{Szekeres}(1965)]{Szekeres1965}
{Szekeres}, P.
\newblock {The Gravitational Compass}.
\newblock {\em \jmp} {\bf 1965}, {\em 6},~1387--1391.
\newblock
  doi:{\changeurlcolor{black}\href{https://doi.org/10.1063/1.1704788}{\detokenize{10.1063/1.1704788}}}.

\bibitem[{Ellis} \em{et~al.}(1990){Ellis}, {Bruni}, and {Hwang}]{Ellis1990}
{Ellis}, G.F.R.; {Bruni}, M.; {Hwang}, J.
\newblock {Density-gradient-vorticity relation in perfect-fluid
  Robertson-Walker perturbations}.
\newblock {\em \prd} {\bf 1990}, {\em 42},~1035--1046.
\newblock
  doi:{\changeurlcolor{black}\href{https://doi.org/10.1103/PhysRevD.42.1035}{\detokenize{10.1103/PhysRevD.42.1035}}}.

\bibitem[{Zel'manov}(1956{\natexlab{a}})]{ZelManov1956a}
{Zel'manov}, A.L.
\newblock {Chronometric invariants and frames of reference in the general
  theory of relativity}.
\newblock {\em Dokl.~Akad.~Nauk.~USSR} {\bf 1956}, {\em 107},~805.

\bibitem[{Zel'manov}(1956{\natexlab{b}})]{ZelManov1956b}
{Zel'manov}, A.L.
\newblock {Chronometric Invariants and Frames of Reference in the General
  Theory of Relativity}.
\newblock {\em Sov.~Phys.~Doklady} {\bf 1956}, {\em 1},~227.

\bibitem[{Cattaneo-Gasperini}(1961)]{Cattaneo-Gasperini1961}
{Cattaneo-Gasperini}, I.
\newblock {G\'{e}om\'{e}trie -- Projections naturelles des tenseurs de courbure
  d'une vari\'{e}t\'{e} $V_{n+1}$ m\'{e}trique hyperbolique normale}.
\newblock {\em C.\,R.~Acad.~Sci.\,(Paris)} {\bf 1961}, {\em 252},~3722.

\bibitem[{Ferrarese}(1963)]{Ferrarese1963}
{Ferrarese}, G.
\newblock {Contributi alla tecnica delle proiezioni in una variet\`{a} a
  metrica iperbolica normale}.
\newblock {\em Rendic. di Matem.} {\bf 1963}, {\em 22},~147.

\bibitem[{Ferrarese}(1965)]{Ferrarese1965}
{Ferrarese}, G.
\newblock {Proj et\`{a} di secondo ordine di un generico riferimento fisico in
  Relativit\`{a} generale}.
\newblock {\em Rendic. di Matem.} {\bf 1965}, {\em 24},~57.

\bibitem[{Lottermoser}(1988)]{Lottermoser1988}
{Lottermoser}, M.
\newblock {\"{U}ber den Newtonschen Grenzwert der Allgemeinen
  Relativit\"{a}tstheorie und die relativistische Erweiterung Newtonscher
  Anfangsdaten}.
\newblock PhD thesis, Ludwig Maximilian University of Munich,  1988.

\bibitem[{Barrow} \em{et~al.}(2007){Barrow}, {Maartens}, and
  {Tsagas}]{Barrow2007}
{Barrow}, J.D.; {Maartens}, R.; {Tsagas}, C.G.
\newblock {Cosmology with inhomogeneous magnetic fields}.
\newblock {\em \physrep} {\bf 2007}, {\em 449},~131--171,
  \href{https://arxiv.org/abs/astro-ph/0611537}{{\normalfont
  [arXiv:astro-ph/0611537]}}.
\newblock
  doi:{\changeurlcolor{black}\href{https://doi.org/10.1016/j.physrep.2007.04.006}{\detokenize{10.1016/j.physrep.2007.04.006}}}.

\bibitem[{Wald}(1984)]{Wald1984}
{Wald}, R.M.
\newblock {\em {General Relativity}}; Chicago: University of Chicago Press,
  1984.

\bibitem[{Poisson}(2004)]{Poisson2004}
{Poisson}, E.
\newblock {\em {A Relativist's Toolkit : the Mathematics of Black-hole
  Mechanics}}; Cambridge, UK: Cambridge University Press,  2004.

\bibitem[{Friedmann}(1922)]{Friedmann1922}
{Friedmann}, A.
\newblock {{\"U}ber die Kr{\"u}mmung des Raumes}.
\newblock {\em Zeitschrift fur Physik} {\bf 1922}, {\em 10},~377--386.
\newblock
  doi:{\changeurlcolor{black}\href{https://doi.org/10.1007/BF01332580}{\detokenize{10.1007/BF01332580}}}.

\bibitem[{Friedmann}(1999)]{Friedmann1999}
{Friedmann}, A.
\newblock {On the Curvature of Space}.
\newblock {\em \gregr} {\bf 1999}, {\em 31},~1991.
\newblock
  doi:{\changeurlcolor{black}\href{https://doi.org/10.1023/A:1026751225741}{\detokenize{10.1023/A:1026751225741}}}.

\bibitem[{Clarkson}(2000)]{Clarkson2000}
{Clarkson}, C.A.
\newblock {On the Observational Characteristics of Inhomogeneous Cosmologies:
  Undermining the Cosmological Principle}.
\newblock PhD thesis, University of Glasgow,  2000.

\bibitem[{Stephani}(1967{\natexlab{a}})]{Stephani1967a}
{Stephani}, H.
\newblock {{\"U}ber L{\"o}sungen der Einsteinschen Feldgleichungen, die sich in
  einen f{\"u}nfdimensionalen flachen Raum einbetten lassen}.
\newblock {\em \cmaph} {\bf 1967}, {\em 4},~137--142.
\newblock
  doi:{\changeurlcolor{black}\href{https://doi.org/10.1007/BF01645757}{\detokenize{10.1007/BF01645757}}}.

\bibitem[{Stephani}(1967{\natexlab{b}})]{Stephani1967b}
{Stephani}, H.
\newblock {Konform flache Gravitationsfelder}.
\newblock {\em \cmaph} {\bf 1967}, {\em 5},~337--342.
\newblock
  doi:{\changeurlcolor{black}\href{https://doi.org/10.1007/BF01646448}{\detokenize{10.1007/BF01646448}}}.

\bibitem[{Krasinski}(1997)]{Krasinski1997}
{Krasinski}, A.
\newblock {\em {Inhomogeneous Cosmological Models}}; Cambridge, UK: Cambridge
  University Press,  1997.

\bibitem[{Raychaudhuri}(1955)]{Raychaudhuri1955}
{Raychaudhuri}, A.
\newblock {Relativistic Cosmology. I}.
\newblock {\em \phrv} {\bf 1955}, {\em 98},~1123--1126.
\newblock
  doi:{\changeurlcolor{black}\href{https://doi.org/10.1103/PhysRev.98.1123}{\detokenize{10.1103/PhysRev.98.1123}}}.

\bibitem[{Danehkar}(2020)]{Danehkar2020}
{Danehkar}, A.
\newblock {Gravitational fields of the magnetic-type}.
\newblock {\em \ijmpd} {\bf 2020}, {\em 29},~2043001,
  \href{https://arxiv.org/abs/2006.13287}{{\normalfont
  [arXiv:gr-qc/2006.13287]}}.
\newblock
  doi:{\changeurlcolor{black}\href{https://doi.org/10.1142/S0218271820430014}{\detokenize{10.1142/S0218271820430014}}}.

\bibitem[{Challinor} and {Lasenby}(1999)]{Challinor1999}
{Challinor}, A.; {Lasenby}, A.
\newblock {Cosmic Microwave Background Anisotropies in the Cold Dark Matter
  Model: A Covariant and Gauge-invariant Approach}.
\newblock {\em \apj} {\bf 1999}, {\em 513},~1--22,
  \href{https://arxiv.org/abs/astro-ph/9804301}{{\normalfont
  [arXiv:astro-ph/9804301]}}.
\newblock
  doi:{\changeurlcolor{black}\href{https://doi.org/10.1086/306841}{\detokenize{10.1086/306841}}}.

\bibitem[{Tsagas} \em{et~al.}(2008){Tsagas}, {Challinor}, and
  {Maartens}]{Tsagas2008}
{Tsagas}, C.G.; {Challinor}, A.; {Maartens}, R.
\newblock {Relativistic cosmology and large-scale structure}.
\newblock {\em \physrep} {\bf 2008}, {\em 465},~61--147,
  \href{https://arxiv.org/abs/0705.4397}{{\normalfont
  [arXiv:astro-ph/0705.4397]}}.
\newblock
  doi:{\changeurlcolor{black}\href{https://doi.org/10.1016/j.physrep.2008.03.003}{\detokenize{10.1016/j.physrep.2008.03.003}}}.

\bibitem[{Carloni} and {Dunsby}(2007)]{Carloni2007}
{Carloni}, S.; {Dunsby}, P.K.S.
\newblock {Evolution of tensor perturbations in scalar-tensor theories of
  gravity}.
\newblock {\em \prd} {\bf 2007}, {\em 75},~064012,
  \href{https://arxiv.org/abs/gr-qc/0612133}{{\normalfont
  [arXiv:gr-qc/0612133]}}.
\newblock
  doi:{\changeurlcolor{black}\href{https://doi.org/10.1103/PhysRevD.75.064012}{\detokenize{10.1103/PhysRevD.75.064012}}}.

\bibitem[{Challinor}(2000)]{Challinor2000}
{Challinor}, A.
\newblock {Microwave background anisotropies from gravitational waves: the 1 +
  3 covariant approach}.
\newblock {\em \cqgra} {\bf 2000}, {\em 17},~871--889,
  \href{https://arxiv.org/abs/astro-ph/9906474}{{\normalfont
  [arXiv:astro-ph/9906474]}}.
\newblock
  doi:{\changeurlcolor{black}\href{https://doi.org/10.1088/0264-9381/17/4/309}{\detokenize{10.1088/0264-9381/17/4/309}}}.

\bibitem[{Hogan} and {Ellis}(1997)]{Hogan1997}
{Hogan}, P.A.; {Ellis}, G.F.R.
\newblock {Propagation of information by electromagnetic and gravitational
  waves in cosmology}.
\newblock {\em \cqgra} {\bf 1997}, {\em 14},~A171--A188.
\newblock
  doi:{\changeurlcolor{black}\href{https://doi.org/10.1088/0264-9381/14/1A/015}{\detokenize{10.1088/0264-9381/14/1A/015}}}.

\bibitem[{Dunsby} \em{et~al.}(1997){Dunsby}, {Bassett}, and
  {Ellis}]{Dunsby1997a}
{Dunsby}, P.K.S.; {Bassett}, B.A.C.C.; {Ellis}, G.F.R.
\newblock {Covariant analysis of gravitational waves in a cosmological
  context}.
\newblock {\em \cqgra} {\bf 1997}, {\em 14},~1215--1222,
  \href{https://arxiv.org/abs/gr-qc/9811092}{{\normalfont
  [arXiv:gr-qc/9811092]}}.
\newblock
  doi:{\changeurlcolor{black}\href{https://doi.org/10.1088/0264-9381/14/5/023}{\detokenize{10.1088/0264-9381/14/5/023}}}.

\bibitem[{Ellis} and {Hogan}(1997)]{Ellis1997a}
{Ellis}, G.F.R.; {Hogan}, P.A.
\newblock {The Electromagnetic Analogue of Some Gravitational Perturbations in
  Cosmology}.
\newblock {\em \gregr} {\bf 1997}, {\em 29},~235--244.
\newblock
  doi:{\changeurlcolor{black}\href{https://doi.org/10.1023/A:1010244212803}{\detokenize{10.1023/A:1010244212803}}}.

\bibitem[{Bardeen}(1980)]{Bardeen1980}
{Bardeen}, J.M.
\newblock {Gauge-invariant cosmological perturbations}.
\newblock {\em \prd} {\bf 1980}, {\em 22},~1882--1905.
\newblock
  doi:{\changeurlcolor{black}\href{https://doi.org/10.1103/PhysRevD.22.1882}{\detokenize{10.1103/PhysRevD.22.1882}}}.

\bibitem[{Harrison}(1967)]{Harrison1967}
{Harrison}, E.R.
\newblock {Normal Modes of Vibrations of the Universe}.
\newblock {\em \rvmp} {\bf 1967}, {\em 39},~862--882.
\newblock
  doi:{\changeurlcolor{black}\href{https://doi.org/10.1103/RevModPhys.39.862}{\detokenize{10.1103/RevModPhys.39.862}}}.

\bibitem[{Danehkar}(2019)]{Danehkar2019}
{Danehkar}, A.
\newblock {Electric-magnetic duality in gravity and higher-spin fields}.
\newblock {\em Frontiers in Physics} {\bf 2019}, {\em 6},~146.
\newblock
  doi:{\changeurlcolor{black}\href{https://doi.org/10.3389/fphy.2018.00146}{\detokenize{10.3389/fphy.2018.00146}}}.

\bibitem[{van Elst} and {Ellis}(1996)]{vanElst1996}
{van Elst}, H.; {Ellis}, G.F.R.
\newblock {The covariant approach to LRS perfect fluid spacetime geometries}.
\newblock {\em \cqgra} {\bf 1996}, {\em 13},~1099--1127,
  \href{https://arxiv.org/abs/gr-qc/9510044}{{\normalfont
  [arXiv:gr-qc/9510044]}}.
\newblock
  doi:{\changeurlcolor{black}\href{https://doi.org/10.1088/0264-9381/13/5/023}{\detokenize{10.1088/0264-9381/13/5/023}}}.

\bibitem[{van Elst} \em{et~al.}(1997){van Elst}, {Uggla}, {Lesame}, {Ellis},
  and {Maartens}]{vanElst1997a}
{van Elst}, H.; {Uggla}, C.; {Lesame}, W.M.; {Ellis}, G.F.R.; {Maartens}, R.
\newblock {Integrability of irrotational silent cosmological models}.
\newblock {\em \cqgra} {\bf 1997}, {\em 14},~1151--1162,
  \href{https://arxiv.org/abs/gr-qc/9611002}{{\normalfont
  [arXiv:gr-qc/9611002]}}.
\newblock
  doi:{\changeurlcolor{black}\href{https://doi.org/10.1088/0264-9381/14/5/018}{\detokenize{10.1088/0264-9381/14/5/018}}}.

\bibitem[{Maartens} \em{et~al.}(1997){Maartens}, {Lesame}, and
  {Ellis}]{Maartens1997}
{Maartens}, R.; {Lesame}, W.M.; {Ellis}, G.F.R.
\newblock {Consistency of dust solutions with div H=0}.
\newblock {\em \prd} {\bf 1997}, {\em 55},~5219--5221,
  \href{https://arxiv.org/abs/gr-qc/9703080}{{\normalfont
  [arXiv:gr-qc/9703080]}}.
\newblock
  doi:{\changeurlcolor{black}\href{https://doi.org/10.1103/PhysRevD.55.5219}{\detokenize{10.1103/PhysRevD.55.5219}}}.

\bibitem[{Matarrese} \em{et~al.}(1994){Matarrese}, {Pantano}, and
  {Saez}]{Matarrese1994}
{Matarrese}, S.; {Pantano}, O.; {Saez}, D.
\newblock {General relativistic dynamics of irrotational dust: Cosmological
  implications}.
\newblock {\em \prl} {\bf 1994}, {\em 72},~320--323,
  \href{https://arxiv.org/abs/astro-ph/9310036}{{\normalfont
  [arXiv:astro-ph/9310036]}}.
\newblock
  doi:{\changeurlcolor{black}\href{https://doi.org/10.1103/PhysRevLett.72.320}{\detokenize{10.1103/PhysRevLett.72.320}}}.

\bibitem[{Croudace} \em{et~al.}(1994){Croudace}, {Parry}, {Salopek}, and
  {Stewart}]{Croudace1994}
{Croudace}, K.M.; {Parry}, J.; {Salopek}, D.S.; {Stewart}, J.M.
\newblock {Applying the Zel'dovich Approximation to General Relativity}.
\newblock {\em \apj} {\bf 1994}, {\em 423},~22.
\newblock
  doi:{\changeurlcolor{black}\href{https://doi.org/10.1086/173787}{\detokenize{10.1086/173787}}}.

\bibitem[{Bruni} \em{et~al.}(1995){Bruni}, {Matarrese}, and
  {Pantano}]{Bruni1995}
{Bruni}, M.; {Matarrese}, S.; {Pantano}, O.
\newblock {Dynamics of Silent Universes}.
\newblock {\em \apj} {\bf 1995}, {\em 445},~958,
  \href{https://arxiv.org/abs/astro-ph/9406068}{{\normalfont
  [arXiv:astro-ph/9406068]}}.
\newblock
  doi:{\changeurlcolor{black}\href{https://doi.org/10.1086/175755}{\detokenize{10.1086/175755}}}.

\bibitem[{Kofman} and {Pogosyan}(1995)]{Kofman1995}
{Kofman}, L.; {Pogosyan}, D.
\newblock {Dynamics of Gravitational Instability Is Nonlocal}.
\newblock {\em \apj} {\bf 1995}, {\em 442},~30,
  \href{https://arxiv.org/abs/astro-ph/9403029}{{\normalfont
  [arXiv:astro-ph/9403029]}}.
\newblock
  doi:{\changeurlcolor{black}\href{https://doi.org/10.1086/175419}{\detokenize{10.1086/175419}}}.

\bibitem[{Hui} and {Bertschinger}(1996)]{Hui1996}
{Hui}, L.; {Bertschinger}, E.
\newblock {Local Approximations to the Gravitational Collapse of Cold Matter}.
\newblock {\em \apj} {\bf 1996}, {\em 471},~1,
  \href{https://arxiv.org/abs/astro-ph/9508114}{{\normalfont
  [arXiv:astro-ph/9508114]}}.
\newblock
  doi:{\changeurlcolor{black}\href{https://doi.org/10.1086/177948}{\detokenize{10.1086/177948}}}.

\bibitem[{Heckmann} and {Sch{\"u}cking}(1959)]{Heckmann1959}
{Heckmann}, O.H.L.; {Sch{\"u}cking}, E.
\newblock {Newtonsche und Einsteinsche Kosmologie.}
\newblock {\em Handbuch der Physik} {\bf 1959}, {\em 53},~489.
\newblock
  doi:{\changeurlcolor{black}\href{https://doi.org/10.1007/978-3-642-45932-0\_13}{\detokenize{10.1007/978-3-642-45932-0\_13}}}.

\bibitem[{York}(1979)]{York1979}
{York}, J.~W., J.
\newblock {Kinematics and dynamics of general relativity}.
\newblock  {Sources of Gravitational Radiation}; {Smarr}, L.L., Ed. Cambridge:
  Cambridge University Press,  1979, pp. 83--126.

\bibitem[{Stephani} \em{et~al.}(2003){Stephani}, {Kramer}, {MacCallum},
  {Hoenselaers}, and {Herlt}]{Stephani2003}
{Stephani}, H.; {Kramer}, D.; {MacCallum}, M.; {Hoenselaers}, C.; {Herlt}, E.
\newblock {\em {Exact solutions of Einstein's field equations}}; Cambridge, UK:
  Cambridge University Press,  2003.
\newblock
  doi:{\changeurlcolor{black}\href{https://doi.org/10.1017/CBO9780511535185}{\detokenize{10.1017/CBO9780511535185}}}.

\bibitem[{Ehlers} and {Buchert}(2009)]{Ehlers2009}
{Ehlers}, J.; {Buchert}, T.
\newblock {On the Newtonian limit of the Weyl tensor}.
\newblock {\em \gregr} {\bf 2009}, {\em 41},~2153--2158,
  \href{https://arxiv.org/abs/0907.2645}{{\normalfont
  [arXiv:gr-qc/0907.2645]}}.
\newblock
  doi:{\changeurlcolor{black}\href{https://doi.org/10.1007/s10714-009-0855-1}{\detokenize{10.1007/s10714-009-0855-1}}}.

\bibitem[{Heckmann}(1961)]{Heckmann1961}
{Heckmann}, O.
\newblock {On the possible influence of a general rotation on the expansion of
  the universe}.
\newblock {\em \aj} {\bf 1961}, {\em 66},~599.
\newblock
  doi:{\changeurlcolor{black}\href{https://doi.org/10.1086/108470}{\detokenize{10.1086/108470}}}.

\bibitem[{Ellis} and {van Elst}(1999)]{Ellis1999}
{Ellis}, G.F.R.; {van Elst}, H.
\newblock {Deviation of Geodesics in FLRW Spacetime Geometries}.
\newblock  On Einstein's Path: essays in honor of Engelbert Schucking;
  {Harvey}, A., Ed. Berlin: Springer,  1999, p. 203,
  \href{https://arxiv.org/abs/gr-qc/9709060}{{\normalfont
  [arXiv:gr-qc/9709060]}}.
\newblock
  doi:{\changeurlcolor{black}\href{https://doi.org/10.1007/978-1-4612-1422-9_14}{\detokenize{10.1007/978-1-4612-1422-9_14}}}.

\bibitem[{Dadhich}(2000)]{Dadhich2000}
{Dadhich}, N.
\newblock {Electromagnetic Duality in General Relativity}.
\newblock {\em \gregr} {\bf 2000}, {\em 32},~1009--1023,
  \href{https://arxiv.org/abs/gr-qc/9909067}{{\normalfont
  [arXiv:gr-qc/9909067]}}.
\newblock
  doi:{\changeurlcolor{black}\href{https://doi.org/10.1023/A:1001913409254}{\detokenize{10.1023/A:1001913409254}}}.

\bibitem[{Tsagas}(2005)]{Tsagas2005}
{Tsagas}, C.G.
\newblock {Resonant amplification of magnetic seed fields by gravitational
  waves in the early universe}.
\newblock {\em \prd} {\bf 2005}, {\em 72},~123509,
  \href{https://arxiv.org/abs/astro-ph/0508556}{{\normalfont
  [arXiv:astro-ph/0508556]}}.
\newblock
  doi:{\changeurlcolor{black}\href{https://doi.org/10.1103/PhysRevD.72.123509}{\detokenize{10.1103/PhysRevD.72.123509}}}.

\bibitem[{Kodama} and {Sasaki}(1984)]{Kodama1984}
{Kodama}, H.; {Sasaki}, M.
\newblock {Cosmological Perturbation Theory}.
\newblock {\em Prog.~Theor.~Phys.~Sup.} {\bf 1984}, {\em 78},~1.
\newblock
  doi:{\changeurlcolor{black}\href{https://doi.org/10.1143/PTPS.78.1}{\detokenize{10.1143/PTPS.78.1}}}.

\bibitem[{Ellis} and {Tsagas}(2002)]{Ellis2002}
{Ellis}, G.F.; {Tsagas}, C.G.
\newblock {Relativistic approach to nonlinear peculiar velocities and the
  Zeldovich approximation}.
\newblock {\em \prd} {\bf 2002}, {\em 66},~124015,
  \href{https://arxiv.org/abs/astro-ph/0209143}{{\normalfont
  [arXiv:astro-ph/0209143]}}.
\newblock
  doi:{\changeurlcolor{black}\href{https://doi.org/10.1103/PhysRevD.66.124015}{\detokenize{10.1103/PhysRevD.66.124015}}}.

\bibitem[{King} and {Ellis}(1973)]{King1973}
{King}, A.R.; {Ellis}, G.F.R.
\newblock {Tilted homogeneous cosmological models}.
\newblock {\em \cmaph} {\bf 1973}, {\em 31},~209--242.
\newblock
  doi:{\changeurlcolor{black}\href{https://doi.org/10.1007/BF01646266}{\detokenize{10.1007/BF01646266}}}.

\bibitem[{Giovannini}(2005)]{Giovannini2005}
{Giovannini}, M.
\newblock {Cosmological perturbations for imperfect fluids}.
\newblock {\em \cqgra} {\bf 2005}, {\em 22},~5243--5269,
  \href{https://arxiv.org/abs/astro-ph/0504655}{{\normalfont
  [arXiv:astro-ph/0504655]}}.
\newblock
  doi:{\changeurlcolor{black}\href{https://doi.org/10.1088/0264-9381/22/24/004}{\detokenize{10.1088/0264-9381/22/24/004}}}.

\bibitem[{Maartens}(1998)]{Maartens1998a}
{Maartens}, R.
\newblock {Covariant velocity and density perturbations in quasi-Newtonian
  cosmologies}.
\newblock {\em \prd} {\bf 1998}, {\em 58},~124006,
  \href{https://arxiv.org/abs/astro-ph/9808235}{{\normalfont
  [arXiv:astro-ph/9808235]}}.
\newblock
  doi:{\changeurlcolor{black}\href{https://doi.org/10.1103/PhysRevD.58.124006}{\detokenize{10.1103/PhysRevD.58.124006}}}.

\bibitem[{Israel} and {Stewart}(1979)]{Israel1979}
{Israel}, W.; {Stewart}, J.M.
\newblock {Transient relativistic thermodynamics and kinetic theory}.
\newblock {\em \anphy} {\bf 1979}, {\em 118},~341--372.
\newblock
  doi:{\changeurlcolor{black}\href{https://doi.org/10.1016/0003-4916(79)90130-1}{\detokenize{10.1016/0003-4916(79)90130-1}}}.

\bibitem[{Ellis} and {MacCallum}(1969)]{Ellis1969}
{Ellis}, G.F.R.; {MacCallum}, M.A.H.
\newblock {A class of homogeneous cosmological models}.
\newblock {\em \cmaph} {\bf 1969}, {\em 12},~108--141.
\newblock
  doi:{\changeurlcolor{black}\href{https://doi.org/10.1007/BF01645908}{\detokenize{10.1007/BF01645908}}}.

\bibitem[{MacCallum} and {Ellis}(1970)]{MacCallum1970}
{MacCallum}, M.A.H.; {Ellis}, G.F.R.
\newblock {A class of homogeneous cosmological models: II. Observations}.
\newblock {\em \cmaph} {\bf 1970}, {\em 19},~31--64.
\newblock
  doi:{\changeurlcolor{black}\href{https://doi.org/10.1007/BF01645496}{\detokenize{10.1007/BF01645496}}}.

\bibitem[{MacCallum}(1971)]{MacCallum1971}
{MacCallum}, M.A.H.
\newblock {A class of homogeneous cosmological models III: Asymptotic
  behaviour}.
\newblock {\em \cmaph} {\bf 1971}, {\em 20},~57--84.
\newblock
  doi:{\changeurlcolor{black}\href{https://doi.org/10.1007/BF01646733}{\detokenize{10.1007/BF01646733}}}.

\bibitem[{Ellis} \em{et~al.}(2001){Ellis}, {van Elst}, and
  {Maartens}]{Ellis2001}
{Ellis}, G.F.R.; {van Elst}, H.; {Maartens}, R.
\newblock {General relativistic analysis of peculiar velocities}.
\newblock {\em \cqgra} {\bf 2001}, {\em 18},~5115--5123,
  \href{https://arxiv.org/abs/gr-qc/0105083}{{\normalfont
  [arXiv:gr-qc/0105083]}}.
\newblock
  doi:{\changeurlcolor{black}\href{https://doi.org/10.1088/0264-9381/18/23/308}{\detokenize{10.1088/0264-9381/18/23/308}}}.

\bibitem[{Maartens} and {Wolvaardt}(1994)]{Maartens1994}
{Maartens}, R.; {Wolvaardt}, F.P.
\newblock {Exact non-equilibrium solutions of the Einstein--Boltzmann
  equations}.
\newblock {\em \cqgra} {\bf 1994}, {\em 11},~203--225.
\newblock
  doi:{\changeurlcolor{black}\href{https://doi.org/10.1088/0264-9381/11/1/021}{\detokenize{10.1088/0264-9381/11/1/021}}}.

\bibitem[{Newman} and {Penrose}(1962)]{Newman1962}
{Newman}, E.; {Penrose}, R.
\newblock {An Approach to Gravitational Radiation by a Method of Spin
  Coefficients}.
\newblock {\em \jmp} {\bf 1962}, {\em 3},~566--578.
\newblock
  doi:{\changeurlcolor{black}\href{https://doi.org/10.1063/1.1724257}{\detokenize{10.1063/1.1724257}}}.

\bibitem[{Stewart} and {Ellis}(1968)]{Stewart1968}
{Stewart}, J.M.; {Ellis}, G.F.R.
\newblock {Solutions of Einstein's Equations for a Fluid Which Exhibit Local
  Rotational Symmetry}.
\newblock {\em \jmp} {\bf 1968}, {\em 9},~1072--1082.
\newblock
  doi:{\changeurlcolor{black}\href{https://doi.org/10.1063/1.1664679}{\detokenize{10.1063/1.1664679}}}.

\bibitem[{MacCallum}(1973)]{MacCallum1973}
{MacCallum}, M.A.H.
\newblock {Cosmological models from a geometric point of view}. In {\em
  Carg\'{e}se Lectures in Physics}; New York: Gordon and Breach,  1973; pp.
  61-174,  \href{https://arxiv.org/abs/2001.11387}{{\normalfont
  [arXiv:gr-qc/2001.11387]}}.

\bibitem[{MacCallum}(1998)]{MacCallum1998}
{MacCallum}, M.A.H.
\newblock {Integrability in Tetrad Formalisms and Conservation in Cosmology}.
\newblock  Current Topics in Mathematical Cosmology; {Rainer}, M.; {Schmidt},
  H.J., Eds.,  1998, p. 133,
  \href{https://arxiv.org/abs/gr-qc/9806003}{{\normalfont
  [arXiv:gr-qc/9806003]}}.

\bibitem[{Papapetrou}(1970)]{Papapetrou1970}
{Papapetrou}, A.
\newblock {Les \'equations suppl\'ementaires dans le formalisme de
  {Newman-Penrose}}.
\newblock {\em Annales de l'I.H.P. Physique th\'eorique} {\bf 1970}, {\em
  13},~271--286.

\bibitem[{Papapetrou}(1971{\natexlab{a}})]{Papapetrou1971a}
{Papapetrou}, A.
\newblock {Relativit\'{e} -- Quelques remarques sur le formalisme de
  Newman-Penrose}.
\newblock {\em C.\,R.~Acad.~Sci.\,(Paris): S\'{e}rie A} {\bf 1971}, {\em
  272},~1537.

\bibitem[{Papapetrou}(1971{\natexlab{b}})]{Papapetrou1971b}
{Papapetrou}, A.
\newblock {Relativit\'{e} -- Les relations identiques entre les \'{e}quation du
  formalisme de Newman-Penrose}.
\newblock {\em C.\,R.~Acad.~Sci.\,(Paris) S\'{e}rie A} {\bf 1971}, {\em
  272},~1613.

\bibitem[{Edgar}(1977)]{Edgar1977}
{Edgar}, S.B.
\newblock {A Comparison between the Flat Space and General Theory of
  Gravitation}.
\newblock PhD thesis, University of London,  1977.

\bibitem[{Edgar}(1980)]{Edgar1980}
{Edgar}, S.B.
\newblock {The structure of tetrad formalisms in general relativity: The
  general case}.
\newblock {\em \gregr} {\bf 1980}, {\em 12},~347--362.
\newblock
  doi:{\changeurlcolor{black}\href{https://doi.org/10.1007/BF00764473}{\detokenize{10.1007/BF00764473}}}.

\bibitem[{Wainwright} and {Ellis}(1997)]{Wainwright1997}
{Wainwright}, J.; {Ellis}, G.F.R., Eds.
\newblock {\em {Dynamical Systems in Cosmology}}. Cambridge: Cambridge
  University Press,  1997.

\bibitem[{de Felice} and {Clarke}(1992)]{deFelice1992}
{de Felice}, F.; {Clarke}, C.J.S.
\newblock {\em {Relativity on Curved Manifolds}}; Cambridge, UK: Cambridge
  University Press,  1992.

\bibitem[{Lesame} \em{et~al.}(1995){Lesame}, {Dunsby}, and {Ellis}]{Lesame1995}
{Lesame}, W.M.; {Dunsby}, P.K.S.; {Ellis}, G.F.R.
\newblock {Integrability conditions for irrotational dust with a purely
  electric Weyl tensor: A tetrad analysis}.
\newblock {\em \prd} {\bf 1995}, {\em 52},~3406--3415,
  \href{https://arxiv.org/abs/astro-ph/9410005}{{\normalfont
  [arXiv:astro-ph/9410005]}}.
\newblock
  doi:{\changeurlcolor{black}\href{https://doi.org/10.1103/PhysRevD.52.3406}{\detokenize{10.1103/PhysRevD.52.3406}}}.

\bibitem[{Velden}(1997)]{Velden1997}
{Velden}, T.
\newblock {Dynamics of pressure-free matter in general relativity}.
\newblock Master's thesis, University of Bielefeld /Albert Einstein Institute,
  AEI,  1997.

\bibitem[{van Elst} and {Uggla}(1997)]{vanElst1997}
{van Elst}, H.; {Uggla}, C.
\newblock {General relativistic 1+3 orthonormal frame approach}.
\newblock {\em \cqgra} {\bf 1997}, {\em 14},~2673--2695,
  \href{https://arxiv.org/abs/gr-qc/9603026}{{\normalfont
  [arXiv:gr-qc/9603026]}}.
\newblock
  doi:{\changeurlcolor{black}\href{https://doi.org/10.1088/0264-9381/14/9/021}{\detokenize{10.1088/0264-9381/14/9/021}}}.

\bibitem[{van Elst} and {Ellis}(1998)]{vanElst1998a}
{van Elst}, H.; {Ellis}, G.F.R.
\newblock {Causal propagation of geometrical fields in relativistic cosmology}.
\newblock {\em \prd} {\bf 1998}, {\em 59},~024013,
  \href{https://arxiv.org/abs/gr-qc/9810058}{{\normalfont
  [arXiv:gr-qc/9810058]}}.
\newblock
  doi:{\changeurlcolor{black}\href{https://doi.org/10.1103/PhysRevD.59.024013}{\detokenize{10.1103/PhysRevD.59.024013}}}.

\bibitem[{Schouten}(1954)]{Schouten1954}
{Schouten}, J.A.
\newblock {\em {Ricci-Calculus: An Introduction to Tensor Analysis and Its
  Geometrical Applications}}; Berlin: Springer,  1954.
\newblock
  doi:{\changeurlcolor{black}\href{https://doi.org/10.1007/978-3-662-12927-2}{\detokenize{10.1007/978-3-662-12927-2}}}.

\bibitem[{Heckmann} and {Sch{\"u}cking}(1962)]{Heckmann1962}
{Heckmann}, O.; {Sch{\"u}cking}, E.
\newblock In {\em {Gravitation: An Introduction to Current Research}};
  {Witten}, L., Ed.; New York: Wiley,  1962; chapter~11.

\bibitem[{Estabrook} and {Wahlquist}(1964)]{Estabrook1964}
{Estabrook}, F.B.; {Wahlquist}, H.D.
\newblock {Dyadic Analysis of Space-Time Congruences}.
\newblock {\em \jmp} {\bf 1964}, {\em 5},~1629--1644.
\newblock
  doi:{\changeurlcolor{black}\href{https://doi.org/10.1063/1.1931200}{\detokenize{10.1063/1.1931200}}}.

\bibitem[{Wahlquist} and {Estabrook}(1966)]{Wahlquist1966}
{Wahlquist}, H.D.; {Estabrook}, F.B.
\newblock {Rigid Motions in Einstein Spaces}.
\newblock {\em \jmp} {\bf 1966}, {\em 7},~894--905.
\newblock
  doi:{\changeurlcolor{black}\href{https://doi.org/10.1063/1.1931225}{\detokenize{10.1063/1.1931225}}}.

\bibitem[{Ellis} \em{et~al.}(1983{\natexlab{a}}){Ellis}, {Matravers}, and
  {Treciokas}]{Ellis1983a}
{Ellis}, G.F.R.; {Matravers}, D.R.; {Treciokas}, R.
\newblock {Anisotropic solutions of the Einstein-Boltzmann equations: I.
  General formalism}.
\newblock {\em \anphy} {\bf 1983}, {\em 150},~455--486.
\newblock
  doi:{\changeurlcolor{black}\href{https://doi.org/10.1016/0003-4916(83)90023-4}{\detokenize{10.1016/0003-4916(83)90023-4}}}.

\bibitem[{Ellis} \em{et~al.}(1983{\natexlab{b}}){Ellis}, {Treciokas}, and
  {Matravers}]{Ellis1983b}
{Ellis}, G.F.R.; {Treciokas}, R.; {Matravers}, D.R.
\newblock {Anisotropic solutions of the Einstein-Boltzmann equations. II. Some
  exact properties of the equations}.
\newblock {\em \anphy} {\bf 1983}, {\em 150},~487--503.
\newblock
  doi:{\changeurlcolor{black}\href{https://doi.org/10.1016/0003-4916(83)90024-6}{\detokenize{10.1016/0003-4916(83)90024-6}}}.

\bibitem[{Clarkson} and {Barrett}(2003)]{Clarkson2003}
{Clarkson}, C.A.; {Barrett}, R.K.
\newblock {Covariant perturbations of Schwarzschild black holes}.
\newblock {\em \cqgra} {\bf 2003}, {\em 20},~3855--3884,
  \href{https://arxiv.org/abs/gr-qc/0209051}{{\normalfont
  [arXiv:gr-qc/0209051]}}.
\newblock
  doi:{\changeurlcolor{black}\href{https://doi.org/10.1088/0264-9381/20/18/301}{\detokenize{10.1088/0264-9381/20/18/301}}}.

\bibitem[{Clarkson} \em{et~al.}(2004){Clarkson}, {Marklund}, {Betschart}, and
  {Dunsby}]{Clarkson2004}
{Clarkson}, C.A.; {Marklund}, M.; {Betschart}, G.; {Dunsby}, P.K.S.
\newblock {The Electromagnetic Signature of Black Hole Ring-Down}.
\newblock {\em \apj} {\bf 2004}, {\em 613},~492--505,
  \href{https://arxiv.org/abs/astro-ph/0310323}{{\normalfont
  [arXiv:astro-ph/0310323]}}.
\newblock
  doi:{\changeurlcolor{black}\href{https://doi.org/10.1086/422497}{\detokenize{10.1086/422497}}}.

\bibitem[{Betschart} and {Clarkson}(2004)]{Betschart2004}
{Betschart}, G.; {Clarkson}, C.A.
\newblock {Scalar field and electromagnetic perturbations on locally
  rotationally symmetric spacetimes}.
\newblock {\em \cqgra} {\bf 2004}, {\em 21},~5587--5607,
  \href{https://arxiv.org/abs/gr-qc/0404116}{{\normalfont
  [arXiv:gr-qc/0404116]}}.
\newblock
  doi:{\changeurlcolor{black}\href{https://doi.org/10.1088/0264-9381/21/23/018}{\detokenize{10.1088/0264-9381/21/23/018}}}.

\bibitem[{Clarkson}(2007)]{Clarkson2007}
{Clarkson}, C.
\newblock {Covariant approach for perturbations of rotationally symmetric
  spacetimes}.
\newblock {\em \prd} {\bf 2007}, {\em 76},~104034,
  \href{https://arxiv.org/abs/0708.1398}{{\normalfont
  [arXiv:gr-qc/0708.1398]}}.
\newblock
  doi:{\changeurlcolor{black}\href{https://doi.org/10.1103/PhysRevD.76.104034}{\detokenize{10.1103/PhysRevD.76.104034}}}.

\bibitem[{Greenberg}(1970)]{Greenberg1970}
{Greenberg}, P.J.
\newblock {The general theory of space-like congruences with an application to
  vorticity in relativistic hydrodynamics}.
\newblock {\em J.~Math.~Anal.~Appl.} {\bf 1970}, {\em 30},~128--143.
\newblock
  doi:{\changeurlcolor{black}\href{https://doi.org/10.1016/0022-247X(70)90188-5}{\detokenize{10.1016/0022-247X(70)90188-5}}}.

\bibitem[{Tsamparlis} and {Mason}(1983)]{Tsamparlis1983}
{Tsamparlis}, M.; {Mason}, D.P.
\newblock {On spacelike congruences in general relativity}.
\newblock {\em \jmp} {\bf 1983}, {\em 24},~1577--1593.
\newblock
  doi:{\changeurlcolor{black}\href{https://doi.org/10.1063/1.525852}{\detokenize{10.1063/1.525852}}}.

\bibitem[{Mason} and {Tsamparlis}(1985)]{Mason1985}
{Mason}, D.P.; {Tsamparlis}, M.
\newblock {Spacelike conformal Killing vectors and spacelike congruences}.
\newblock {\em \jmp} {\bf 1985}, {\em 26},~2881--2901.
\newblock
  doi:{\changeurlcolor{black}\href{https://doi.org/10.1063/1.526714}{\detokenize{10.1063/1.526714}}}.

\bibitem[{Tsamparlis}(1992)]{Tsamparlis1992}
{Tsamparlis}, M.
\newblock {Geometrization of a general collineation}.
\newblock {\em \jmp} {\bf 1992}, {\em 33},~1472--1479.
\newblock
  doi:{\changeurlcolor{black}\href{https://doi.org/10.1063/1.529724}{\detokenize{10.1063/1.529724}}}.

\bibitem[{Ellis} \em{et~al.}(1985){Ellis}, {Nel}, {Maartens}, {Stoeger}, and
  {Whitman}]{Ellis1985}
{Ellis}, G.F.R.; {Nel}, S.D.; {Maartens}, R.; {Stoeger}, W.R.; {Whitman}, A.P.
\newblock {Ideal observational cosmology.}
\newblock {\em \physrep} {\bf 1985}, {\em 124},~315--417.
\newblock
  doi:{\changeurlcolor{black}\href{https://doi.org/10.1016/0370-1573(85)90030-4}{\detokenize{10.1016/0370-1573(85)90030-4}}}.

\bibitem[{Stoeger} \em{et~al.}(1995){Stoeger}, {Maartens}, and
  {Ellis}]{Stoeger1995}
{Stoeger}, W.R.; {Maartens}, R.; {Ellis}, G.F.R.
\newblock {Proving Almost-Homogeneity of the Universe: an Almost
  Ehlers-Geren-Sachs Theorem}.
\newblock {\em \apj} {\bf 1995}, {\em 443},~1.
\newblock
  doi:{\changeurlcolor{black}\href{https://doi.org/10.1086/175496}{\detokenize{10.1086/175496}}}.

\bibitem[{Kristian} and {Sachs}(1966)]{Kristian1966}
{Kristian}, J.; {Sachs}, R.K.
\newblock {Observations in Cosmology}.
\newblock {\em \apj} {\bf 1966}, {\em 143},~379.
\newblock
  doi:{\changeurlcolor{black}\href{https://doi.org/10.1086/148522}{\detokenize{10.1086/148522}}}.

\bibitem[{Ehlers} and {Newman}(2000)]{Ehlers2000}
{Ehlers}, J.; {Newman}, E.T.
\newblock {The theory of caustics and wave front singularities with physical
  applications}.
\newblock {\em \jmp} {\bf 2000}, {\em 41},~3344--3378,
  \href{https://arxiv.org/abs/gr-qc/9906065}{{\normalfont
  [arXiv:gr-qc/9906065]}}.
\newblock
  doi:{\changeurlcolor{black}\href{https://doi.org/10.1063/1.533316}{\detokenize{10.1063/1.533316}}}.

\bibitem[{Jordan} \em{et~al.}(1961){Jordan}, {Ehlers}, and {Sachs}]{Jordan1961}
{Jordan}, P.; {Ehlers}, J.; {Sachs}, R.
\newblock {Beitr\"{a}ge zur Theorie der reinen Gravitationsstrahlung, Strenge
  L\"{o}sungen der Feldgleichungen der Allgemeinen Relativit\"{a}tstheorie II}.
\newblock {\em Akad. Wiss. Lit. Mainz, Abhandl. Math.-Nat. Kl.} {\bf 1961},
  {\em 1},~1--62.

\bibitem[{Jordan} \em{et~al.}(2013){Jordan}, {Ehlers}, and {Sachs}]{Jordan2013}
{Jordan}, P.; {Ehlers}, J.; {Sachs}, R.K.
\newblock {Republication of: Contributions to the theory of pure gravitational
  radiation. Exact solutions of the field equations of the general theory of
  relativity II}.
\newblock {\em \gregr} {\bf 2013}, {\em 45},~2691--2753.
\newblock
  doi:{\changeurlcolor{black}\href{https://doi.org/10.1007/s10714-013-1590-1}{\detokenize{10.1007/s10714-013-1590-1}}}.

\bibitem[{Schneider} \em{et~al.}(1992){Schneider}, {Ehlers}, and
  {Falco}]{Schneider1992}
{Schneider}, P.; {Ehlers}, J.; {Falco}, E.E.
\newblock {\em {Gravitational Lenses}}; Berlin: Springer-Verlag,  1992.
\newblock
  doi:{\changeurlcolor{black}\href{https://doi.org/10.1007/978-3-662-03758-4}{\detokenize{10.1007/978-3-662-03758-4}}}.

\bibitem[{Holz} and {Wald}(1998)]{Holz1998}
{Holz}, D.E.; {Wald}, R.M.
\newblock {New method for determining cumulative gravitational lensing effects
  in inhomogeneous universes}.
\newblock {\em \prd} {\bf 1998}, {\em 58},~063501,
  \href{https://arxiv.org/abs/astro-ph/9708036}{{\normalfont
  [arXiv:astro-ph/9708036]}}.
\newblock
  doi:{\changeurlcolor{black}\href{https://doi.org/10.1103/PhysRevD.58.063501}{\detokenize{10.1103/PhysRevD.58.063501}}}.

\bibitem[{Iorio}(2001)]{Iorio2001}
{Iorio}, L.
\newblock {Satellite Gravitational Orbital Perturbations and the
  Gravitomagnetic Clock Effect}.
\newblock {\em \ijmpd} {\bf 2001}, {\em 10},~465--476,
  \href{https://arxiv.org/abs/gr-qc/0007014}{{\normalfont
  [arXiv:gr-qc/0007014]}}.
\newblock
  doi:{\changeurlcolor{black}\href{https://doi.org/10.1142/S0218271801000925}{\detokenize{10.1142/S0218271801000925}}}.

\bibitem[{Mashhoon} \em{et~al.}(2001){Mashhoon}, {Iorio}, and
  {Lichtenegger}]{Mashhoon2001}
{Mashhoon}, B.; {Iorio}, L.; {Lichtenegger}, H.
\newblock {On the gravitomagnetic clock effect}.
\newblock {\em \phla} {\bf 2001}, {\em 292},~49--57,
  \href{https://arxiv.org/abs/gr-qc/0110055}{{\normalfont
  [arXiv:gr-qc/0110055]}}.
\newblock
  doi:{\changeurlcolor{black}\href{https://doi.org/10.1016/S0375-9601(01)00776-9}{\detokenize{10.1016/S0375-9601(01)00776-9}}}.

\bibitem[{Ruggiero} and {Tartaglia}(2002)]{Ruggiero2002}
{Ruggiero}, M.L.; {Tartaglia}, A.
\newblock {Gravitomagnetic effects}.
\newblock {\em Nuovo Cimento B} {\bf 2002}, {\em 117},~743,
  \href{https://arxiv.org/abs/gr-qc/0207065}{{\normalfont
  [arXiv:gr-qc/0207065]}}.

\bibitem[{Iorio} \em{et~al.}(2002){Iorio}, {Lichtenegger}, and
  {Mashhoon}]{Iorio2002}
{Iorio}, L.; {Lichtenegger}, H.; {Mashhoon}, B.
\newblock {An alternative derivation of the gravitomagnetic clock effect}.
\newblock {\em \cqgra} {\bf 2002}, {\em 19},~39--49,
  \href{https://arxiv.org/abs/gr-qc/0107002}{{\normalfont
  [arXiv:gr-qc/0107002]}}.
\newblock
  doi:{\changeurlcolor{black}\href{https://doi.org/10.1088/0264-9381/19/1/303}{\detokenize{10.1088/0264-9381/19/1/303}}}.

\bibitem[{Iorio} and {Lichtenegger}(2005)]{Iorio2005}
{Iorio}, L.; {Lichtenegger}, H.I.M.
\newblock {On the possibility of measuring the gravitomagnetic clock effect in
  an Earth space-based experiment}.
\newblock {\em \cqgra} {\bf 2005}, {\em 22},~119--132,
  \href{https://arxiv.org/abs/gr-qc/0210030}{{\normalfont
  [arXiv:gr-qc/0210030]}}.
\newblock
  doi:{\changeurlcolor{black}\href{https://doi.org/10.1088/0264-9381/22/1/008}{\detokenize{10.1088/0264-9381/22/1/008}}}.

\bibitem[{Lichtenegger} \em{et~al.}(2006){Lichtenegger}, {Iorio}, and
  {Mashhoon}]{Lichtenegger2006}
{Lichtenegger}, H.; {Iorio}, L.; {Mashhoon}, B.
\newblock {The gravitomagnetic clock effect and its possible observation}.
\newblock {\em \anp} {\bf 2006}, {\em 518},~868--876,
  \href{https://arxiv.org/abs/gr-qc/0211108}{{\normalfont
  [arXiv:gr-qc/0211108]}}.
\newblock
  doi:{\changeurlcolor{black}\href{https://doi.org/10.1002/andp.200610214}{\detokenize{10.1002/andp.200610214}}}.

\bibitem[{Braginskii} \em{et~al.}(1977){Braginskii}, {Caves}, and
  {Thorne}]{Braginskii1977}
{Braginskii}, V.B.; {Caves}, C.M.; {Thorne}, K.S.
\newblock {Laboratory experiments to test relativistic gravity}.
\newblock {\em \prd} {\bf 1977}, {\em 15},~2047--2068.
\newblock
  doi:{\changeurlcolor{black}\href{https://doi.org/10.1103/PhysRevD.15.2047}{\detokenize{10.1103/PhysRevD.15.2047}}}.

\bibitem[{Peng}(1983)]{Peng1983}
{Peng}, H.
\newblock {On calculation of magnetic-type gravitation and experiments}.
\newblock {\em \gregr} {\bf 1983}, {\em 15},~725--735.
\newblock
  doi:{\changeurlcolor{black}\href{https://doi.org/10.1007/BF01031880}{\detokenize{10.1007/BF01031880}}}.

\bibitem[{Mashhoon} \em{et~al.}(1989){Mashhoon}, {Paik}, and
  {Will}]{Mashhoon1989}
{Mashhoon}, B.; {Paik}, H.J.; {Will}, C.M.
\newblock {Detection of the gravitomagnetic field using an orbiting
  superconducting gravity gradiometer. Theoretical principles}.
\newblock {\em \prd} {\bf 1989}, {\em 39},~2825--2838.
\newblock
  doi:{\changeurlcolor{black}\href{https://doi.org/10.1103/PhysRevD.39.2825}{\detokenize{10.1103/PhysRevD.39.2825}}}.

\bibitem[{Peng}(1990)]{Peng1990}
{Peng}, H.
\newblock {A new approach to studying local gravitomagnetic effects on a
  superconductor.}
\newblock {\em \gregr} {\bf 1990}, {\em 22},~609--617.
\newblock
  doi:{\changeurlcolor{black}\href{https://doi.org/10.1007/BF00755981}{\detokenize{10.1007/BF00755981}}}.

\bibitem[{Li} and {Torr}(1991)]{Li1991}
{Li}, N.; {Torr}, D.G.
\newblock {Effects of a gravitomagnetic field on pure superconductors}.
\newblock {\em \prd} {\bf 1991}, {\em 43},~457--459.
\newblock
  doi:{\changeurlcolor{black}\href{https://doi.org/10.1103/PhysRevD.43.457}{\detokenize{10.1103/PhysRevD.43.457}}}.

\bibitem[{Tartaglia} and {Ruggiero}(2004)]{Tartaglia2004}
{Tartaglia}, A.; {Ruggiero}, M.L.
\newblock {Gravito-electromagnetism versus electromagnetism}.
\newblock {\em \ejph} {\bf 2004}, {\em 25},~203--210,
  \href{https://arxiv.org/abs/gr-qc/0311024}{{\normalfont
  [arXiv:gr-qc/0311024]}}.
\newblock
  doi:{\changeurlcolor{black}\href{https://doi.org/10.1088/0143-0807/25/2/007}{\detokenize{10.1088/0143-0807/25/2/007}}}.

\bibitem[{Ummarino} and {Gallerati}(2017)]{Ummarino2017}
{Ummarino}, G.A.; {Gallerati}, A.
\newblock {Superconductor in a weak static gravitational field}.
\newblock {\em \epjc} {\bf 2017}, {\em 77},~549,
  \href{https://arxiv.org/abs/1710.01267}{{\normalfont
  [arXiv:gr-qc/1710.01267]}}.
\newblock
  doi:{\changeurlcolor{black}\href{https://doi.org/10.1140/epjc/s10052-017-5116-y}{\detokenize{10.1140/epjc/s10052-017-5116-y}}}.

\bibitem[{Ummarino} and {Gallerati}(2020)]{Ummarino2020}
{Ummarino}, G.A.; {Gallerati}, A.
\newblock {Josephson AC effect induced by weak gravitational field}.
\newblock {\em \cqgra} {\bf 2020}, {\em 37},~217001,
  \href{https://arxiv.org/abs/2009.04967}{{\normalfont
  [arXiv:gr-qc/2009.04967]}}.
\newblock
  doi:{\changeurlcolor{black}\href{https://doi.org/10.1088/1361-6382/abb57b}{\detokenize{10.1088/1361-6382/abb57b}}}.

\bibitem[{Owen} \em{et~al.}(2011){Owen}, {Brink}, {Chen}, {Kaplan}, {Lovelace},
  {Matthews}, {Nichols}, {Scheel}, {Zhang}, {Zimmerman}, and
  {Thorne}]{Owen2011}
{Owen}, R.; {Brink}, J.; {Chen}, Y.; {Kaplan}, J.D.; {Lovelace}, G.;
  {Matthews}, K.D.; {Nichols}, D.A.; {Scheel}, M.A.; {Zhang}, F.; {Zimmerman},
  A.;  et~al.
\newblock {Frame-Dragging Vortexes and Tidal Tendexes Attached to Colliding
  Black Holes: Visualizing the Curvature of Spacetime}.
\newblock {\em \prl} {\bf 2011}, {\em 106},~151101,
  \href{https://arxiv.org/abs/1012.4869}{{\normalfont
  [arXiv:gr-qc/1012.4869]}}.
\newblock
  doi:{\changeurlcolor{black}\href{https://doi.org/10.1103/PhysRevLett.106.151101}{\detokenize{10.1103/PhysRevLett.106.151101}}}.

\bibitem[{Nichols} \em{et~al.}(2011){Nichols}, {Owen}, {Zhang}, {Zimmerman},
  {Brink}, {Chen}, {Kaplan}, {Lovelace}, {Matthews}, {Scheel}, and
  {Thorne}]{Nichols2011}
{Nichols}, D.A.; {Owen}, R.; {Zhang}, F.; {Zimmerman}, A.; {Brink}, J.; {Chen},
  Y.; {Kaplan}, J.D.; {Lovelace}, G.; {Matthews}, K.D.; {Scheel}, M.A.;  et~al.
\newblock {Visualizing spacetime curvature via frame-drag vortexes and tidal
  tendexes: General theory and weak-gravity applications}.
\newblock {\em \prd} {\bf 2011}, {\em 84},~124014,
  \href{https://arxiv.org/abs/1108.5486}{{\normalfont
  [arXiv:gr-qc/1108.5486]}}.
\newblock
  doi:{\changeurlcolor{black}\href{https://doi.org/10.1103/PhysRevD.84.124014}{\detokenize{10.1103/PhysRevD.84.124014}}}.

\bibitem[{Zhang} \em{et~al.}(2012){Zhang}, {Zimmerman}, {Nichols}, {Chen},
  {Lovelace}, {Matthews}, {Owen}, and {Thorne}]{Zhang2012}
{Zhang}, F.; {Zimmerman}, A.; {Nichols}, D.A.; {Chen}, Y.; {Lovelace}, G.;
  {Matthews}, K.D.; {Owen}, R.; {Thorne}, K.S.
\newblock {Visualizing spacetime curvature via frame-drag vortexes and tidal
  tendexes. II. Stationary black holes}.
\newblock {\em \prd} {\bf 2012}, {\em 86},~084049,
  \href{https://arxiv.org/abs/1208.3034}{{\normalfont
  [arXiv:gr-qc/1208.3034]}}.
\newblock
  doi:{\changeurlcolor{black}\href{https://doi.org/10.1103/PhysRevD.86.084049}{\detokenize{10.1103/PhysRevD.86.084049}}}.

\bibitem[{Nichols} \em{et~al.}(2012){Nichols}, {Zimmerman}, {Chen}, {Lovelace},
  {Matthews}, {Owen}, {Zhang}, and {Thorne}]{Nichols2012}
{Nichols}, D.A.; {Zimmerman}, A.; {Chen}, Y.; {Lovelace}, G.; {Matthews}, K.D.;
  {Owen}, R.; {Zhang}, F.; {Thorne}, K.S.
\newblock {Visualizing spacetime curvature via frame-drag vortexes and tidal
  tendexes. III. Quasinormal pulsations of Schwarzschild and Kerr black holes}.
\newblock {\em \prd} {\bf 2012}, {\em 86},~104028,
  \href{https://arxiv.org/abs/1208.3038}{{\normalfont
  [arXiv:gr-qc/1208.3038]}}.
\newblock
  doi:{\changeurlcolor{black}\href{https://doi.org/10.1103/PhysRevD.86.104028}{\detokenize{10.1103/PhysRevD.86.104028}}}.

\bibitem[{Thorne}(1980)]{Thorne1980}
{Thorne}, K.S.
\newblock {Multipole expansions of gravitational radiation}.
\newblock {\em \rvmp} {\bf 1980}, {\em 52},~299--340.
\newblock
  doi:{\changeurlcolor{black}\href{https://doi.org/10.1103/RevModPhys.52.299}{\detokenize{10.1103/RevModPhys.52.299}}}.

\bibitem[{Pirani}(1965)]{Pirani1965}
{Pirani}, F.A.E.
\newblock {Introduction to Gravitational Radiation Theory}. In {\em {Lectures
  on General Relativity}}; {Trautmann}, A.; {Pirani}, F.A.E.; {Bondi}, H.,
  Eds.; Brandeis Summer Institute in Theoretical Physics,  1965.

\bibitem[{Sachs}(1961)]{Sachs1961}
{Sachs}, R.
\newblock {Gravitational Waves in General Relativity. VI. The Outgoing
  Radiation Condition}.
\newblock {\em Proc.~R.~Soc.~London~Ser.~A} {\bf 1961}, {\em 264},~309--338.
\newblock
  doi:{\changeurlcolor{black}\href{https://doi.org/10.1098/rspa.1961.0202}{\detokenize{10.1098/rspa.1961.0202}}}.

\bibitem[{Gebbie} and {Ellis}(2000)]{Gebbie2000}
{Gebbie}, T.; {Ellis}, G.F.R.
\newblock {1+3 covariant cosmic microwave background anisotropies. I. Algebraic
  relations for mode and multipole expansions.}
\newblock {\em \anphy} {\bf 2000}, {\em 282},~285--320,
  \href{https://arxiv.org/abs/astro-ph/9804316}{{\normalfont
  [arXiv:astro-ph/9804316]}}.
\newblock
  doi:{\changeurlcolor{black}\href{https://doi.org/10.1006/aphy.2000.6033}{\detokenize{10.1006/aphy.2000.6033}}}.

\bibitem[{Treciokas}(1972)]{Treciokas1972}
{Treciokas}, R.
\newblock {Relativistic kinetic theory}.
\newblock PhD thesis, Department of Applied Mathematics and Theoretical
  Physics, University of Cambridge,  1972.

\end{thebibliography}
\end{document}